\documentclass[11pt, draftclsnofoot, onecolumn]{IEEEtran}

\usepackage{array}
\usepackage{textcomp}
\usepackage{stfloats}
\usepackage{url}
\usepackage{verbatim}
\usepackage{graphicx}
\usepackage{cite}
\usepackage{microtype}
\usepackage{graphicx}
\usepackage{subfigure}
\usepackage{booktabs} 
\usepackage{amsmath,amsfonts,amssymb,amsbsy,bm,paralist,theorem,ifthen,color}
\usepackage{algorithmic}
\usepackage{algorithm}
\usepackage{hyperref}
\usepackage{multirow}
\usepackage[table]{xcolor}
\usepackage{pifont}
\newcommand\Dc{\ensuremath{\mathcal{D}}}

\newcommand\Ac{\ensuremath{{\mathcal{A}}}}

\newcommand\Qc{\ensuremath{{\mathcal{Q}}}}
\newcommand\Oc{\ensuremath{{\mathcal{O}}}}

\newcommand\xb{\ensuremath{{\bf x}}}

\newcommand\yb{\ensuremath{{\bf y}}}

\newcommand\ab{\ensuremath{{\bf a}}}
\newcommand\Bb{\ensuremath{{\bf B}}}
\newcommand\bb{\ensuremath{{\bf b}}}

\newcommand\cb{\ensuremath{{\bf c}}}

\newcommand\Qbar{\ensuremath{{\bar Q}}}
\newcommand\Ebar{\ensuremath{{\bar E}}}

\newcommand\vb{\ensuremath{{\bf v}}}

\newcommand\zb{\ensuremath{{\bf z}}}

\newcommand\thetab{\ensuremath{{\bm \theta}}}

\newcommand\omegabar{\ensuremath{{\bar \omega}}}

\newcommand\zerob{\ensuremath{{\bm 0}}}

\newcommand\E{\ensuremath{{\mathbb{E}}}}

\newcommand\oneb{\ensuremath{{\bf 1}}}

\newcommand\Prob{\ensuremath{{\rm Prob}}}

\newcommand\Var{\ensuremath{{\rm Var}}}

\newcommand\Rbb{\ensuremath{{\mathbb{R}}}}
\newcommand\st{\ensuremath{{\rm s.t.}}}

\newcommand{\wt}{\widetilde}

\newcommand{\ol}{\overline}
\newcommand{\ul}{\underline}

\newtheorem{Lemma}{Lemma}

\newtheorem{Theorem}{Theorem}

\newtheorem{Corollary}{Corollary}

\newtheorem{Rmk}{Remark}
\newtheorem{assumption}{Assumption}

\begin{document}

\title{Communication-Efficient Federated Learning by Quantized Variance Reduction for Heterogeneous Wireless Edge Networks}

\author{Shuai Wang,~\IEEEmembership{Member,~IEEE}, 
        ~Yanqing Xu,~\IEEEmembership{Member,~IEEE},
        ~Chaoqun You,~\IEEEmembership{Member,~IEEE}, \\
        ~Mingjie Shao, ~\IEEEmembership{Member,~IEEE},
        ~and Tony Q. S. Quek,~\IEEEmembership{Fellow,~IEEE}

\thanks{Shuai Wang is with the National Key Laboratory of Wireless Communications,
University of Electronic Science and Technology of China, Chengdu,
611731, China (e-mail: shuaiwang@uestc.edu.cn).\protect}
\thanks{Yanqing Xu is with the Shenzhen Research Institute of Big Data and School of Science and Engineering, The Chinese University of Hong Kong, Shenzhen 518172, China (e-mail: xuyanqing@cuhk.edu.cn).}
\thanks{Chaoqun You is with the School of Computer Science and Technology, Fudan University, China 200433 (e-mail: chaoqunyou@gmail.com).}
\thanks{Mingjie Shao is with the Key Laboratory of Systems and Control, Institute of Systems Science, Academy of Mathematics and Systems Science (AMSS), Chinese Academy of Sciences (CAS), Beijing, 100149, China (email: mingjieshao@amss.ac.cn).}
\thanks{Tony Q. S. Quek is with the Information Systems Technology and Design, Singapore University of Technology and Design,
Singapore 487372 (e-mail: tonyquek@sutd.edu.sg).}
}



\maketitle

\begin{abstract}
Federated learning (FL) has been recognized as a viable solution for local-privacy-aware collaborative model training in wireless edge networks, but its practical deployment is hindered by the high communication overhead caused by frequent and costly server-device synchronization. Notably, most existing communication-efficient FL algorithms fail to reduce the significant inter-device variance resulting from the prevalent issue of device heterogeneity. This variance severely decelerates  algorithm convergence, increasing communication overhead and making it more challenging to achieve a well-performed model. In this paper, we propose a novel communication-efficient FL algorithm, named FedQVR, which relies on a sophisticated variance-reduced scheme to achieve heterogeneity-robustness in the presence of quantized transmission and heterogeneous local updates among active edge devices. Comprehensive theoretical analysis justifies that FedQVR is inherently resilient to device heterogeneity and has a comparable convergence rate even with a small number of quantization bits, yielding significant communication savings. Besides, considering non-ideal wireless channels, we propose FedQVR-E which enhances the convergence of FedQVR by performing joint allocation of bandwidth and quantization bits across devices under constrained transmission delays. Extensive experimental results are also presented to demonstrate the superior performance of the proposed algorithms over their counterparts in terms of both communication efficiency and application performance.
\end{abstract}


\section{Introduction}
\IEEEPARstart{R}{ecent} years have witnessed the unprecedented amount of sensitive data generated by network edge devices, such as smartphones and IoT devices, along with their growing computational capabilities in wireless edge networks \cite{FEDL_2019,ZhaoTWC2022,Age_MEC_2024}. This motivates a heightened need to harness the recent success of machine learning (ML) and artificial intelligence (AI) using these valuable data to power a wide range of intelligent applications for improving users' daily lives, while preserving data privacy\cite{Khan2021CST,Nguyen2021CST,FedCPSL_Shuai_2024, Shuai_FedMA_2021}. Federated learning (FL), as a local-privacy-aware distributed paradigm, is thus proposed and has recently gained prominence for training ML models in wireless edge networks \cite{FL_Ondevice_2016,FL_Beyond_2015,FL_MoblieEdge_2019,MA_CFL_2022}.
It enables the collaborative model learning over distributed data owned by massive edge devices under the orchestration of an edge server (e.g. base station) without the need of knowing the devices' private data. 

In detail, FL typically operates at the network edge and follows an iterative computation-and-communication procedure where the active devices repeatedly compute their local models and transmit them to the edge server which then updates the global model via aggregation and broadcasts it back\cite{Shuai_FedMA_2021}. This salient feature of on-device training while keeping data localized makes FL particularly relevant for mobile and intelligent applications in wireless edge networks \cite{Yang2020TCOM}. Nevertheless, the repetitive and costly synchronization of ML models between the edge server and edge devices in FL can cause significant communication cost, leading to inefficient model training as the communication is much more time-consuming than the computation \cite{FL_Ondevice_2016}. The problem is remarkably pronounced in wireless edge networks, presenting as a significant obstacle for the wide deployment of FL.

Two fundamental properties of wireless edge networks  contribute to the communication inefficiency of FL: prevalent device heterogeneity, and the mismatch between ML model size and the available communication resources. They respectively increase the number of communication rounds required for model training and the communication cost per round. First, the edge devices in wireless edge networks are usually heterogeneous, with non-i.i.d. local datasets  (data heterogeneity)  and  varied computational capabilities (system heterogeneity)\cite{FedProx_2018}\cite{wang2023batch}.
Ideally, if device heterogeneity is absent, one-shot FL can be achieved which only allows one communication round \cite{FedFisher_2024}. In contrast, the non-i.i.d data would cause large inter-device variance among the received local updates during the global aggregation step of FL algorithms, especially the typical FedAvg algorithm, due to the inconsistency among local objectives and optimization directions. It is acknowledged that the inter-device variance is the main reason for the device drift issue which yields significant slowdown of algorithm convergence and dramatic performance deterioration \cite{SCAFFOLD_2020}. In parallel, system heterogeneity results in large variations
in the number of local updates performed by each device\cite{FedProx_2018}, which can also cause inter-device variance, leading to solution bias and convergence slowdown \cite{FedNova_2020}. As a result, the communication burden is aggravated in the presence of device heterogeneity, as a substantial number of extra communication rounds are demanded for attaining a high-quality ML model. 

Second, the wireless links connecting the edge devices to the edge server are typically constrained in communication resources, such as bandwidth and transmit power, and are subject to various channel impairments, such as fading and shadowing. In contrast, the size of ML models exchanged in FL, which are often deep neural networks  containing millions or even billions of parameters \cite{FL_MoblieEdge_2019}, could be very large. Owing to the mismatch and the fact that the devices are usually geographically scattered over the network edge, the communication sources of some edge devices may not be sufficient to support the long-distance and fast transmission of large-size ML models over the wireless links. This naturally yields high communication cost and large transmission latency per round, and even transmission outage due to stringent delay constraints and large-scale channel fading \cite{FedToE2022}.

In response, considerable efforts have been devoted to the development of communication-efficient FL algorithms in wireless edge networks. In particular, one can mitigate the mismatch between the model size and communication resources by transmitting the incomplete (quantized or sparsified) model updates when synchronizing \cite{FedPAQ_2020, FedCOM_2021, UDC_FL_2021, AdaGQ_2023, DAdaQuant_2022, FedVQCS_2023, EP-GAMP_2023}, aiming to reduce the number of information bits used to represent the ML models. Another way is to exploit partial participation and device scheduling or communication resources allocation, optimizing resource consumption under tolerable constraints to mitigate this issue \cite{DeviceS_wenshi_2020, CSBA_LT_2021,ChenTWC2021, FL_URCCWN_2021, SP_FLWN_2020, CostE_FL_2021, AutomatedFL_You_2023}. Besides, the combination of these strategies also plays an important role in improving communication efficiency of FL \cite{FedToE2022, WQFL_2023, GQFL_2023}. However, the prior works are not communication-efficient due to two-fold reasons. To start with, they mostly rely on the FL algorithms, such as FedAvg, that fail to reduce the inter-device variance incurred by device heterogeneity \cite{FedAvg_noniid_2019, FedProx_2018}, having a high communication complexity. In addition, despite communication cost reduction per round, the communication saving strategies adopted by these works, such as model compression, may even increase the inter-device variance \cite{FedPAQ_2020}, compromising algorithm convergence. Their  inefficiency naturally arises as the number of communication rounds could matter more \cite{FL_MoblieEdge_2019}.

To address the data heterogeneity issue, recently, variance-reduced FL algorithms have been
proposed in the FL literature \cite{VRL_SGD_2020, SCAFFOLD_2020, FedDyn_2021, MIME_2021, FedVRA_Shuai_2023}. One of the most popular variance-reduced schemes is SCAFFOLD \cite{SCAFFOLD_2020} in which control variates are introduced to correct the local update direction so as to approximate the global one, thereby mitigating the device drift effect. It has been proven that SCAFFOLD can converge in significantly fewer rounds of communication than FedAvg \cite{SCAFFOLD_2020}. Nevertheless, these methods do not fully resolve the device heterogeneity issue as the system heterogeneity is overlooked. Moreover, SCAFFOLD doubles the communication cost per round for maintaining the capability of inter-device variance-reduction, and may perform poorly when partial participation is applied \cite{FedDyn_2021}. More recent works, such as FEDADAM\cite{AsyncFedOpt_2020} and FedCAMS \cite{FedCAMS_2022}, eliminate the need of double communication cost per round, but sacrifice their variance reduction capabilities.
Therefore, it is still an open problem how to achieve communication efficient FL in wireless edge networks.

\subsection{Contributions}

In this paper, we aim to develop a new client-variance-reduced FL algorithm that is communication-efficient while also providing rigorous theoretical guarantees in wireless edge networks. Specifically, recognizing the significance role of inter-device variance reduction in handing data heterogeneity, we are interested in developing a communication-efficient FL algorithm that are robust against the aforementioned  device heterogeneity, based on the variance-reduced schemes and communication saving strategies, and establishing solid convergence analysis. 

To this end, we propose a novel Federated Quantized Varianced Reduced algorithm, termed FedQVR. It is communication-efficient because it not only achieves robustness to device heterogeneity by owning the capability of inter-device variance reduction in the presence of both data heterogeneity and system heterogeneity, but also enjoys the advantage of communication cost reduction incurred by model quantization. Considering the practical server-device wireless environments, where limited radio resources and deep channel fading may prevent successful model exchange, more communication rounds are required to achieve a desired model, jepardizing  communication efficiency. Based on FedQVR, we also propose an enhanced FL algorithm, termed FedQVR-E, which cogitates on non-ideal wireless conditions.

The contributions of the paper can be expounded from the
following aspects.

\begin{itemize}
    \item We propose the FedQVR algorithm, by judiciously integrating the principles of inter-device variance reduction, quantized uplink transmission, and (time-varying) heterogeneous local updates (HLU) among devices, where the latter is a popular approach for handling the system heterogeneity \cite{FedProx_2018, FedNova_2020}. In particular, FedQVR reduces the overall communication cost by adopting  partial participation and quantized uplink transmission, while allowing edge devices to perform HLU across rounds based on their system constraints. In contrast to SCAFFOLD, FedQVR employs a novel and sophisticated inter-device variance reduction strategy to achieve  heterogeneity-robustness without increasing the communication cost per round (and even significantly decreasing it by quantization), leading to a communication-efficient solution in wireless edge networks.
    
    
    \item We provide a comprehensive theoretical analysis of the proposed FedQVR algorithm. To shed light on its superiority, we firstly demonstrate that, FedQVR owns the capability of inter-device variance reduction, but it is more robust in the presence of partial participation, large amounts of local computation, time-varying HLU, and quantized transmission, facilitating its fast convergence in heterogeneous wireless edge networks. Then, under mild conditions, we build the convergence properties of FedQVR, justifying that it is intrinsically resilient to device heterogeneity, and providing useful insights on how the quantization bits can affect the algorithm convergence. We remark that FedQVR has a sublinear convergence rate for nonconvex objectives, which outperforms or matches the state-of-the-art lower bounds, and enjoys significant communication savings as compared to its counterparts.

    \item 
    We propose the FedQVR-E algorithm based on FedQVR with the aim of improving the algorithm robustness against non-ideal wireless environment between the edge server and  devices. 
    In particular, by assuming a FDMA system setting, FedQVR-E tackles a joint bandwidth and quantization bits allocation problem among the active edge devices in each communication round, aiming to maximize the minimum number of quantization bits allocated to edge devices under the constraint of model transmission delay. This approach enhances the reliability of model exchanges, thereby improving the convergence of FedQVR in practical wireless conditions.
    \item To examine the performance of the proposed algorithms, we apply them to the image classification tasks on the CIFAR-10 and MNIST datasets. The experimental results demonstrate
that they outperform the baseline algorithms under various experimental settings. They not only enjoy a faster convergence and a better application performance, but also are significantly more communication-efficient than the baseline algorithms.

\end{itemize}

{\bf Synopsis}: Section \ref{sec: related_works} presents the related works. The system model and problem formulation are stated in Section \ref{sec: system_model}. Section \ref{sec: FedQVR} presents the proposed FedQVR algorithm and its theoretical analysis. Experiment results are analyzed in Section \ref{sec: simulation}, and conclusions are drawn in Section \ref{sec: conclusion}.

\section{Related works}
\label{sec: related_works}


{\bf Compressed FL:} To reduce the communication cost, one strand of literature has focused on reducing the amount of information bits for each transmission in which the
edge devices send compressed (quantified or sparsified) model parameters to the edge server in every round, aiming to achieve faster communication without heavily deteriorating the algorithm convergence \cite{FedPAQ_2020, FedCOM_2021, UDC_FL_2021, AdaGQ_2023, DAdaQuant_2022,FedVQCS_2023, EP-GAMP_2023}. In this regard, \cite{FedPAQ_2020} proposed a communication-efficient FL method, FedPAQ, which sends the quantized global model in the uplink communication, and then analyzed the effect of quantization error on the algorithm convergence. \cite{FedCOM_2021} proposed a compressed FL algorithm, FedCOM, with tighter convergence rates, while \cite{UDC_FL_2021} proposed a layered quantization strategy for FedAvg where layer-specific quantization levels are assigned to the communicated models. To boost the uplink compression, the works \cite{AdaGQ_2023, DAdaQuant_2022} adopted dynamically adaptive quantization levels across both communication rounds and clients and some works \cite{FedVQCS_2023, EP-GAMP_2023} pursued the combination of compressed sensing and quantization. Besides quantization, model sparsification has been used to reduce communication overhead in FL \cite{FastFL_2021, FedLamp_2023}. Although it is another valuable technique that can achieve similar communication cost reductions, in this paper, we primarily focus on stochastic quantization to provides an effective way to improve communication efficiency.

{\bf Resource allocation in FL:} To mitigate the communication bottleneck caused by the mismatch between communication resources and model size, various studies have focused on optimizing wireless resource scheduling strategies, such as bandwidth allocation, power control, and computation resource management, to accelerate algorithm convergence under resource constraints \cite{DeviceS_wenshi_2020, CSBA_LT_2021, ChenTWC2021, FL_URCCWN_2021, SP_FLWN_2020, CostE_FL_2021, AutomatedFL_You_2023}. For example, \cite{DeviceS_wenshi_2020} examined joint bandwidth allocation and device scheduling to maximize algorithm convergence rate. \cite{FL_URCCWN_2021} proposed a FL algorithm that accounts for unreliable uplink transmissions with bandwidth allocation policies. Similarly, \cite{ChenTWC2021} optimized transmit power, device selection, and bandwidth allocation to minimize FL training loss while considering device energy consumption. Unlike \cite{ChenTWC2021}, \cite{CSBA_LT_2021} explicitly considered resource allocation strategies under long-term energy constraints and uncertain wireless channels. Additionally, some studies \cite{BAA_FEL_2020} explored analog aggregation techniques to reduce uplink latency by allowing simultaneous model uploads over a multi-access wireless channel. Recent works \cite{FedToE2022, WQFL_2023, GQFL_2023} have also optimized wireless resource consumption alongside quantization error to improve communication efficiency. While these approaches are noteworthy, most rely on the FedAvg algorithm due to its simple-yet-good empirical performance. However, FedAvg is known to be inefficient in handling device heterogeneity, where the latter can significantly hinder algorithm convergence \cite{FedAvg_noniid_2019}.

{\bf Heterogeneity-robust FL:} 
To address device heterogeneity, numerous studies have modified the FedAvg algorithm to improve its robustness, thereby reducing the number of required communication rounds. These modifications include penalizing the distance between local and global models \cite{FedProx_2018, FedDyn_2021}, correcting local update directions \cite{VRL_SGD_2020, SCAFFOLD_2020, MIME_2021}, and implementing adaptive global aggregation \cite{AdaptiveFL_2021, FedCAMS_2022, FedNova_2020}. Specifically, to tackle data heterogeneity, many works, including \cite{VRL_SGD_2020, SCAFFOLD_2020, MIME_2021}, attempted to to mitigate the inter-client variance caused by non-i.i.d. data, thereby enhancing convergence. For instance, SCAFFOLD \cite{SCAFFOLD_2020} introduces two control variates to correct local update directions, approximating the global update direction. However, these variance-reduced approaches focus solely on non-i.i.d. data and double the communication cost per round as compared to FedAvg. Additionally, some studies \cite{AdaptiveFL_2021, FedCAMS_2022} introduce adaptive optimization into the global aggregation step, but they still struggle with non-i.i.d. data and overlook system heterogeneity.

To combat system heterogeneity, several works \cite{FedProx_2018, FedNova_2020} permit HLU  across devices, but suffer from slow convergence due to non-i.i.d. data. Recent methods like \cite{FedPD_2021, FedADMM_2022, FedDyn_2021} show resilience to both data and system heterogeneity, but their convergence depends on solving local subproblems to a high degree of accuracy, which is challenging and even infeasible in practice \cite{FedAvg_noniid_2019}. 
Therefore, it is still imperative to develop heterogeneity-robust FL algorithms which can achieve faster convergence in the presence of both data and system heterogeneity. More importantly, there are few attempts to improve communication efficiency by fully exploiting heterogeneity-robust FL, compressed transmission or resource allocation strategies, which is our key focus in this work.

\section{System Model and Problem formulation}
\label{sec: system_model}
Consider a wireless edge network  with an edge server and $N$ mobile devices for ML applications. These devices are geographically scattered around the edge server, and connect to it via wireless links with limited bandwidth and unreliable connections.  Each edge device $i$ owns a local dataset $\Dc_i$, which consists of the data generated or sensed by that device. Specifically, $\Dc_i = \{(\xb_{i, k}, \yb_{i, k})\}_{k = 1}^{n_i}$, where $(\xb_{i, k}, \yb_{i, k}) \in \Rbb^{S} \times \Rbb^C$ is the $k$-th input and output pair in $\Dc_i$. The size of $|\Dc_i|$ is $n_i$, i.e., $|\Dc_i| = n_i$.
The entire dataset is denoted as $\Dc = \cup_{i = 1}^{N} \Dc_i$ with size $ n = \sum_{i = 1}^{n} n_i$. It should be noted that the data generated or sensed by edge devices is usually biased  towards the associated applications or environments, and thus the local datasets, $\Dc_i, i \in [N]$ may follow different data distributions (non-i.i.d.) and their sizes $\{n_i\}_{i}^{N}$ could be unbalanced. These edge devices  may also have varied system constraints including different hardwares (CPU, GPU, and memory size), network configurations (wireless channels, bandwidth, and transmission power), battery states, and the like, resulting in significant variability in computational and transmission capabilities among them.

\subsection{Federated learning}
The objective of FL is to enable the edge server and the edge devices to collaboratively learn a shared and high-quality ML model for certain applications by utilizing the local datasets $\Dc_i, i \in [N]$ while keeping them localized. In particular, we denote $\thetab \in \Rbb^{d}$ as the shared model parameter and introduce the common loss function $f(\thetab; \xb_{i,k}, \yb_{i,k})$ which captures the error of the model $\thetab$ for the input-output data pair $(\xb_{i,k}, \yb_{i,k})$. Then, the total loss function of device $i$ is defined as
\begin{align} \label{eqn: local_loss}
    f_i(\thetab) \triangleq \frac{1}{n_i} \sum_{k = 1}^{n_i} f(\thetab; \xb_{i,k}, \yb_{i,k}). 
\end{align}
Note that the choice of the model $\thetab$ and the local loss $f_i$ depends on the target ML application \cite{FL_MoblieEdge_2019, FedCPSL_Shuai_2024}. For example, for a classification learning task, they respectively could be deep neural network and the cross-entropy loss. 

The whole FL process is to train the shared model parameter $\thetab$ for all edge devices that performs the best across all local datasets. The FL problem can thus be formulated as
\begin{align}\label{eqn: vanilla FL}
	\min_{\substack{\thetab \in \Rbb^d}}~&  f(\thetab) \triangleq \sum_{i = 1}^{N} p_i  f_i(\thetab),
\end{align}where $p_i = \frac{n_i}{n}$ is the weight coefficient associated with edge device $i$. To solve it, most FL algorithms follow the principle of model averaging (MA) which seeks the optimal $\thetab^\star$ in a privacy-preserving manner\cite{Parallel_RSGD_2019}. To be concrete, by introducing $\thetab_i$ for each  device $i$, problem \eqref{eqn: vanilla FL} is equivalently rewritten as 
\begin{align} \label{eqn: FL prob1}
	\min_{\substack{\xb, \xb_i,\\ i \in [N]}}~ \sum_{i = 1}^{N} p_i f_i(\thetab_i), ~~{\rm s.t.}&~ \thetab = \thetab_i,  i \in [N],
\end{align}
where $\thetab_i$ is the local model copy of $\thetab$ owned by edge device $i$ and $\thetab$ represents the global model located at the edge server. The equality constraints in \eqref{eqn: FL prob1} decouple the learning of local models $\thetab_i, i \in [N]$ and the global model $\thetab$ while ensuring that all edge devices and the edge server will finally share the same model. Following MA, the edge server periodically aggregates local models, which are respectively updated by the edge devices through approximately solving $\min f_i(\thetab_i)$ to a certain accuracy, ultimately reaching an accurate global model. 

\subsection{Review of FedAvg and SCAFFOLD}
The most typical FL algorithm in the literature is FedAvg \cite{FedAvg_noniid_2019} which was proposed based on MA while allowing partial participation and local stochastic gradient descent (SGD) \cite{LocalSGD_2019} to improve communication efficiency. In particular, FedAvg solves \eqref{eqn: FL prob1} by repeating the following steps round by round until convergence. That is, at round $r = 0, \ldots$,
\begin{itemize}
	\item {\bf Partial participation by device sampling:} A set of edge devices $\Ac^r$ ($|\Ac^r| = m \leq N$) is sampled  from $[N]$ and the edge server broadcasts the global model $\thetab^r$ to them.
	\item  {\bf Local update via multiple SGD steps:} Each device $i \in \Ac^r$ processes its own data to update local model $\thetab_i^{r+1}$ through $E \geq 1$ consecutive steps of local SGD with $\thetab^r$ as the initialization, and then upload it to the edge server.
	\item {\bf Global aggregation:} The edge server aggregates the received local models $\thetab_i^{r+1}$'s to generate the new global model  $\thetab^{r+1}$, i.e., $\thetab^{r+1} = \frac{1}{m}\sum_{i \in \Ac^r} \thetab_{i}^{r+1}$.
\end{itemize}

Although local SGD can reduce the number of communication rounds \cite{CE_DDNN_2017,Parallel_RSGD_2019}, and partial participation can alleviate the communication burden when the number of edge devices is large and communication bandwidth is limited \cite{FL_Ondevice_2016}, they only work effectively when the variability among local data distributions is highly restricted\cite{FedAvg_noniid_2019}. However, the data distributions of edge devices in wireless edge networks can be significantly differential. This non-i.i.d data would cause the device drift issue in FL. It has been proven \cite{SCAFFOLD_2020, FedDyn_2021} that such issue heavily slows down the convergence speed of FedAvg and is aggravated by partial participation. Besides, to guarantee algorithm convergence, the amount of local SGD steps should be small and carefully controlled \cite{FedAvg_noniid_2019, FedMAJ_2020}, which undermines the power of FedAvg in reducing communication cost. In addition to data heterogeneity, the heterogeneity in the computational speeds of edge devices results in the straggler problem \cite{FedProx_2018, FedDyn_2021}, which causes large model training delays in FedAvg. In short, FedAvg is quite inefficient in handling device heterogeneity, suffering from communication inefficiency.

SCAFFOLD was then proposed to mitigate the device drift issue by locally reducing the inter-device variance caused by non-i.i.d. data \cite{SCAFFOLD_2020}. To this end, it introduces two kinds of control variates $\cb$ and $\cb_i$ for the server and client $i$, respectively, and leverages them to correct the local update directions to approximate the global one. In particular, at round $r$, the participating device $i$ performs $E$ consecutive steps of perturbed SGD in \eqref{eqn: scaffold_xi} to obtains the local model $\thetab_{i}^{r+1}$, and updates $\cb_i$ via \eqref{eqn: scaffold_ci}. The global model $\thetab$ is updated in the same way as that in FedAvg and $\cb$ is updated by \eqref{eqn: scaffold_c} in the edge server.
\begin{align}
	&\thetab_i^{r, t+1} =  \thetab_i^{r, t} - \eta_l(g_i(\thetab_i^{r, t}) +\cb^r - \cb_i^r), ~0 \leq t \leq E - 1,\label{eqn: scaffold_xi}\\
	& \cb_{i}^{r+1} = \cb_{i}^r - \cb^r + \frac{1}{E \eta_l}(\thetab^r - \thetab_{i}^{r+1}), \label{eqn: scaffold_ci} \\
	&\cb^{r+1} = \cb^r + \frac{m}{N}(\cb_{i}^{r+1} - \cb_{i}^r),\label{eqn: scaffold_c} 
\end{align}
where $t$ denotes the local iteration index, $g_i(\thetab_i^{r,t})$ is the stochastic gradient (SG) of $f(\cdot)$ over $\thetab_i^{r,t}$ with a mini-batch of $S$ samples i.i.d drawn from $\Dc_i$, and $\eta_l > 0$ is the local learning rate. It can be derived from the update rules \eqref{eqn: scaffold_ci} and \eqref{eqn: scaffold_c} that $\cb_i^{r+1} = \frac{1}{Q}\sum_{t=0}^{Q - 1} g_i(\thetab_i^{r, t}), \forall i \in \Ac^r$ and $\cb^{r+1} = \frac{1}{N}\sum_{i=1}^{N}\cb_i^{r+1}$. That implies that, SCAFFOLD achieves inter-device variance reduction relying on that $\cb_i$ estimates the local update direction using a simple average of all SGs in the last active round while $\cb$ estimates the global one by averaging all $\cb_i$'s; see more details in \cite{SCAFFOLD_2020}. 

Despite its theoretical superiority, three principal drawbacks hinders SCAFFOLD from being a communication-efficient alternative to FedAvg. First, in contrast to FedAvg, SCAFFOLD requires transmitting back and forth both the change of local model and the control variate and thus communicate twice as many bits as FedAvg per communication round, which greatly increase the communication burden for wireless links connecting edge devices and the edge server. Second, SCAFFOLD exhibits a deficiency in robustness to non-i.i.d. data, especially when partial participation is employed. Recent studies 
\cite{AdaptiveFL_2021, FedVRA_Shuai_2023, MIME_2021} have shown that it may perform poorly in reducing the inter-device variance if a small fraction of edge devices is active per round. This is because the control variate $\cb_i$ of many devices in SCAFFOLD may frequently become stale, and thus the update  \eqref{eqn: scaffold_ci} yields inaccurate estimations to the local update direction, which jeopardizes its variance reduction capability. Third, it still does not take the system heterogeneity into account and certainly fails to tackle it for efficiency improvement.

\subsection{Quantized transmission} \label{sec: quantization}

In our system model, we consider using quantized transmission to relieve the high communication burden over wireless channels. Here we present the quantization scheme which will be adopted in our proposed FedQVR algorithm later. 

In particular, we follow the stochastic quantization method in \cite{FedToE2022, WQFL_2023, LFL_2020} and denote it by $\Qc(\zb, B)$ where $\zb \in \Rbb^d$ is any model parameter to be quantized and $B \geq 1$ is the number of quantization bits. In $\Qc(\zb, B)$, it is assumed that each element of $\zb$ is bounded, i.e., $\ol z \leq [\zb]_j  \leq \ul z, \forall j \in [d]$, where $\ol z \triangleq \max_{j} \{[\zb]_j\}$ and $\ul z \triangleq \min_{j} \{[\zb]_j\}$. The idea of $\Qc(\zb, B)$ is to uniformly divide the interval $[\ul z, \ol z]$ into several sub-intervals and quantize every element of $\zb$ using the "right" boundary of the sub-interval where the element falls into. That is, with $B$ quantization bits, the interval $[\ul z, \ol z]$ is divided into the following $K = 2^B - 1$ sub-intervals:
\begin{align}
    [c_0, c_1], [c_1, c_2], [c_2, c_3], \ldots, [c_{K-1}, c_K],
\end{align}where $c_k = \ul z + k(\ol z - \ul z) / (2^B - 1), k = 0, 1, \ldots, 2^B - 1$. Then, the $i$-th element of $\zb$ falling into the sub-interval $[c_{k-1}, c_k)$ will be quantized as
\begin{align}
    \Qc([\zb]_i, B) = \begin{cases}
	\text{sign}([\zb]_i) \cdot c_{k-1} & {\rm w.p.~} \frac{c_{k} - |[\zb|_i|}{c_{k} - c_{k-1}}, \\
	\text{sign}([\zb]_i) \cdot c_{k} & {\rm w.p.~} \frac{|[\zb|_i| - c_{k-1}}{c_{k} - c_{k-1}},
	\end{cases}
\end{align}where 'w.p.' stands for "with probability'. The quantized model parameter $\Qc(\zb, B)$ is $\Qc(\zb, B)= [\Qc([\zb]_1, B), \ldots, \Qc([\zb]_d, B)]$. In addition, the total number of bits used to represent the quantized model parameter $\Qc(\zb, B)$ is $d(B + 1) + \mu$, where each element is represented by $B$ bits plus one bit for its sign, and $\mu$ bits are needed to represent the bounds $\ol z$ and $\ul z$. 

\section{Proposed FedQVR Algorithm and Its Theoretical Analysis} \label{sec: FedQVR}

In this section, we present the proposed FedQVR algorithm, which is communication-efficient as it not only addresses the device heterogeneity issue without requiring extra transmissions but also utilize the quantized uplink transmission to further relieve the communication burden. We also provide a detailed explanation of FedQVR's robust inter-device variance reduction capability, demonstrating its effectiveness in handling device heterogeneity. 
Moreover, a rigorous theoretical analysis is presented to establish its convergence properties in the presence of device heterogeneity and quantization. 


\vspace{-0.25cm}
\subsection{Algorithm development}
In the development of FedQVR, we follow the MA principle, but introduce the control variates in the edge devices and the edge server with the aim of inter-device variance reduction, effectively combating data heterogeneity. Notably,  we go beyond it with carefully designed update rules to simultaneously account for the system heterogeneity and eliminate additional communication overhead.  In particular, the control variate $\cb_i$ is assigned to each device $i$ and $\cb$ is owned by the edge server. At the beginning of round $r$, we let the edge server uniformly sample a small set of $m$ devices $\Ac^r \subset [N]$ and broadcasts $\thetab_{0}^r = \thetab^r - \frac{1}{\gamma} \cb^r$ to all devices in $\Ac^r$. 
\begin{itemize}
	\item {\bf Local update}:  each client $i \in \Ac^r$ is asked to take $E_i^r$ consecutive steps of SGD, i.e. $\thetab_i^{r, 0} =\thetab_0^r, \thetab_i^{r+1} = \thetab_i^{r, E_i^r}$, and $\forall t = 0, \ldots, E_i^r - 1$, 
	\begin{align}		
		\thetab_i^{r, t+1} = \frac{1}{1 + \gamma \eta} (\thetab_i^{r,t} - \eta(g_i(\thetab_i^{r,t})- \cb_i^r)) +  \frac{ \gamma \eta}{1 + \gamma \eta} \thetab_0^r, \label{eqn: thetai_FedQVR}
	\end{align}where $\eta > 0$ is the stepsize. Then, we quantize the model difference $\thetab_i^{r+1} - \thetab_0^r$ to get $\Delta_i^{r+1} = \Qc(\thetab_i^{r+1} - \thetab_0^r, B_i^r)$, and update the control variate $\cb_i$ by
    \begin{align}
		\cb_i^{r+1} = \cb_i^{r} -\frac{a}{\eta \wt E_i^r}\Delta_i^{r+1}, \forall i \in \Ac^r, \label{eqn: dual_FedVRA}
	\end{align} where $\wt E_i^r \triangleq \frac{1}{\gamma\eta}\big(1- \frac{1}{(1 + \gamma\eta)^{E_i^r}}\big)$ and $a 
 \in (0, 1)$ is a pre-defined parameter. Note that $(\thetab_{i}, \cb_i)$ are set as $\thetab_{i}^{r+1} = \thetab_{0}^r, \cb_i^{r+1} = \cb_i^r, \forall i \notin \Ac^r$. Then, the variable $\Delta_i^{r+1}$ and the scalar $\frac{a}{\eta \wt E_i^r}$ are uploaded to the edge server. 
	\item {\bf Global aggregation}: after receiving $\thetab_i^{r+1} - \thetab_{0}^r$ and $ a / (\eta \wt E_i^r), i \in \Ac^r$, the edge server aggregates them to produce the new global model $\thetab^{r+1}$ and the server control variate $\cb^{r+1}$ by
	\begin{align}
		\cb^{r+1} =&~ \cb^r - \sum\limits_{i \in \Ac^r}  \frac{a p_i}{\eta \wt E_i^r}\Delta_i^{r+1}, \label{eqn: update_c_FedQVR}\\
		\thetab^{r+1} =&~ \thetab_0^r + \frac{N}{m} \sum\limits_{i \in \Ac^r} p_i\Delta_i^{r+1}. \label{eqn: update_theta_FedQVR} 
	\end{align}
\end{itemize}

The detailed steps of FedQVR are summarized in Algorithm \ref{alg: FedAVR}. The properties of FedQVR and its distinctions from SCAFFOLD are stated as follows, underscoring its role as a communication-efficient FL algorithm.

\begin{algorithm}[t!]
	\caption{Proposed FedQVR algorithm}
	\label{alg: FedAVR}
	\begin{algorithmic}[1]
		\STATE {\bfseries Input:} initial values of $\thetab_i^0 = \thetab^0$, $\cb_i^0 = \cb^0=\zerob,  \eta > 0, \gamma > 0, 1 > a > 0, B_i^r = B \geq 1, \forall i, r$.
		\FOR{round $r=0$ {\bfseries to} $R - 1$}
		\STATE {\bfseries \underline{Edge server side:}} sample a set of devices $\Ac^r$ ($|\Ac^r|=m$) randomly without replacement from $[N]$ and broadcast $\thetab_0^{r} = \thetab^r - \frac{1}{\gamma}\cb^r$ to the devices in $\Ac^r$.
		\STATE {\bfseries \underline{Edge device side:}}
		\FOR{device $i = 1$ {\bfseries to} $N$ in parallel}
		\IF{device $i \notin \Ac^r$} \STATE Set $\thetab_i^{r+1} = \thetab_0^r, \cb_i^{r+1} = \cb_i^r$.	
		\ELSE
		\STATE Set $\thetab_i^{r, 0} = \thetab_0^r$.
		\FOR{$t = 0$ {\bfseries to} $E_i^r - 1$}
		\STATE $\thetab_i^{r, t+1} = \frac{1}{1 + \gamma \eta} (\thetab_i^{r,t} - \eta(g_i(\thetab_i^{r,t})- \cb_i^r)) +  \frac{ \gamma \eta}{1 + \gamma \eta} \thetab_0^r$
		\ENDFOR
		\STATE Set $\thetab_i^{r+1} = \thetab_i^{r, E_i^r}$.
		\STATE Compute $\wt E_i^r = \frac{1}{\gamma\eta}\big(1- \frac{1}{(1 + \gamma\eta)^{E_i^r}}\big)$.
        \STATE Compute $ \Delta_i^{r+1} = \Qc(\thetab_i^{r+1} - \thetab_0^r, B_i^r)$.
		\STATE Compute $\cb_i^{r+1} = \cb_i^{r} -\frac{a}{\eta \wt E_i^r} \Delta_i^{r+1}$.
		\STATE Upload $\Delta_i^{r+1}$ and $a / (\eta \wt E_i^r)$ to the server.
		\ENDIF
		\ENDFOR
		\STATE {\bfseries \underline{Edge server side:}} Compute $\cb^{r+1}$ and $\thetab^{r+1}$  by
		\STATE ~~$\cb^{r+1} = \cb^r - \sum\limits_{i \in \Ac^r}  \frac{ap_i}{\eta \wt E_i^r}\Delta_i^{r+1}$,
		\STATE ~~$\thetab^{r+1} = \thetab_0^r + \frac{N}{m}\sum\limits_{i \in \Ac^r} p_i  \Delta_i^{r+1}$,
		\ENDFOR
		\STATE {\bfseries Output:} $\thetab^R$.
	\end{algorithmic}
\end{algorithm}


\begin{itemize}
    \item To deal with system heterogeneity, FedQVR allows heterogeneous local updates (HLU) by using a time-varying $E_i^r$ to denote the number of local SGD steps for device $i$ at round $r$. It should be remarked that the scheme of time-varying HLU across rounds is highly desired in wireless edge computing scenarios where low-latency ML applications are demanded. The benefit of such scheme is that each device can flexibility decide the amount of its local update steps according to its current system state in each round. This flexibility not only resolves the straggler problem in FL but also allows the participating devices with good system states to contribute more for fast convergence, thereby improving communication efficiency. Note that, in practice, the value of $E_i^r$ depends on the number of local data samples, the mini-batch size and the number of epochs, which are predetermined by the data and computational resources of the device $i$ \cite{FedNova_2020,FedProx_2018}.
    \item In contrast to SCAFFOLD, we introduce a new variable $\thetab_{0}$ in the edge server of FedQVR and update the local model $\thetab_{i}$ by a convex combination of the perturbed SGD and the new variable $\thetab_{0}$ at each communication round. The advantages of this alteration are two-fold. First, due to the fact that $\thetab_0^r = \thetab^r - \frac{1}{\gamma} \cb^r$, the update rule in \eqref{eqn: thetai_FedQVR} (line 11 of Algorithm \ref{alg: FedAVR}) yields the effect of inter-device variance reduction by utilizing the control variates $\cb_i^r$ and $\cb^r$ to correct the local update direction. While it seems similar to SCAFFOLD, FedQVR achieves inter-device variance reduction in a robust manner as it not only accommodates the stale devices resulting from partial participation but also accounts from time-varying HLU. This improves the convergence of FedQVR in the presence of device heterogeneity and will be demonstrated in detail in Section \ref{sec: robust_vr}. Second, unlike SCAFFOLD, the introduction of $\thetab_0$ by deducting the server control variate $\cb$ from the global model $\thetab$, eliminates the need of broadcasting the variable $\cb$ in FedQVR, making it  more communication-efficient than SCAFFOLD in the downlink channel.
    \item FedQVR applies the stochastic quantization scheme in Section \ref{sec: quantization} to the local update $\thetab^{r+1} - \thetab_0^r$ at every round $r$, further relieving the communication burden in the uplink channel\footnote{It is worth noting that we only consider the quantization in the uplink transmission since in practice the edge server (i.e., base station) is always powerful enough to provide fast and reliable communications for the downlink broadcast channels.}, which again enhances its communication efficiency. Instead of quantizing the local model, we quantize and then transmit the local model update, which has been shown \cite{FedToE2022} effective to lower down the quantization error due to algorithm convergence. We remark that the quantized local update is also used in the update of control variates $\cb_i^{r+1}$ and $\cb^{r+1}$ at every round $r$ (line 16 and 19 in Algorithm \ref{alg: FedAVR}), which is necessary to maintain the relation $\cb = \sum_{i=1}^{N}p_i\cb_i$ for inter-device variance reduction.
    \item In FedQVR, the control variates $\cb_i$ and $\cb$ are updated in a novel way. In particular, the control variate $\cb_i^{r+1}$ is updated with a new stepsize $a/(\eta \wt E_i^r)$. This novel stepsize enables $\cb_i$ generate a more accurate estimation of the local update direction through fine-grained exploitation of historical local SGs, and thus improves the robustness of FedQVR in dealing with device heterogeneity; see details in Section \ref{sec: robust_vr}. In addition, unlike SCAFFOLD, the control variate $\cb^{r+1}$ is obtained via \eqref{eqn: update_c_FedQVR} (line 21 of Algorithm \ref{alg: FedAVR}) without the need of $\cb_i^{r+1}$ or $\cb_i^{r+1} - \cb_i^r$. By this way, the device $i$ only needs to upload the stepsize $a/(\eta \wt E_i^r)$ to the server instead of $ \cb_i^{r+1}, \forall i \in \Ac^r$; see line 17 of Algorithm \ref{alg: FedAVR}. This again demonstrates the communication efficiency of FedQVR as it solely transmits a quantized local update and a scalar in the uplink channel at every round, resulting in much less communication cost per round than SCAFFOLD and FedAvg.  
\end{itemize}

\subsection{Robust inter-device variance reduction} \label{sec: robust_vr}

We now detailedly show that the proposed FedQVR algorithm indeed achieves inter-device variance reduction and further intuitively demonstrate its robustness against device heterogeneity. Before proceeding, let us make the following assumptions, which are standard and widely adopted in the FL literature, for ease of analysis.

\begin{assumption} \label{assumption: lower-bounded}
	Each local cost function $f_i(\cdot)$ in problem \eqref{eqn: vanilla FL} is lower bounded, i.e., $f_i(\thetab) \geq \underline{f} > {-\infty}$ and $L$-smooth, which implies $\|\nabla f_i(\thetab) - \nabla f_i(\thetab^\prime)\| \leq L\|\thetab - \thetab^\prime\|, \forall \thetab, \thetab^\prime \in \Rbb^d$.
\end{assumption}
\begin{assumption} \label{assumption: SGD_variance}
For a data sample $\xi_i$ uniformly sampled from $\Dc_i$, the resulting stochastic gradients (SG) for problem \eqref{eqn: vanilla FL} is unbiased and have bounded variances, i.e.,
\begin{align}
		& \E[\nabla f_i(\thetab_i; \xi_i)] = \E[\nabla f_i(\thetab_i)], \\
		&\E[\|\nabla f_i(\thetab_i; \xi_i) - \nabla f_i(\thetab_i)\|^2] \leq \sigma^2, 
	\end{align}where $\sigma > 0$ is a constant.
\end{assumption}

\begin{assumption} \label{assumption: quantization}
    For the stochastic quantization function $\Qc(,B_i^r)$ and the local model change $\thetab_i^{r+1} - \thetab_0^r$ at round $r$ of FedQVR, it holds that
    \begin{align}
       & \E[\Qc(\thetab_i^{r+1} - \thetab_0^r; B_i^r)] = \E[\thetab_i^{r+1} - \thetab_0^r], \\
     & \E[\|\Qc(\thetab_i^{r+1} - \thetab_0^r, B_i^r) - (\thetab_i^{r+1} - \thetab_0^r)\|^2] \notag \\
     \leq& \omega_i^r \E[\|\thetab_i^{r+1} - \thetab_0^r\|^2], \label{assumption: quan_bd}
    \end{align}where $0 \leq \omega_{i}^r \leq 1$.
\end{assumption}

Note that, by assumption \ref{assumption: SGD_variance}, the SG $g_i(\thetab_i)$ is unbiased with variance bounded by $\frac{\sigma^2}{S}$. We also remark that Assumption \ref{assumption: quantization} usually holds as it is known \cite{FedToE2022, WQFL_2023, LFL_2020} that, for $ \zb_i^r = \thetab_i^{r+1} - \thetab_0^r$, the quantization $\Qc(\zb_i^r, B_i^r)$ is an unbiased and contractive compressor with 
\begin{align}
    \omega_{i}^r = \frac{\sum_{j=1}^{d}(\max\{[\zb_i^r]_k\} - \min\{[\zb_i^r]_k\})}{4(2^{B_i^r} - 1)\|\zb_i^r\|^2}. \label{eqn: quan_error}
\end{align}
The value of $\omega_i^r$ depends on not only the variance among the elements of $\thetab_i^{r+1} - \thetab_0^r$ but also the number of quantization bits used to represent each of its elements. For a given model update $ \thetab_i^{r+1} - \thetab_0^r$, the inequality \eqref{assumption: quan_bd} indicates that the quantization error decreases with the increase of $B_i^r$ (the decrease of $\omega_i^r$), vanishes as $B_i^r \rightarrow \infty$ ($\omega_i^r \rightarrow 0$). 

To show FedQVR's capability of inter-device variance reduction, we delve into the updates of $\thetab_i$ and $\cb_i$ with the help of $\thetab_0$, which yields Lemma \ref{lem: local_update}. The proof can be found in the supplementary material.

\begin{Lemma} \label{lem: local_update}
	For any round $r \geq 0$ and device $i \in \Ac^r$, it holds that $\forall t = 0, \ldots, E_i^r - 1$,
	\begin{align}
		&\thetab_i^{r, t+1}
		=  \frac{1}{1 + \gamma\eta} (\thetab_i^{r, t} - \eta( g_i(\thetab_i^{r,t}) + \cb^r - \cb_i^{r})) + \frac{\gamma\eta}{1 + \gamma\eta} \thetab^r, \label{lem: local_update_xi}\\
		&\E[\cb_i^{r+1}]
		= (1 - a)\E[\cb_i^r] + a \E[G_i^r], \cb^{r+1} = \sum_{i=1}^{N}p_i \cb_i^{r+1}, \label{lem: local_update_dual}
	\end{align}
	where $\bb_i^r \triangleq [b_i^{r, 0}, b_i^{r, 1}, \ldots, b_i^{r, E_i^r-1}]^\top \in \Rbb^{E_i^r}$, $b_i^{r, t} =(\frac{1}{1 + \gamma \eta})^{E_i^r - t}$, $G_i^r = \sum_{t= 0}^{E_i^r - 1}\frac{b_i^{r, t} }{\|\bb_i^r\|_1} g_i(\thetab_i^{r,t})$.
\end{Lemma}

\begin{Lemma} \label{lem: opt_bound1}
	For any round $r \geq 0$ and device i, if $0 < a < \min\{\frac{1}{\omega_i^r}, 1\}$, then it holds that
	\begin{align}
		&\E[\|f_i(\thetab^{r+1}) - \cb_i^{r+1}\|^2] \notag \\
  \leq & \bigg(1 - \frac{ma(1-a\omega_i^r)}{2N - ma(1-a\omega_i^r)}\bigg) \E[\|f_i(\thetab^{r}) - \cb_i^{r}\|^2]  \notag \\
& + \frac{2NL^2}{ma(1-a\omega_i^r)} \E[\|\thetab^{r+1} - \thetab^r\|^2] \notag \\
&+ \frac{2m a(a\omega_i^r+1)L^2}{2N - ma(1-a\omega_i^r)}\E\bigg[\sum_{t= 0}^{E_i^r - 1}\frac{b_i^{r,t}}{\|\bb_i^r\|_1}\| \wt \thetab_i^{r,t} - \thetab^r\|^2\bigg] \notag \\
&+ \frac{2(1 + \omega_i^r)ma^2\sigma^2\|\bb_i^r\|_2^2}{(2N - ma(1-a\omega_i^r))S\|\bb_i^r\|_1^2}, \label{lem: opt_bda}
\end{align} where $\thetab_i^{r,t}$ is a virtual model generated by assuming that device $i$ is active at round $r$, and $\bb_i^r$ is defined in Lemma \ref{lem: local_update}.
\end{Lemma}

One can see from \eqref{lem: local_update_dual} that, on expectation, the control variate $\cb_i^{r+1} $ accumulates the historical normalized averaging SGs $G_i^r$. As $\cb = \sum_{i = 1}^N p_i \cb_i$, the control variate $\cb^r$ stands for the accumulation of the global normalized averaging SG $\sum_{i=1}^{N} p_iG_i^r$. Thus, $\cb_{i}$ attempts to estimate the local update direction through accumulating previous local gradients while $\cb$ attempts to estimate the global update direction by averaging all $\cb_i$‘s, which shares a similar spirit as the inter-device variance reduction strategy in SCAFFOLD. 

To further elaborate this point, we present Lemma \ref{lem: opt_bound1} which characterizes the dynamics of the estimation error of $\cb_i$ to the local update direction as algorithm proceeds. In detail, Lemma \ref{lem: opt_bound1} tells that the estimation error of $\cb_i$ is intimately related to the variance of the global update, the local model drift, the quantization error and the SG noise. More importantly, the estimation error will gradually decrease under mild conditions. To gain the insight, one can simplify 
the inequality \eqref{lem: opt_bda} as 
\begin{align}
    &\E[\|f_i(\thetab^{r+1}) - \cb_i^{r+1}\|^2] \notag \\
    \leq&  (1 - \xi_i) \E[\|f_i(\thetab^{r}) - \cb_i^{r}\|^2] + H_i^r, \label{eqn: approx_bd}
\end{align}where $ 0 < \frac{ma(1-a\omega_i^r)}{2N - ma(1-a\omega_i^r)} < \xi_i < 1$ and  $H_i^r$ is the summation of the other terms in the right hand size (RHS) of \ref{lem: opt_bda}. Intuitively, as FedQVR approaches a stationary point of problem \eqref{eqn: vanilla FL} with a small enough learning rate $\eta$ and sufficient large mini-batch size $S$, we obtain a small value of $H_i^r$. Thus, by \eqref{eqn: approx_bd}, the estimation error will decrease as $r$ increases, which confirms the approximation effect of $\cb_i$ to the local update direction. In view of this, one can also conclude that the term $ \cb^r - \cb_i^{r}$ in \eqref{lem: local_update_xi} acts as a gradient correction term so that 
$g_i(\xb_i^{r,t}) + \cb^r - \cb_i^{r}$ approximates the global update direction. This implies that FedQVR actually reduces the inter-device variance caused by device heterogeneity through sophisticated local and global update rules.

It is worth-noting that the inter-device variance reduction in FedQVR is inherently robust in the sense that it really accommodates the key challenges for inter-device variance reduction including partial participation, large amounts of local computation in the presence of non-i.i.d. data, time-varying HLU, and quantized transmission, thereby facilitating fast convergence and communication efficiency in wireless edge networks. Specifically, these challenges would lead to incorrect approximations of $\cb_i^r$ to the local update direction in existing variance reduced schemes, e.g. SCAFFOLD, but FedQVR is capable of addressing them. On one hand, as indicated by \eqref{lem: local_update_xi} in Lemma \ref{lem: local_update}, one-step update of $\thetab_i^{r, t+1}$ in FedQVR consists of a perturbed  SGD (corrected by $ \cb^r - \cb_i^{r}$) and the ensuing convex combination with the global model $\thetab^r$. Such an update rule can benefit the stable and fast convergence of FedCVR by preventing the local model from deviating too much from the global one. This is particularly advantageous when  $\cb^r$  and $\cb_{i}^r$ have incorrect approximations and thus the perturbed SGD goes out of the way to the global update direction \cite{FedProx_2018}.


On the other hand, FedQVR adopts a novel "two-layer" accumulation of all historical local SGs for the update of $\cb_{i}$, which again improves the robustness of FedQVR. In particular, one can see from \eqref{lem: local_update_dual} in Lemma \ref{lem: local_update} that the update of $\cb_{i}^{r+1}$ considers the exponentially weighted average of local SGs in all previous active rounds. This is different from SCAFFOLD which treats all local SGs fairly and merely utilize the local SGs of the last active round in the update of $\cb_i$. We point out that the control variate $\cb_i$ in FedQVR is more likely to produce a good approximation to the local update direction. To understand this, let us firstly set $a = 0$, and then $\cb_i^{r+1} = G_i^r$. In $G_i^r$, the SGs of different iteration $t$'s own different weights and more fresh SGs (that are closer to the iteration $E_i^r$ of round $r$) contribute more, which makes sense as fresh SGs contains more information of each round $r$. As $\gamma\eta \approx 0$, $G_i^r$ becomes the simple average and $\cb_{i}^{r+1}$ reduces to that in SCAFFOLD. In addition, if $a \ne 0$, the update of $\cb^{r+1}$ considers both the current normalized averaging SGs of round $r$ and the historical ones, and the parameter $a$ controls the weight of the former relative to the latter (which are hidden in $\cb_{i}^r$). Obviously, this novel strategy benefits the approximation capability of $\cb_i$ through a flexible and advanced exploitation of all historical local SGs.

\subsection{Convergence analysis}
The following theorem delineates the convergence conditions for the proposed FedQVR algorithm. The proof can be found in the supplementary material.

\begin{Theorem}\label{thm: FedQVR}
Suppose that the parameters $\gamma, \eta$ and $a$ satisfy $\forall r, 0 < a < \min\{\frac{1}{\omegabar^r}, 1\}$, 
\begin{small}
    \begin{align}
        &\eta \leq \min\bigg\{\frac{1}{2\gamma\Ebar^r\sqrt{N(1+\omegabar^r)}}, \frac{m\bar a\sqrt{m}}{3\sqrt{(1+\omegabar^r)N(2N - m\bar a)}}\bigg\}, \label{thm: cond1}\\
        & \gamma \geq \max\bigg\{8L, L\sqrt{\frac{30(a\omegabar^r+3)}{(1-a\omegabar^r)} - 4}, \frac{2L\sqrt{N(2N-m\bar a)}}{m\bar a} \bigg\}, \label{thm: cond2}
    \end{align}
    \end{small}where $\Ebar^r \triangleq \max\limits_{i} \wt E_i^r$, $\omegabar^r \triangleq \max\limits_{i} \omega_i^r$, and $\bar a \triangleq a(1-a\omegabar^r)$. Then, under Assumption  \ref{assumption: lower-bounded} and \ref{assumption: SGD_variance}, we have
\begin{align}
&\frac{1}{R}\sum_{r= 0}^{R-1}	\E[\|\nabla f(\thetab^r)\|^2]  \notag \\
\leq & \frac{4\gamma(P^0 - \ul f)}{R} + \frac{51  a\sigma^2 }{4SR }\sum_{r = 0}^{R - 1}\sum_{i=1}^{N} \frac{p_i(1+\omega_i^r)\|\bb_i^r\|_2^2}{(1-a\omegabar^r) \|\bb_i^r\|_1^2} \notag 
\end{align}
\begin{align}
  &+ \frac{2\gamma \eta^2\sigma^2 L}{mSR}\sum_{r = 0}^{R}\sum_{i=0}^{N}p_i (1+ \omega_i^r) \|\bb_i^r\|_2^2\notag \\
  & + \frac{51N^2\eta^2\sigma^2L^2}{2m^3a^2SR}\sum_{r = 0}^{R}\sum_{i=0}^{N} \frac{p_i(1+ \omega_i^r)\|\bb_i^r\|_2^2 }{(1-a\omegabar^r)^2} \notag \\
&+ \frac{\gamma^2 L^2 \eta^2\sigma^2}{8SR}\sum_{r = 0}^{R - 1}\sum_{i=1}^{N}\frac{p_i(59\omegabar^r + 145)\|\bb_i^r\|_2^2}{1-a\omegabar^r}, \label{thm:  bd6}
\end{align}
where $P^0 \triangleq f(\xb^0) + \frac{C_0^0}{N}p_i\sum_{i=1}^{N}\|\nabla f_i(\xb^0) - \cb_{i}^0\|^2$; the term $C_0^r$ is independent of $R$ and $S$, and is defined in the supplementary material.
\end{Theorem}

As shown in \eqref{thm:  bd6} of Theorem \ref{thm: FedQVR}, the upper bound of $\frac{1}{R}\sum_{r= 0}^{R-1}	\E[\|\nabla f(\thetab^r)\|^2]$ comprises five parts. The first term is directly related to the total number of communication rounds $R$ and is vanishing as $R \rightarrow \infty$. All remaining items in the RHS of \eqref{thm:  bd6} are attributed to the variance of local SGs $\sigma^2$, and can vanish to 0 if $\sigma = 0$, where full gradient is applied in the local update of FedQVR. It can also be seen that these terms intimately relate to the local stepsize $\eta$, the parameter $a$, the mini-batch size $S$, or the quantization error related parameters $\omega_i^r, \forall i,r$.  


\begin{Rmk}
    It is noteworthy that the convergence bound in \eqref{thm:  bd6} is obtained without relying on any assumptions regarding device heterogeneity. Moreover, none of the terms in the RHS of \eqref{thm:  bd6} are related to the challenge of device heterogeneity. Therefore, one can conclude from Theorem \ref{thm: FedQVR} that the proposed FedQVR algorithm is resilient to both heterogeneous local data distributions and HLU among edge devices. 
\end{Rmk}

\begin{Rmk}
    The convergence bound in \eqref{thm:  bd6} depends on the quantization error related parameters $\omega_i^r, \forall i,r$, and thus the convergence of FedQVR can be  influenced by the quantization levels adopted by the edge devices. In particular, one can observe  that the parameters $\omega_i^r, \forall i,r$ exist in all terms in the RHS of \eqref{thm:  bd6} except the first one. More importantly, a larger $\omegabar^r$ can increase the value of these terms, thereby deteriorating the algorithm convergence. This suggests that, for any round $r$, the increase of the minimum quantization bits among edge devices, i.e. $\min\limits_{i} B_i^r$, can boost the convergence of FedQVR, 
    because of \eqref{eqn: quan_error}.

\end{Rmk}

Theorem \ref{thm: FedQVR} implies that the convergence rate of FedQVR is determined by the total number of communication rounds $R$, the step size $\eta$ and the mini-batch size $S$. This is because the terms in the RHS of \eqref{thm:  bd6} can be made arbitrarily small or gradually decrease as the algorithm proceeds by adjusting these parameters. As suggested by Corollary \ref{corly: uniform_conv_rate}, with carefully chosen $S$, the proposed FedQVR algorithm achieves a sublinear convergence rate, which matches the result for general non-convex FL\cite{SCAFFOLD_2020}\cite{AdaptiveFL_2021}. Moreover, Corollary \ref{corly: com_complexity} presents the communication complexity of FedQVR and the corresponding communication cost for achieving $\epsilon$-accuracy to problem \eqref{eqn: vanilla FL}, and it is proved in the supplementary material.

\begin{Corollary} \label{corly: uniform_conv_rate}
	Given $S = \sqrt{R}$, then, under the same setting as Theorem \ref{thm: FedQVR}, the sequence $\{\thetab^r\}$ of FedQVR satisfies
		\begin{align}
			&\frac{1}{R} \sum_{r = 0}^{R - 1}\E[\|\nabla f(\thetab^r)\|^2] =
			\mathcal{O}\bigg(\frac{1}{R}+ \frac{\sigma^2}{\sqrt{R}}\bigg).
		\end{align}
\end{Corollary}



\begin{Rmk} The convergence of FedQVR can be improved by further choosing $\eta = \Oc( \frac{1}{\sqrt{R}})$ or $S = \Oc(R)$. Note that the last five terms in the RHS of \eqref{thm:  bd6} are dominant in deciding the convergence rate of FedQVR while four of them depend on $\eta^2$. It is easy to find that using $\eta = \Oc(\frac{1}{\sqrt{R}})$ leads to a better convergence rate of $\mathcal{O}(\frac{1}{R}+ \frac{\sigma^2}{R^{3/2}} + \frac{\sigma^2}{\sqrt{R}})$. Moreover, the convergence rate of FedQVR can be improved to $\mathcal{O}(\frac{1}{R})$ by choosing $S = \Oc(R)$. In practice, this condition on $S$ is mild assuming that the local data sizes of edge devices are not large or the total number of communication rounds consumed is moderate \cite{FedBCD_2021}. 

\end{Rmk}


\begin{Corollary} \label{corly: com_complexity}
	There exist a finite and small $\eta$ and a moderate  $S$ so that, the communication complexity  of FedQVR to reach $\epsilon$-accuracy, i.e., $\|\nabla f(\thetab^r)\|^2 \leq \epsilon  $ for some $r \in (1, R)$, is of the order $\Oc(\frac{1}{\epsilon})$, and accordingly, the communication cost (the total number of bits uploaded to the server\footnote{Only the uplink communication cost is considered since it is the primary
bottleneck when the number of edge devices is large.}) is of the order $\Oc(\frac{m(d(B+1) + \mu)}{\epsilon})$.
\end{Corollary}

\begin{Rmk}
As discussed in \cite{FedADMM_2022} and \cite{FedDyn_2021}, for non-convex FL, the best order of communication complexity achieving $\epsilon$-accuracy owned by existing FL algorithms, especially the variance-reduced ones, is $\Oc(\frac{1}{\epsilon})$. As shown in Corollary \ref{corly: com_complexity}, the proposed FedQVR algorithm enjoys the same favorable dependency on solution accuracy w.r.t. the communication complexity in the presence of HLU and quantized uplink transmission. Compared with existing FL algorithms which mostly exchange complete model updates during communication, FedQVR saves communication significantly by having quantized uplink transmission, even with a small number of quantization bits; see Sec. \ref{sec: perform_FedQVR}. Note that the communication complexity of FedAvg and SCAFFOLD is of the order $\Oc(\frac{1}{\epsilon^2})$, which is inferior to FedQVR. In Section \ref{sec: simulation}, we will further demonstrate that FedQVR not only has faster convergence but also is more communication-efficient than its counterparts.
\end{Rmk}

\section{An Enhanced FedQVR Algorithm}

As seen, the FedQVR algorithm can effectively reduce the overall communication cost by a novel and sophisticated variance reduction scheme and quantized unplink transmission. However, it assumes a perfect wireless environment, where the models can be successfully exchanged in each communication round.
Unfortunately, this assumption is impractical in real-world scenarios, where wireless channels can experience deep fading and communication is constrained by limited transmission power, bandwidth, and transmission delay. These imperfections can result in frequent failures during model exchange from edge devices to server, ultimately degrading the reliability and communication efficiency of FedQVR. This motivates us to devise an enhanced FedQVR algorithm, named FedQVR-E, which incorporates a flexible radio resource allocation scheme to combat the non-ideal wireless environment and enhance the convergence of FedQVR.

As indicated by Theorem \ref{thm: FedQVR}, the convergence rate of FedQVR can be significantly influenced by the quantization errors of the edge devices.  
Therefore, we focus on scenarios involving bandwidth-limited frequency division multiple access (FDMA) networks, where we delve into the joint allocation of bandwidth and quantization bits for the edge devices.
The goal is to maximize the minimum number of quantization bits allocated to the active devices while adhering to the devices' imposed delay constraints for uploading their locally quantized model updates to the server during each communication round. 

\subsubsection{Wireless Model Description and Problem Formulation}
Consider a bandwidth-limited FDMA system, the uplink transmission rate of client $i$ at round $r$ can be modeled by
\begin{align}
	R_i^r = W_i^r \log \left(1 + \frac{P_i |h_i^r|^2}{W_i^r \sigma_n^2}\right),
\end{align}
where $h_i^r$ and $W_i^r$ are the channel and the allocated bandwidth for device $i$ at round $r$, respectively, $P_i$ is the transmission power of device $i$, and $\sigma_n^2$ is the power spectrum density of the additive white Gaussian noise. 

Remind that each entry of the model is quantized into $B_i^r$ bits. Thus, the total number of bits for representing a model is given by 
\begin{align}
	\hat B_i^r = d (B_i^r + 1) + \mu^r, \forall i,r
\end{align}
 where $\mu^r$ is the number of bits to signfy both the lower and upper bounds in the quantization process, as shown in Sec. \ref{sec: quantization}. Then, given the number of quantization bits, $B_i^r$, the transmission delay of device $i$ at round $r$ is given by 
\begin{align}
T_i^r = \frac{\hat B_i^r}{R_i^r}.
\end{align}

We aim to maximize the quantization bits for all selected edge devices by jointly optimizing the quantization bits and bandwidth allocation in each communication round. In particular, we adopt the $\alpha$-fairness criterion to ensure that users with poor channel conditions also have the opportunity to participate in model training, thereby promoting fairness and inclusivity in the learning process. The associated problem can be formulated as
\begin{subequations} \label{eqn: maxmin rra}
	\begin{align}
		\max_{B_i^r, W_i^r} &~ \sum_{i=1}^m \frac{(B_i^r)^{1-\alpha}}{1-\alpha} \\
		\st & \sum_{i \in \Ac^r} W_i^r \leq W_{total}, 0 \leq W_i^r, \forall i,r,\\
		& T_i^r \leq \tau_i, \forall i,r,\\
		& B_i^r \in \mathbb{Z}^+, \forall i,r,
	\end{align}
\end{subequations}
where $\alpha \geq 0$ and $\neq 1$ is the preset fairness coefficient, $W_{total}$ is the total bandwidth, $\tau_i$ is the tolerable transmission delay of device $i$ for each round, and $\mathbb{Z}^+$ denotes the set of the positive integers. 

\subsubsection{Solving the Radio Resource Allocation Problem}

Since problem \eqref{eqn: maxmin rra} can be decoupled for different communication rounds, we can resort to the joint bandwidth and quantization bits allocation for a single communication round. Specifically, at round $r$,  the problem to be solved is given by 
\begin{subequations} \label{eqn: maxmin rra2}
	\begin{align}
		\max_{B_i^r, W_i^r} &~ \sum_{i=1}^m \frac{(B_i^r)^{1-\alpha}}{1-\alpha} \\
		\st ~&  \sum_{i \in \Ac^r} W_i^r \leq W_{total}, 0 \leq W_i^r, \forall i,\\
		& \hat B_i^r \leq \tau_i W_i^r \log \left(1 + \frac{P_i |h_i^r|^2}{W_i^r \sigma_n^2}\right), \forall i, \label{eqn: rra2_constraint}\\
            & \hat B_i^r = d (B_i^r + 1) + \mu^r, \forall i,r \\
		& B_i^r \in \mathbb{Z}^+, \forall i. \label{eqn: integer quantization variable}
	\end{align}
\end{subequations}
It is not difficult to verify that the non-convexity of problem \eqref{eqn: maxmin rra2} only comes from the integer variables $B_i^r$. By relaxing \eqref{eqn: integer quantization variable} to $B_i^r \geq 0$, one can verify that the resultant approximated problem is convex\footnote{By checking the second-order derivatives of the objective function and the right hand sides of constraints \eqref{eqn: rra2_constraint} in problem \eqref{eqn: maxmin rra2}, we can see that they are concave functions. Since the remaining constraints are also convex, the approximated problem, which replaces \eqref{eqn: integer quantization variable} with the constraints $B_i^r \geq 0, \forall i$, is convex.}, and thus can be solved by convex solvers, e.g., \texttt{CVXPY}. Once the solution is obtained, $B_i^r$ is floored to its nearest integer $\lfloor B_i^r \rfloor$.

The details of the FedQVR-E algorithm are presented in Algorithm \ref{alg: FedAVR}. At round $r$ of FedQVR-E, the edge server first randomly samples $m$ edge devices to participate in the model training. Then, problem \eqref{eqn: maxmin rra2} is solved to obtain the number of quantization bits $B_i^r$ and bandwidth allocation $W_i^r$ for these devices. Note that some devices may be allocated a very small $B_i^r$ due to the constraints \eqref{eqn: rra2_constraint} and the experienced deep fading from time to time. In this case, the edge server will delete the edge devices, whose number of quantization bits is less than a predefined lower bound $ 1\leq \ul B \in \mathbb{Z}^+$, from the set of selected devices $\Ac^r$. This is to guarantee that the quantized model updates transmitted by the active devices have the least amount of information for stable algorithm convergence. Then, each active device performs local update by following that in FedQVR, quantizes the model update $\Delta_i^{r+1}$ according to the solution of problem \eqref{eqn: maxmin rra2}, and uploads it to the edge server for the global update steps. 

\begin{Rmk} {\rm 
    While the dynamic resource allocation strategy introduced by the proposed FedQVR-E algorithm can enhance the algorithm robustness and performance, it involves trade-offs. For instance, dynamic optimization may increase computational complexity, particularly in networks with rapid bandwidth fluctuations. Additionally, adopting an $\alpha$-fairness criterion ensures fairness among devices with poor channel conditions but may occasionally reduce overall resource efficiency. Future work could explore hybrid strategies, such as predictive resource management or adaptive fallback mechanisms, to address these trade-offs and further improve performance under highly dynamic and resource-constrained scenarios.}
\end{Rmk}

\begin{algorithm}
	\caption{Proposed FedQVR-E algorithm}
	\label{alg: FedAVR}
	\begin{algorithmic}[1]
		\STATE {\bfseries Input:} initial values of $\xb_i^0 = \xb^0$, $\cb_i^0 = \cb^0=\zerob, \forall i, \eta > 0, \gamma > 0, 1 > a > 0, \ul B \geq 1, \tau_i, W_{total}, P_i,\forall i$.
		\FOR{round $r=0$ {\bfseries to} $R - 1$}
		\STATE {\bfseries \underline{Edge server side:}} sample devices $\Ac^r$ ($|\Ac^r|=m$) randomly without replacement from $[N]$ and broadcast $\thetab_0^{r} = \thetab^r - \frac{1}{\gamma}\cb^r$ to the devices in $\Ac^r$.
            \STATE {Solve problem \eqref{eqn: maxmin rra2} to obtain the bandwidth and quantization bits allocation for the sampled devices, and sends the results to the sampled devices.}
            \STATE Remove device $i$ from $\Ac^r$ if  $B_i^r < \ul B, \forall i \in \Ac^r$.
		\STATE {\bfseries \underline{Edge device side:}}
		\FOR{device $i = 1$ {\bfseries to} $N$ in parallel}
		\STATE{Repeat steps 6-18 based on the quantization bits allocation and each device uploads the model updates $\Delta_i^{r+1}$ and $a / (\eta \wt E_i^r)$ using the allocated bandwidth.}
		\ENDFOR
		\STATE {\bfseries \underline{Edge server side:}} Compute $\cb^{r+1}$ and $\thetab^{r+1}$  by
		\STATE ~~$\cb^{r+1} = \cb^r - \frac{1}{N}\sum\limits_{i \in \Ac^r}  \frac{a p_i}{\eta \wt E_i^r}\Delta_i^{r+1}$,
		\STATE ~~$\thetab^{r+1} = \thetab_0^r + \frac{N}{m} \sum\limits_{i \in \Ac^r}  p_i\Delta_i^{r+1}$.
		\ENDFOR
		\STATE {\bfseries Output:} $\thetab^R$.
	\end{algorithmic}
\end{algorithm}


\section{Experiment Results}
\label{sec: simulation}

In this section, we will examine the performance of the proposed algorithms by comparing them against numerous baseline FL algorithms under various experimental settings. 

\subsection{Experiment setup}

In the experiments, we assume that the edge server (i.e., base station) is located at the cell center and $N = 100$ edge devices are uniformly distributed within the cell. The proposed algorithms are deployed so that the server can coordinate the devices to train the ML models (e.g. deep neural networks) using the datasets.

{\bf Datasets}: The popular CIFAR-10 \cite{CIFAR10_Krizhevsky09} and MNIST datasets \cite{website_MNIST} are considered for evaluation, which are widely used in the FL literature for performance evaluation. The
CIFAR-10 dataset contains $50K$ training images of handwritten digits and $10K$ test ones, while the MNIST dataset has $60K$ training images and $10K$ test ones. To facilitate the FL process, the local datasets of edge devices are generated from CIFAR-10 \cite{CIFAR10_Krizhevsky09} or MNIST in a {\bf Non-IID} fashion. Specifically, we adopt the partition method in \cite{FedAvg_noniid_2019} to distribute the datasets among edge devices, in which each device is allocated data samples
of only a few class labels. Note that the number of class labels $k \geq 1$ in each edge device determines the non-i.i.d. degree. A smaller $k$ corresponds to a higher degree of non-i.i.d. data, and $k$ is set as $2$, leading to highly non-i.i.d. local datasets. 

{\bf DNN Models}: Different deep neural networks (DNN) are respectively 
adopted for the CIFAR-10 and MNIST datasets. Specifically,
the DNN structure for CIFAR-10 consists of four convolutional
layers, four pooling layers, and five fully connected layers, leading to the number of its parameters $d = 2513418$.  For the MNIST dataset, the DNN structure comprises three fully connected layers, having $d = 199210$. 

{\bf Communication cost}: During the FL process, we define the communication cost at any round $r \geq 0$ as the accumulated number of bits transmitted to the edge server. Note that only the uplink communication cost is considered since it is the primary bottleneck and the downlink cost is usually negligible compared to the uplink cost in practice. Therefore, at round $r$, the communication cost of FedQVR is $\sum_{k = 0}^{r} m(d(B_i^r + 1) + \mu^r)$. In the quantization process as \cite{FedToE2022}, the elements of the local model updates belonging to the same layer share the same upper and lower bounds $\ul z$ and $\ol z$. By this way, $\mu^r = 9 * 2 * 2 * 32 = 1152$\footnote{We assume that each model parameter is represented as a 32-bit floating-point number, and apply quantization to the weights and biases of each layer using distinct upper and lower bounds tailored to their respective ranges.} when the CIFAR-10 dataset is considered, while $\mu^r = 6 * 2 * 32 = 384$ when the MNIST dataset is considered. On the other hand, the communication cost of FL algorithms without quantization is $32md r$.

{\bf Baseline algorithms}: Two categories of baseline FL algorithms are considered for comparison with the proposed algorithms. The first category includes FedAvg \cite{FedAvg_noniid_2019}, SCAFFOLD \cite{SCAFFOLD_2020}, FedDyn\cite{FedDyn_2021}, and FEDADAM \cite{AdaptiveFL_2021}, most of which are the state-of-the-art FL algorithms dealing with the data heterogeneity. These methods focus on enhancing learning performance, such as convergence speed and model accuracy, thereby directly influencing communication efficiency by reducing the number of communication rounds for convergence. The second category contains FedPAQ\cite{FedPAQ_2020},  FedCOMGATE\cite{FedCOM_2021}, and 
 FedCAMS\cite{FedCAMS_2022}. They are popular FL algorithms leveraging model compression to reduce the overall communication cost. In particular, they aim to minimize the data size exchanged during communication while maintaining or even improving communication complexity. Note that we modify some of these compressed FL algorithms such that they utilize the same quantization method as FedQVR for fair comparison. Together, these two categories provide a comprehensive comparison framework, highlighting the strengths and limitations of methods that address either device heterogeneity to improve communication complexity or communication cost per round. This dual perspective allows for a thorough evaluation of our proposed algorithms under diverse scenarios.

{\bf Parameter setting}: For all algorithms under test, the mini-batch size $S = 50$ and the learning rate $\eta = 0.01$. In each communication round, we only sample $|\Ac^r| = 10$ edge devices which perform $E = 2$ local epochs by default. To simulate the case of (time-varying) HLU, the number of local epochs for participating devices are chosen randomly from $[1, 5]$ at each communication round. Other algorithm specific parameters are tuned individually. The penalty parameter $\alpha$ of FedDyn is set as $0.1$, and $\gamma$ of FedQVR and FedQVR-E are set as $0.3$ by default. SCAFFOLD takes the global learning rate $\eta_g = 1$ while FEDADAM, FedCOMGATE, FedCAMS respectively choose $\eta_g = 0.3$,  $\eta_g = 0.2$ and $\eta_g = 0.2$. The parameter $B_i^r$ is set as $2$ for the algorithms, including FedPAQ, FedCOMGATE, FedCAMS and FedQVR, by default. Finally, all algorithms stop when $500$ communication rounds are achieved.

\begin{figure}
	\centering
	\subfigure[CIFAR-10]
 {\includegraphics[width=7cm]{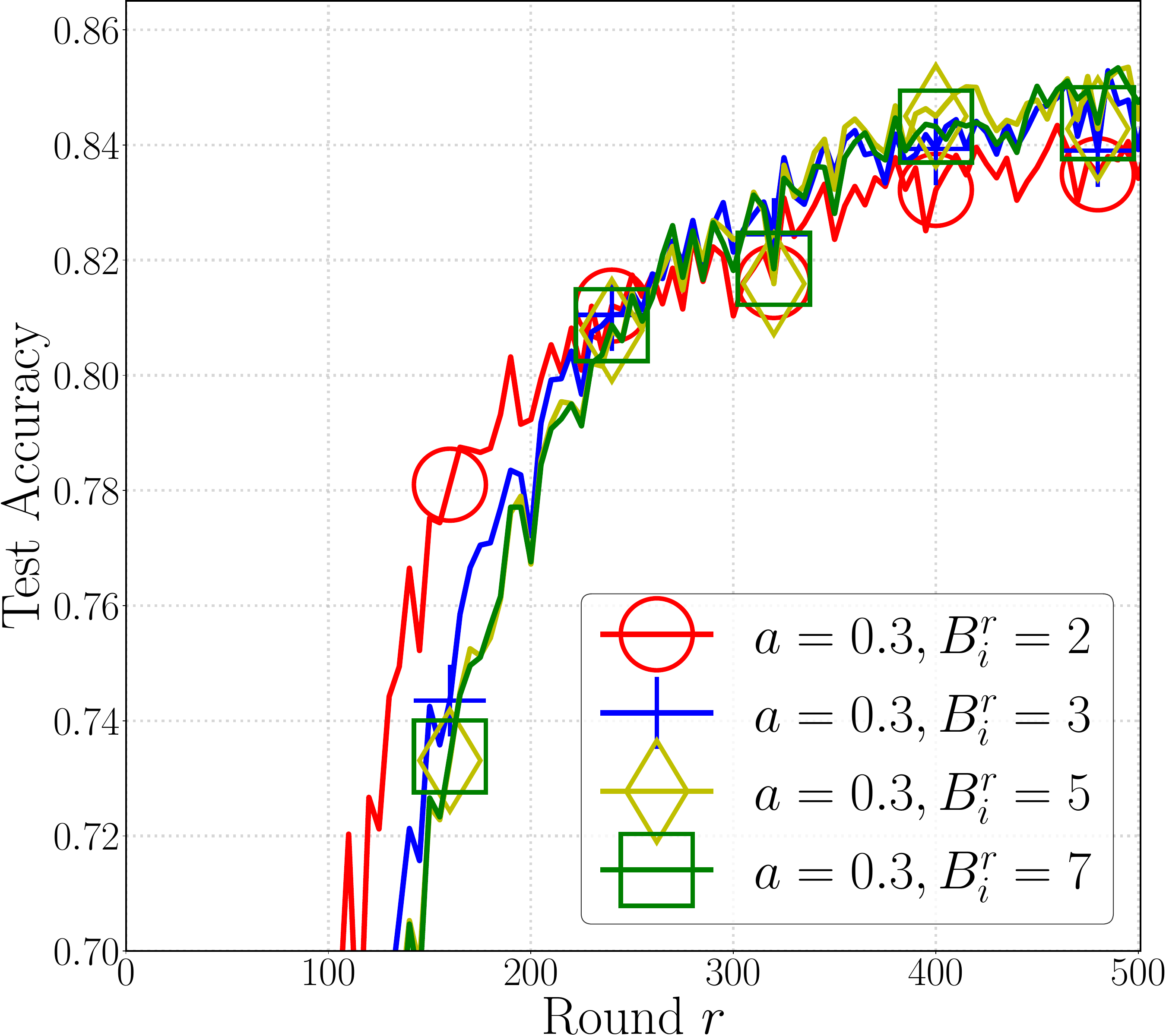} \label{fig: effects1}} 
        \subfigure[CIFAR-10]{\includegraphics[width=7cm]{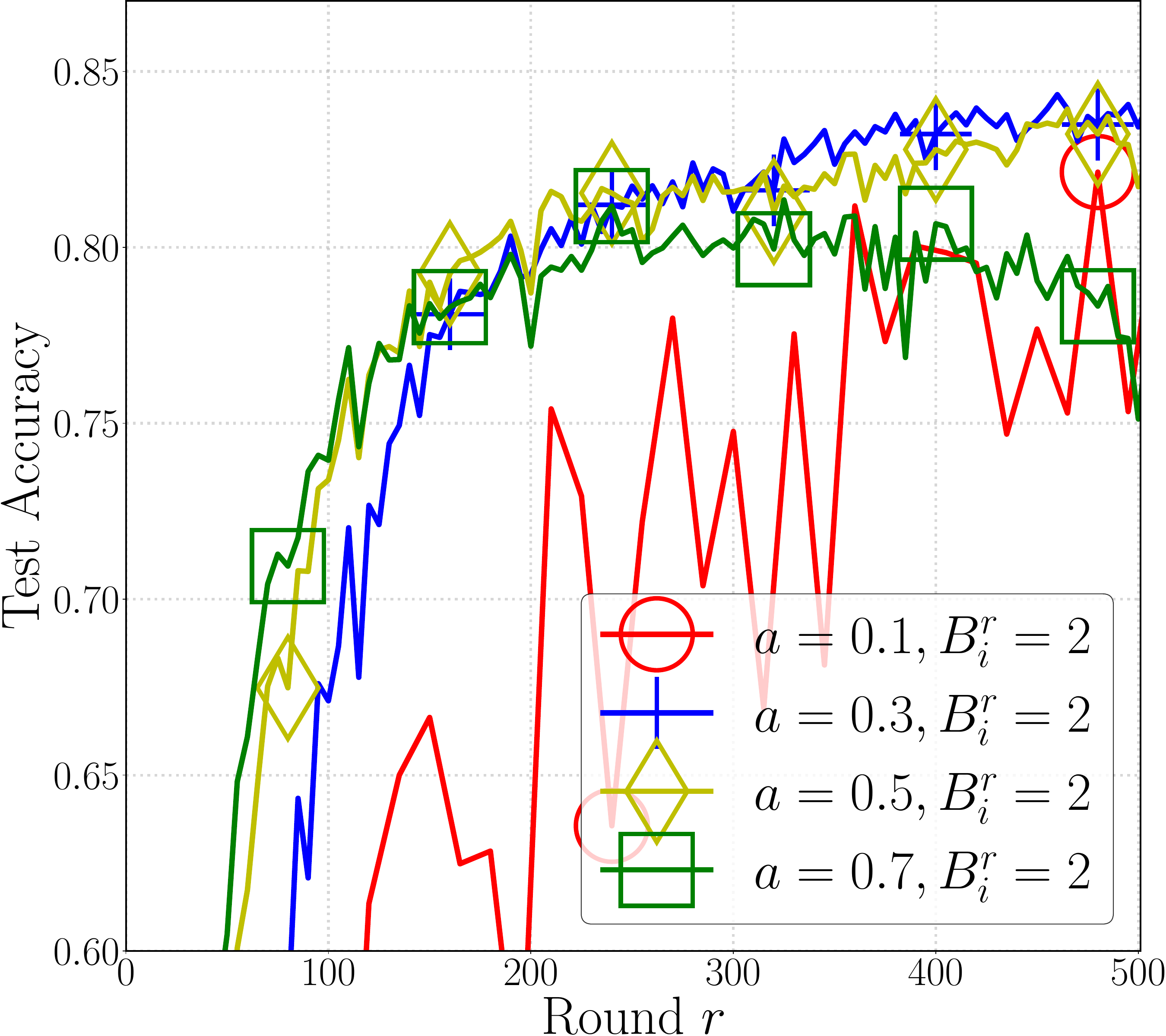} \label{fig: effects2}} 
	\centering
	\subfigure[MNIST]{\includegraphics[width=7cm]{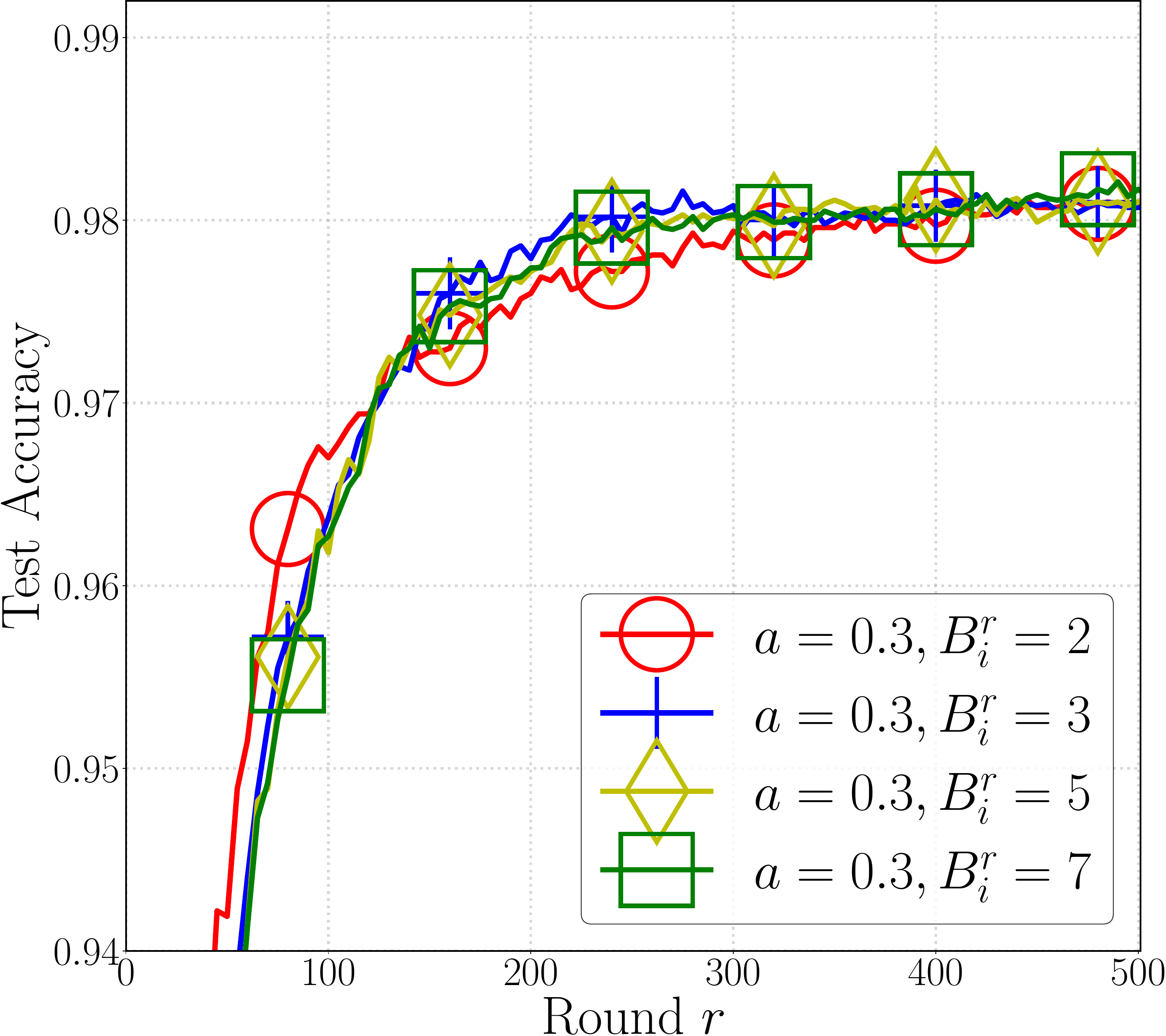} \label{fig: effects3}} 
        \subfigure[MNIST]{\includegraphics[width=7cm]{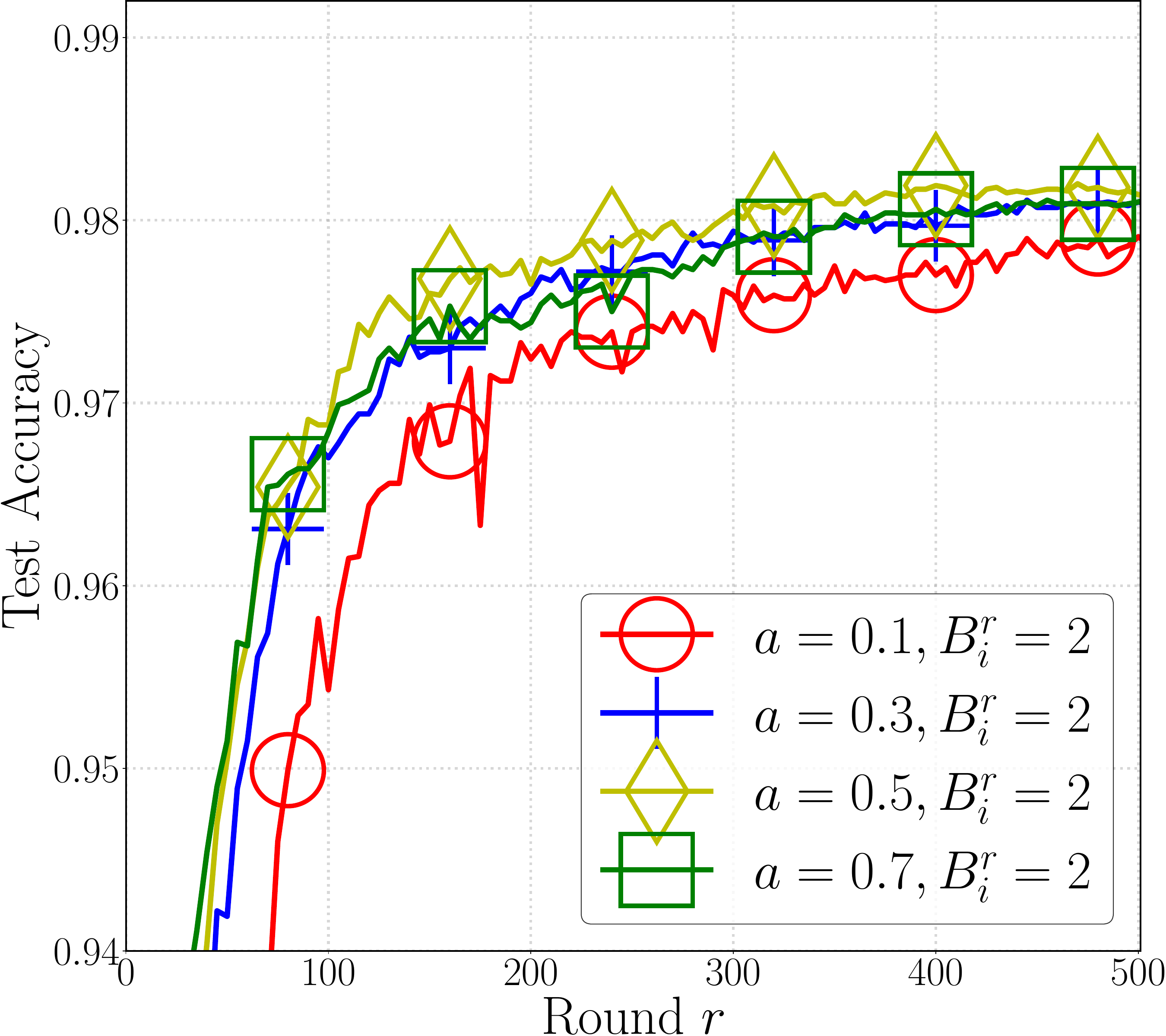} \label{fig: effects4}} 
	\centering\caption{Convergence performance of the proposed FedQVR algorithm with various choices of $a$ and $B_i^r$ } 
 \label{fig: effects}
\end{figure}

\subsection{Performance evaluation of FedQVR}
\label{sec: perform_FedQVR}

In this subsection, the performance of the proposed FedQVR algorithm is evaluated.

\subsubsection{Effects of $a$ and $B_i^r$}
Fig. \ref{fig: effects} presents the convergence performance of
FedQVR with different choices of constant $a$
and $B_i^r$ on both
CIFAR-10 and MNIST datasets. One can observe from Fig. \ref{fig: effects1}
and \ref{fig: effects3} that for constant $a$, the value of quantization bits $B_i^r$ does have an effect on the convergence performance of FedQVR. In particular, a smaller $B_i^r$ may lead to a little bit faster convergence in the early stage of FedQVR but achieve a lower test accuracy. Increasing $B_i^r$ can stablize the algorithm convergence and gradually reach a better application performance. This aligns with the known fact that overly aggressive quantization can introduce noise, which may slow convergence or affect model accuracy, while increasing the number of quantization bits preserves update fidelity, enhancing accuracy but at the expense of increased communication costs. It is worth-noting that FedQVR affords a small  $B_i^r = 2$ without much performance deteroritation, making it communication-efficient. On the other hand,
one can see from Fig. \ref{fig: effects2} and \ref{fig: effects4} that,
increasing $a$ properly (less than $0.5$) can not only speed up the convergence but also achieve a better test performance, while a large $a$ (greater than $0.5$) may not be preferred. The above results
show that proper values of $a$ and $B_i^r$
would boost the convergence of FedQVR.

\begin{figure}
	\centering
	\subfigure[CIFAR-$10$]
 {\includegraphics[width=7cm]{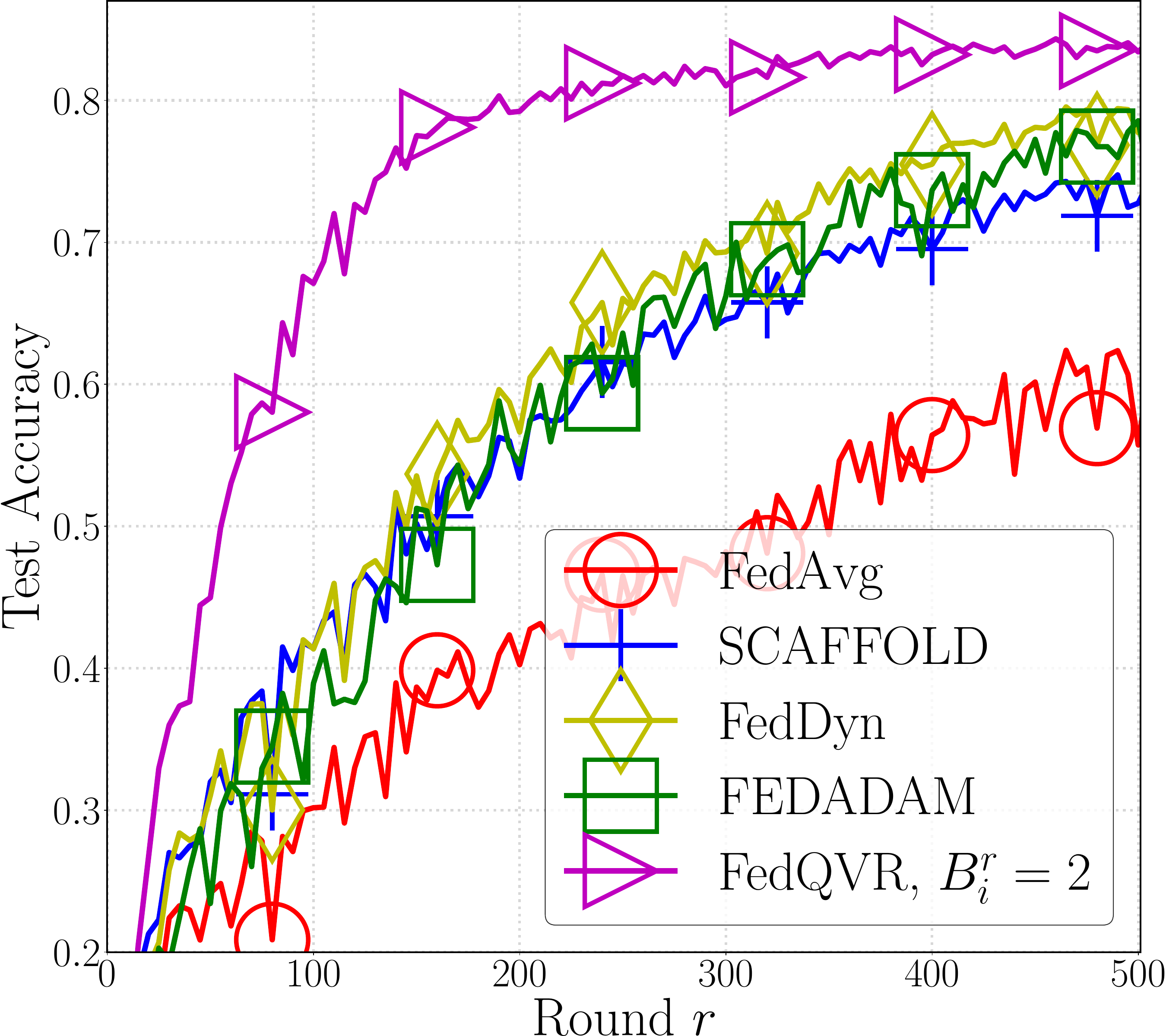} \label{fig: comb1_cifar_rounds}}
        \subfigure[CIFAR-$10$]{\includegraphics[width=7cm]{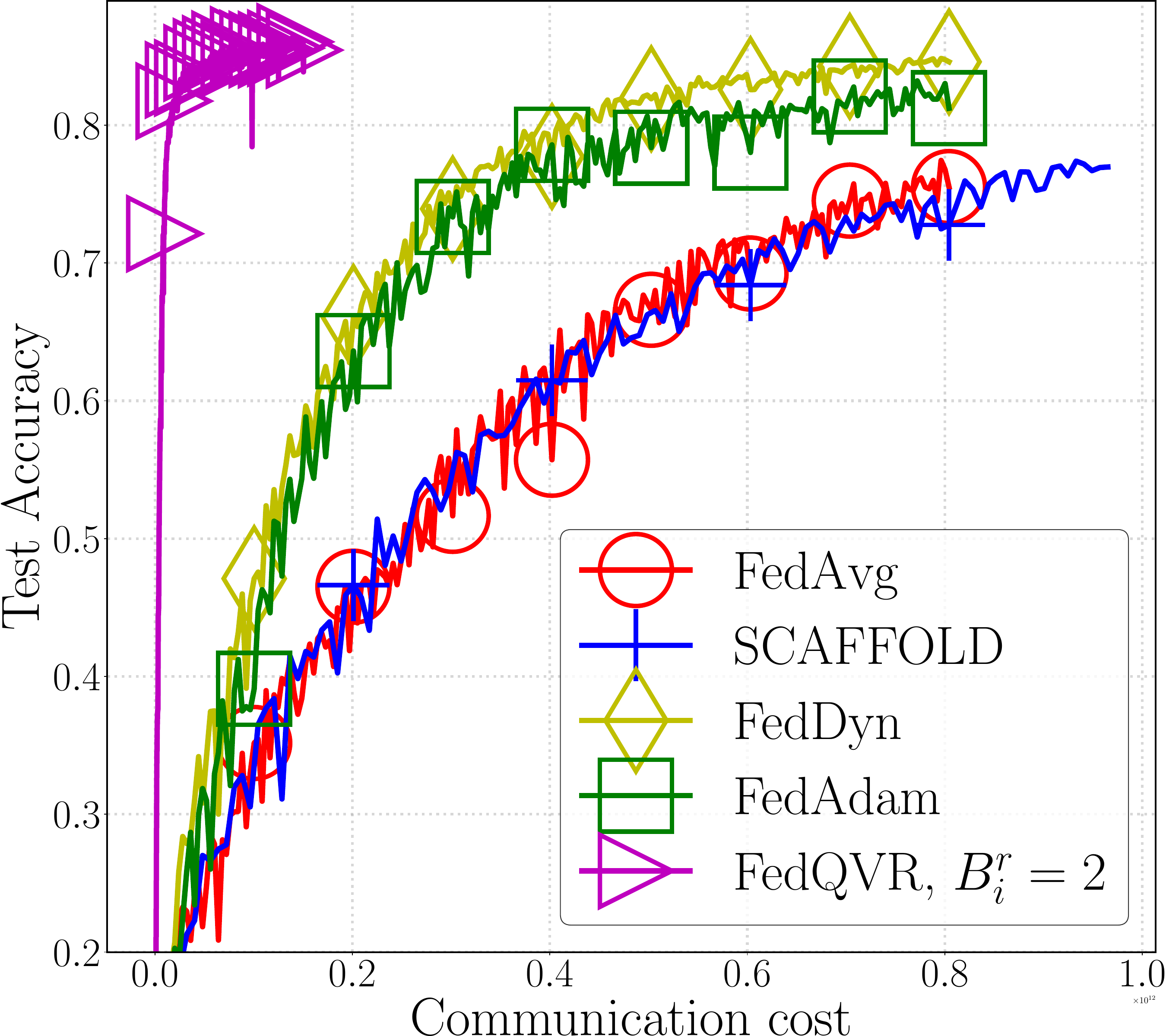}\label{fig: comb1_cifar_cost}}
	\centering
	\subfigure[MNIST]{\includegraphics[width=7cm]{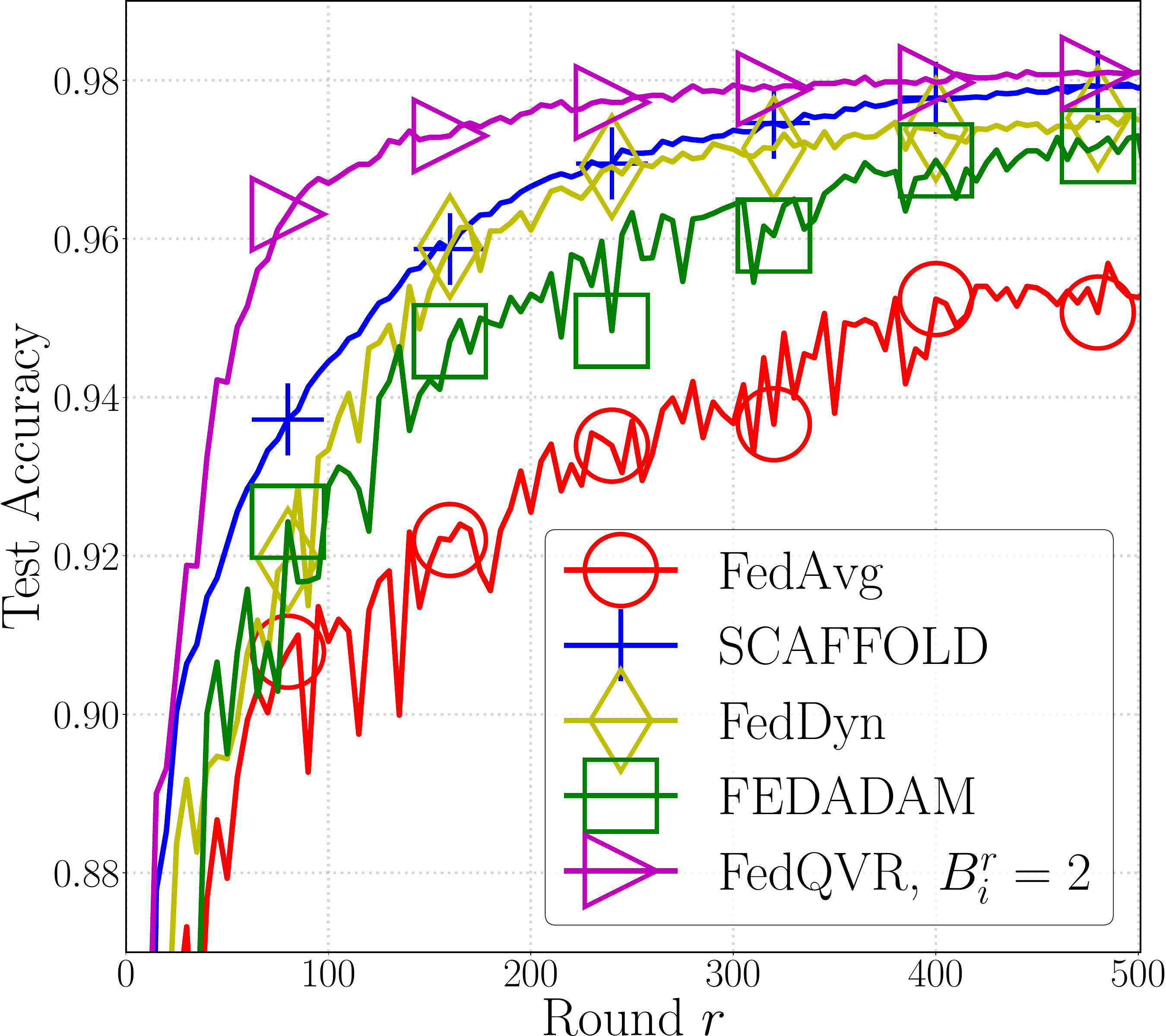}\label{fig: comb1_mnist_rounds}}
        \subfigure[MNIST]{\includegraphics[width=7cm]{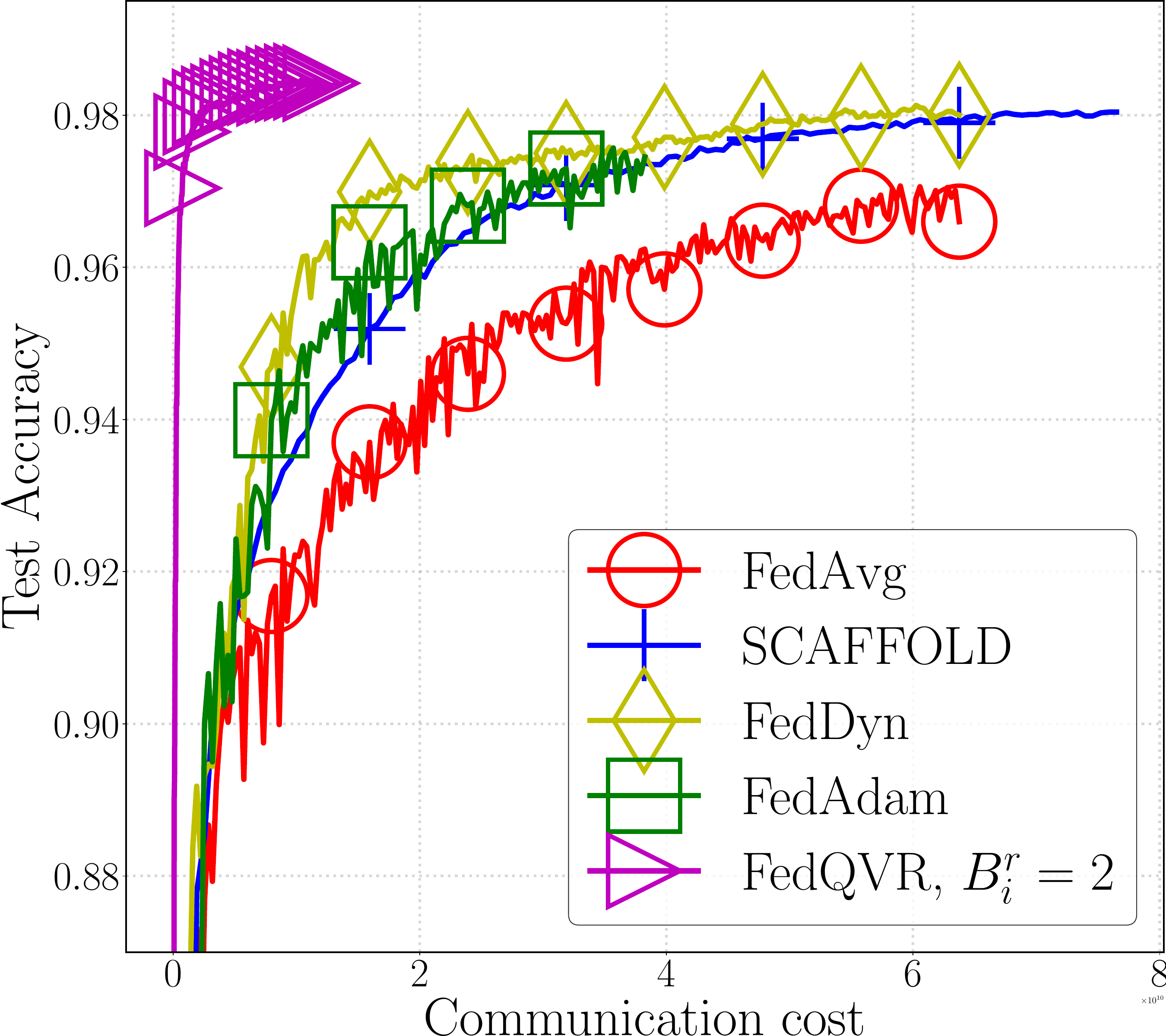} \label{fig: comb1_mnist_cost}}
	\centering\caption{Performance comparison between the proposed FedQVR algorithm and baseline algorithms without quantization} 
 \label{fig: comparison_baseline1}
\end{figure}

\begin{figure}
	\centering
	\subfigure[CIFAR-$10$, $B_i^r = 2$]
 {\includegraphics[width=7cm]{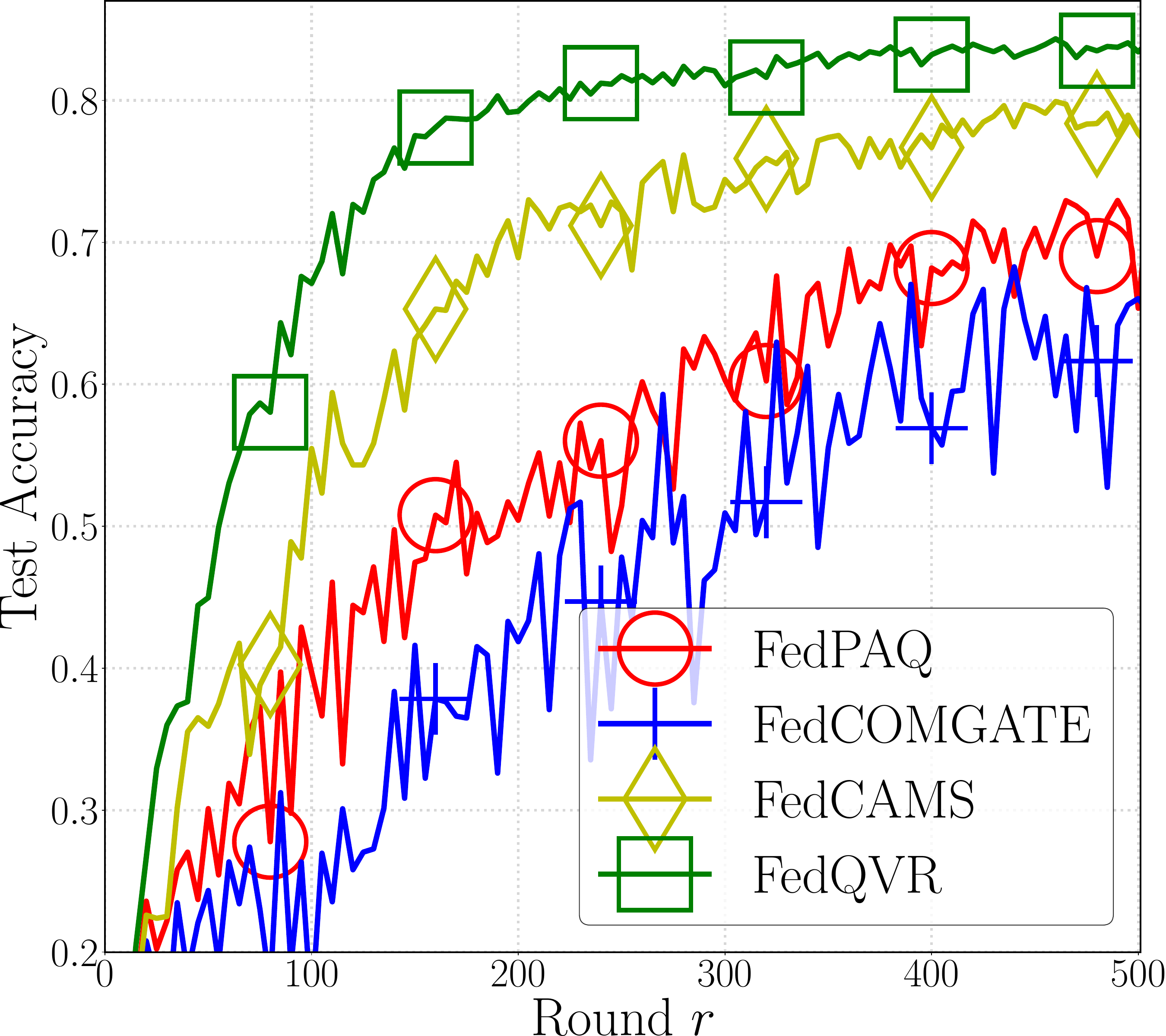} \label{fig: comb2_cifar_rounds}}
        \subfigure[MNIST, $B_i^r = 2$]{\includegraphics[width=7cm]{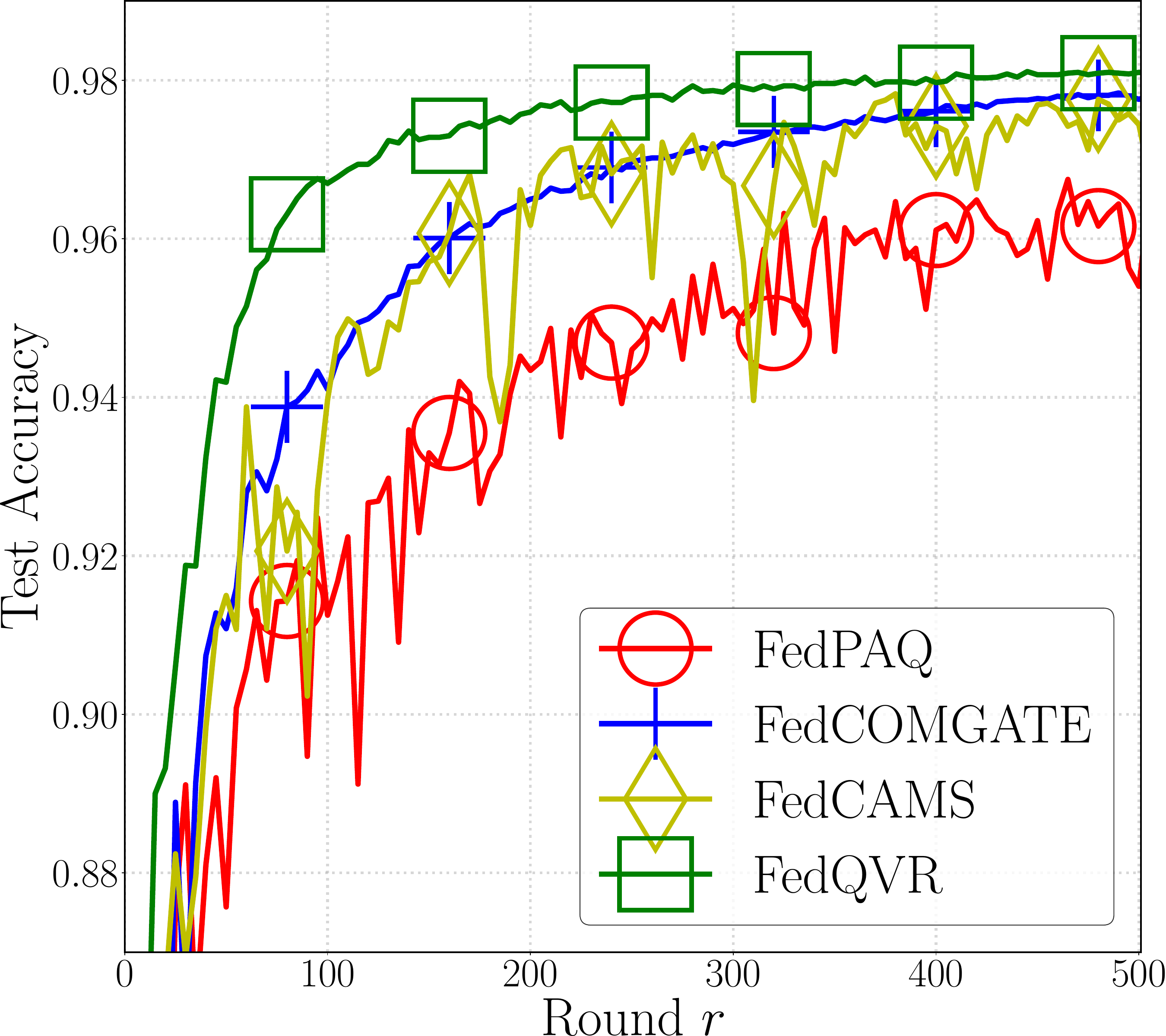} \label{fig: comb2_cifar_cost} }
	\centering\caption{Performance comparison between the proposed FedQVR algorithm and baseline algorithms with quantization} 
 \label{fig: comparison_baseline2}
\end{figure}

\subsubsection{Performance comparison with baseline algorithms}

In Fig. \ref{fig: comparison_baseline1}, we compare FedQVR with the baseline algorithms in the first category, the four algorithms without quantized transmission, on the non-i.i.d. CIFAR-10 and MNIST datasets. Note that the parameters $a$ and $B_i^r$ are respectively set as $0.3$ and $2$ in FedQVR. One can see from Fig. \ref{fig: comb1_cifar_rounds} that on the CIFAR-10 dataset, FedQVR significantly outperforms these baseline algorithms in terms of both convergence speed and performance. This is attributed to the robustness of FedQVR to the non-i.i.d. data even in the presence of quantized uplink transmission, implying the communication-efficiency of FedQVR. To furter verify this, Fig. \ref{fig: comb1_cifar_cost} re-plots the same result with respect to the communication cost. One can observe that FedQVR is quite effective in the communication cost reduction since it can reach a higher test accuracy (e.g., $80\%$) at the expense of significantly less communication cost than the others. To look into the reason for communication efficiency improvement, one can find that the communication cost of FedQVR (with $B_i^r=2$) per round is roughly one-tenth of that of the baseline algorithms in Fig. \ref{fig: comb1_cifar_cost}. As FedQVR converge much faster, its communication cost consumed for achieving a desirable test accuracy is certainly significantly less, proving its superiority in communication efficiency. The same trend can be observed from the results on the MNIST datasets in Fig. \ref{fig: comb1_mnist_rounds} and \ref{fig: comb1_mnist_cost}. Besides, it should be noted that, although SCAFFOLD convergence faster than FedAvg on the CIFAR-10 dataset, it has almost  the same communication cost as FedAvg  as it doubles the communication cost per round.

Fig. \ref{fig: comparison_baseline2} depicts the comparison results between FedQVR and the baseline algorithms in the second category, the four algorithms with quantized transmission, on the non-i.i.d. CIFAR-10 and MNIST datasets. One can observe from Fig. \ref{fig: comparison_baseline2} that, on both datasets, FedQVR significantly outperforms these FL algorithms in terms of both convergence speed and application performance. We remark that FedCAMS performs much better than the other baseline algorithms in Fig. \ref{fig: comparison_baseline2}. This is because it adopts the global adaptive optimization technique to speed up algorithm convergence and the error feedback technique to remedy the adverse effect brought by quantization errors. However, the convergence performance of FedQVR is still superior to FedCAMS, thanks to its capability of robust inter-device variance reduction.

\subsubsection{Influence of $m$ and HLU}
\begin{figure}
	\centering
	\subfigure[CIFAR-10]
 {\includegraphics[width=7cm]{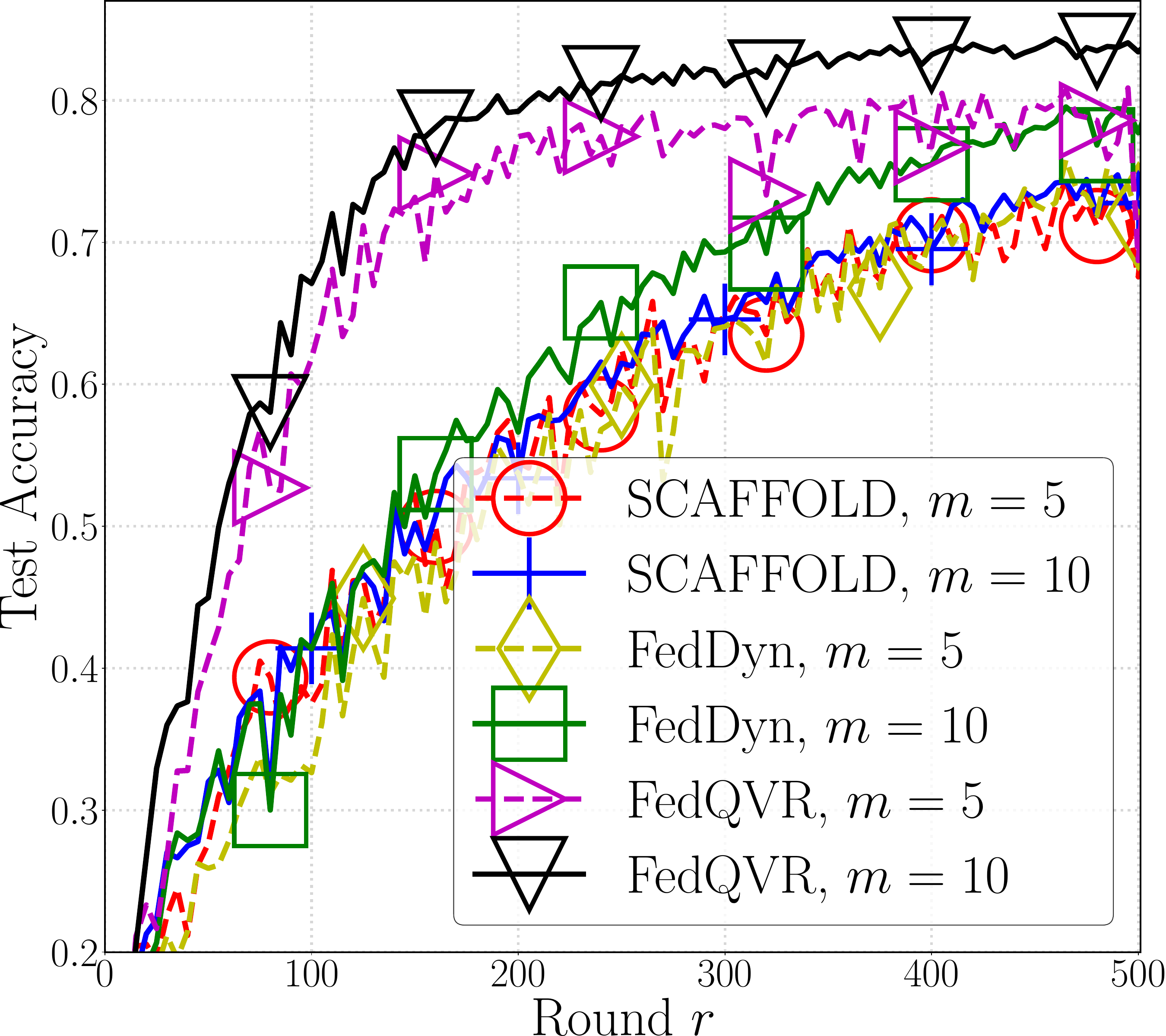} \label{fig: mHLU1}} 
        \subfigure[CIFAR-10, HLU]{\includegraphics[width=7cm]{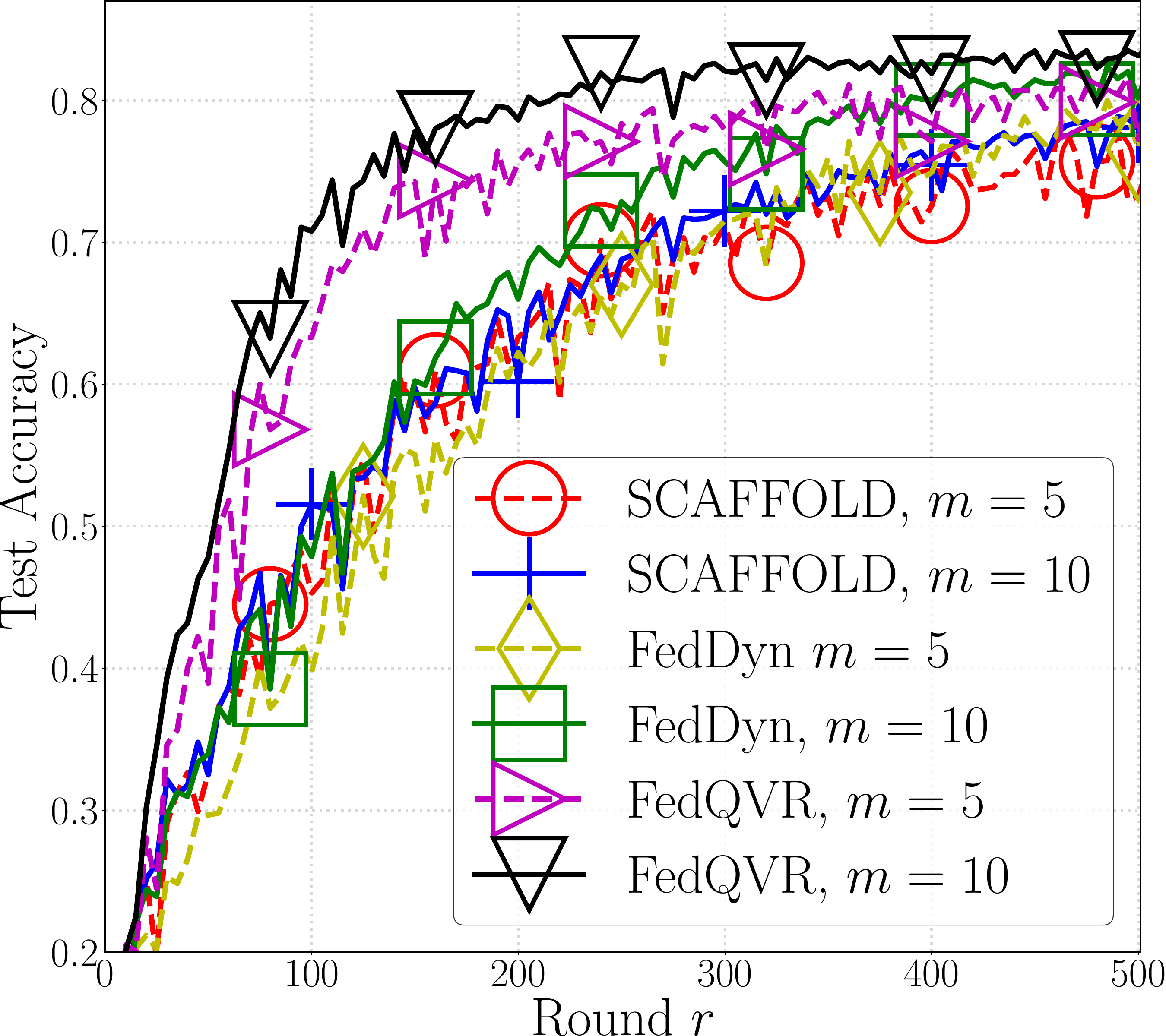} \label{fig: mHLU2}} 
	\centering
	\subfigure[CIFAR-10]{\includegraphics[width=7cm]{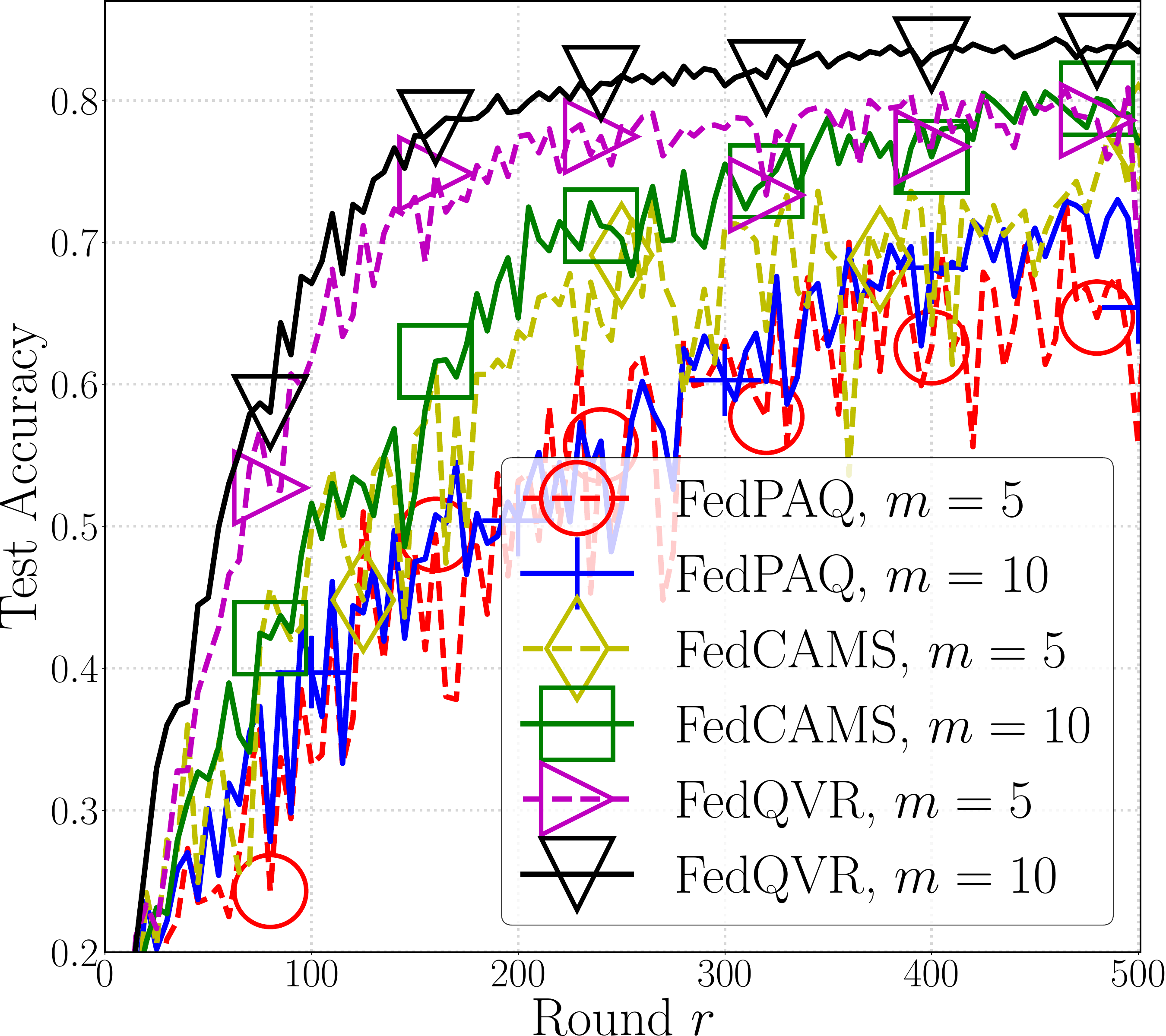} \label{fig: mHLU3}} 
        \subfigure[CIFAR-10, HLU]{\includegraphics[width=7cm]{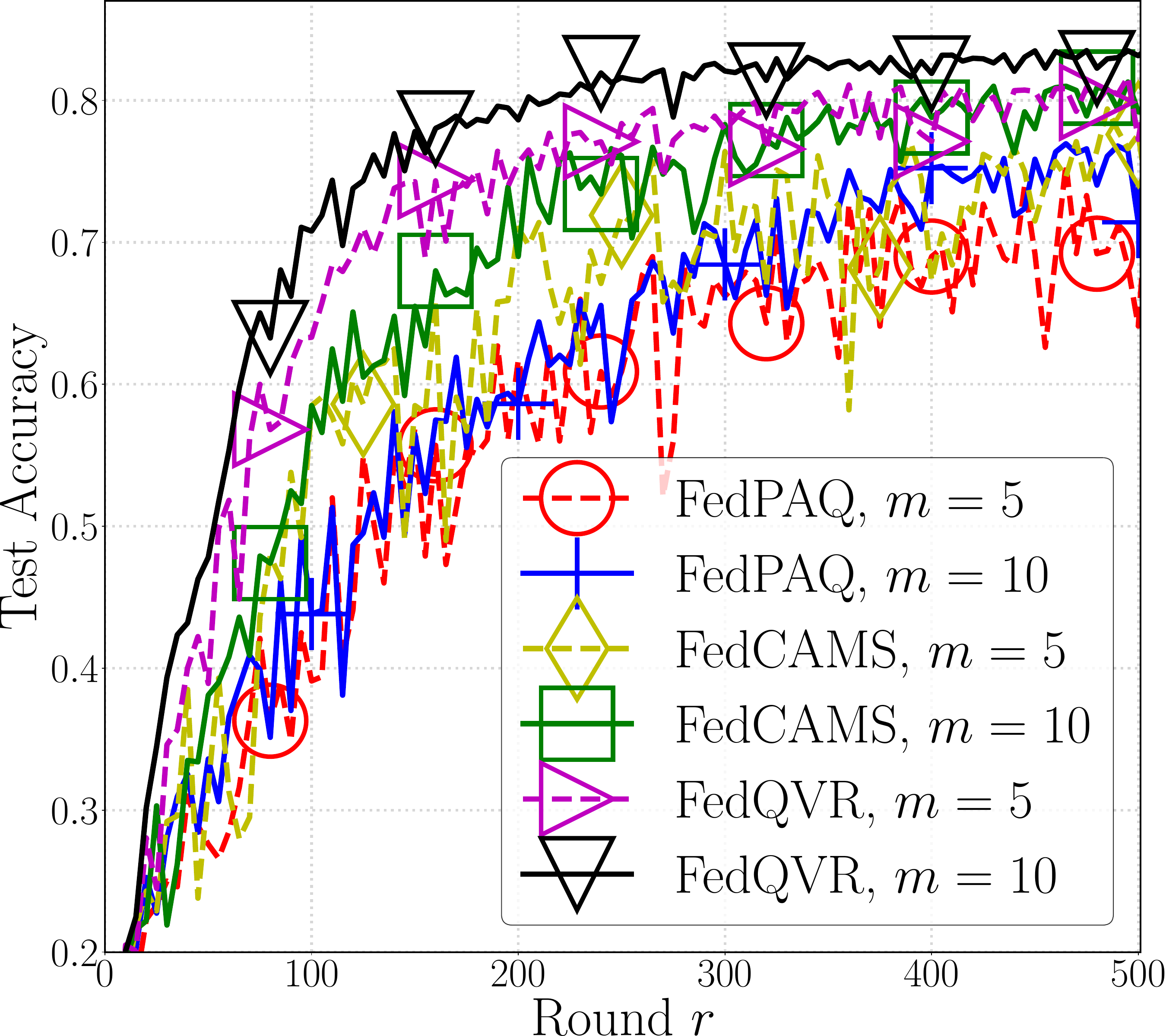} \label{fig: mHLU4}} 
	\centering\caption{Performance comparison between the proposed FedQVR algorithm and  baseline algorithms with different $m$ or under the setting of HLU } 
 \label{fig: mHLU}
\end{figure}

\begin{table}
	\footnotesize
 
	\setlength{\tabcolsep}{0.4mm}
	\caption{Performance comparison between FedQVR and the baseline algorithms. "$\#$R$-$XX"(resp. "$\#$C$-$XX") denotes the number of communication rounds (resp. communication cost) for achieving XX$\%$ accuracy. "---" means that the algorithm crashes.} 
 \centering
	\label{table: comp_results}
	\begin{tabular}{l|c|c|c|c|c|c}
		\toprule
		\midrule
		\rowcolor{gray!50} 
		Cases                                                                      & Algorithm&  ACC(\%) & \#R-70 & \#R-80 &  \#C-70($10^9$)  & \#C-80($10^9$)  \\ \midrule
		&FedAvg   &   60.90      &   $>500$    &   $>500$      & $> 402.1$  & $> 402.1 $ \\ \cline{2-7} 
		& SCAFFOLD  &   72.76   &     362   &  $>500$    &  $582.3 $   &  $> 804.3 $ \\ \cline{2-7} 
		&FedDyn &   77.72     &   290    &    496    &  $ 233.2 $    & $398.9 $  \\ \cline{2-7} 
		&FEDADAM    &   78.57       &    305   &    $>500$     &  $245.3 $   & $> 402.1 $   \\ \cline{2-7} 
  	&FedPAQ   &   73.41      &    361   &   $>500$      &  $27.22 $  &  $> 37.71  $ \\ \cline{2-7} 
       	&FedCOMGATE  &  68.91      & 489     &  $>500$     &  $ 36.88  $  &  $> 37.71 $ \\ \cline{2-7} 
        &FedCAMS  &  77.00       &   187   &  413     & $ 14.10 $   & $ 31.15 $ \\ \cline{2-7} 
		\multirow{-8}{*}{CIFAR-10}                                                         & FedQVR & {\bf 83.42}     &  {\bf 102}    &   {\bf 190}   & {\bf 7.692}  & {\bf 14.33}       \\ \midrule
  		&FedAvg   &   70.09       &   438    &  $>500$     &  $ 320.1  $ & $> 402.1$ \\ \cline{2-7} 
		& SCAFFOLD  &   78.11   &    254   &   $>500$       & $ 823.6  $ &   $> 804.3 $  \\ \cline{2-7} 
		&FedDyn &   80.18     &   218     &    378   &$ 175.3  $  &$ 304.0  $     \\ \cline{2-7} 
		&FEDADAM    &   80.75       &    225   &    436   & $ 181.0  $    &  $350.7 $  \\ \cline{2-7} 
  	&FedPAQ   &    77.23    &  286   &  $>500$      & $21.57 $   &   $> 37.71  $  \\ \cline{2-7} 
       	&FedCOMGATE  &   ---     &   ---   &  ---      &---   &    ---\\ \cline{2-7} 
        &FedCAMS  &    79.90     & 161     &   353   &   $ 12.14  $   &  $ 26.62 $ \\ \cline{2-7} 
		\multirow{-8}{*}{\begin{tabular}[c]{@{}c@{}}CIFAR-10\\ + HLU\end{tabular}}                                                         & FedQVR &  {\bf 83.34}     &    {\bf 95}   &  {\bf 197}   &  {\bf 7.315} & {\bf 14.86}  \\ \midrule
  	\rowcolor{gray!50} 
		Cases                                                                      & Algorithm&  ACC(\%) & \#R-95 & \#R-97 & \#C-95($10^8$) & \#C-97($10^8$) \\ \midrule
		& FedAvg    &   95.26      &   361    & $>500$       &  $ 230.1  $  & $ > 318.7  $  \\ \cline{2-7} 
		& SCAFFOLD   &   97.11      &   118    &   233   & $ 150.4  $   &$ 297.1 $\\ \cline{2-7} 
		& FedDyn  &    97.50     &  128    &   248   &  $ 81.59 $  & $ 158.1 $\\ \cline{2-7} 
		& FEDADAM    &  97.30     & 164   &   389      & $ 104.5  $ &  $ 248.0  $ \\ \cline{2-7} 
  	& FedPAQ    &    96.50    &     219   &   $>500$      & $ 13.10  $  & $> 29.90 $\\ \cline{2-7} 
    	& FedCOMGATE   &  97.76   &  116   &    253   &$ 6.937  $  & $ 15.13 $  \\ \cline{2-7} 
    &  FedCAMS    &   97.44   & 98      &  207     & $ 5.860  $  & $ 12.38  $\\ \cline{2-7} 
		\multirow{-8}{*}{MNIST} & FedQVR  &   {\bf 98.10}      &  {\bf 56}   &  {\bf 123}   & {\bf 3.350}     & {\bf 7.356} \\ \midrule
		& FedAvg    &   95.89     &  241     &  $>500$      & $ 153.6  $   & $ > 318.7 $  \\ \cline{2-7} 
		& SCAFFOLD   &   97.62    &   157    &   304    &  $ 200.2  $    & $387.6 $ \\ \cline{2-7} 
		& FedDyn  &  97.81     &  102     &    189   &  $ 65.02  $  &  $ 119.8  $ \\ \cline{2-7} 
		& FEDADAM    &  97.49    &  124     &  293     &  $79.05$   &  $187.8  $ \\ \cline{2-7} 
  	& FedPAQ    &   96.79     &  192     &  474     &  $ 11.48 $   &$28.35 $ \\ \cline{2-7} 
    	& FedCOMGATE   &   ---     &  ---     &   ---   & ---&  --- \\ \cline{2-7} 
    &  FedCAMS    &   97.73     &  104     &    165   & 6.219   &  9.867 \\ \cline{2-7} 
		\multirow{-8}{*}{ \begin{tabular}[c]{@{}c@{}}MNIST\\ + HLU\end{tabular}} & FedQVR  &   {\bf 98.12}      &   {\bf 59}   &   {\bf 116}     & {\bf 3.528}     & {\bf 6.937}  
  \\\bottomrule
	\end{tabular} 
\end{table} 

In Fig. \ref{fig: mHLU}, we present the convergence performance of FedQVR and some baseline algorithms on the non-i.i.d. CIFAR-10 dataset with a smaller number of active devices $m = 5$ or the setting of HLU. First of all, one can observe from Fig. \ref{fig: mHLU} that, FedQVR with smaller $m = 5$, performs slightly worse than FedQVR, but still outperform the baseline algorithms, even though they use $m = 10$. We should remark that the robustness of FedQVR to the number of active devices can greatly benefit  communication cost reduction, which again demonstrates its communication efficiency. Then, as shown in Fig. \ref{fig: mHLU2} and \ref{fig: mHLU4}, the proposed FedQVR algorithm is also resilient to HLU as it exhibits almost the same convergence behavior as the case without HLU in Fig. \ref{fig: mHLU1} and \ref{fig: mHLU2}. It can  be seen that, on all cases, FedQVR consistently performs better than the others in terms of convergence speed, application performance, and has a much smoother convergence curve, showing its stability against various experimental settings.

Table \ref{table: comp_results} summarizes the detailed results of FedQVR and baseline algorithms for above settings, including test accuracy achieved, the number of communication rounds required for a certain accuracy, and the corresponding communication costs. One can observe that FedQVR always performs the best and
achieves higher test accuracy than the baseline algorithms with significantly less communication costs. In particular, on the CIFAR-10 dataset, FedQVR converges almost 1
time faster than FedDyn and FedCAMS, and at least three times faster than the rest. The same trend can also be observed on the MNIST dataset. 
More importantly, to achieve a certain test accuracy, FedQVR needs much lower communication cost, confirming its high communication efficiency. For instance, as observed from row $9$ of Table \ref{table: comp_results}, FedQVR consumes the communication cost of $7.692 \times 10^9$ to achieve $70\% $ accuracy on the CIFAR-10 dataset while FedCAMS needs the communication cost of $14.10 \times 10^9$ and the rest requires the communication cost of more than $27 \times 10^9$.

\begin{figure}
	\centering
	\subfigure[CIFAR-10]
 {\includegraphics[width=7cm]{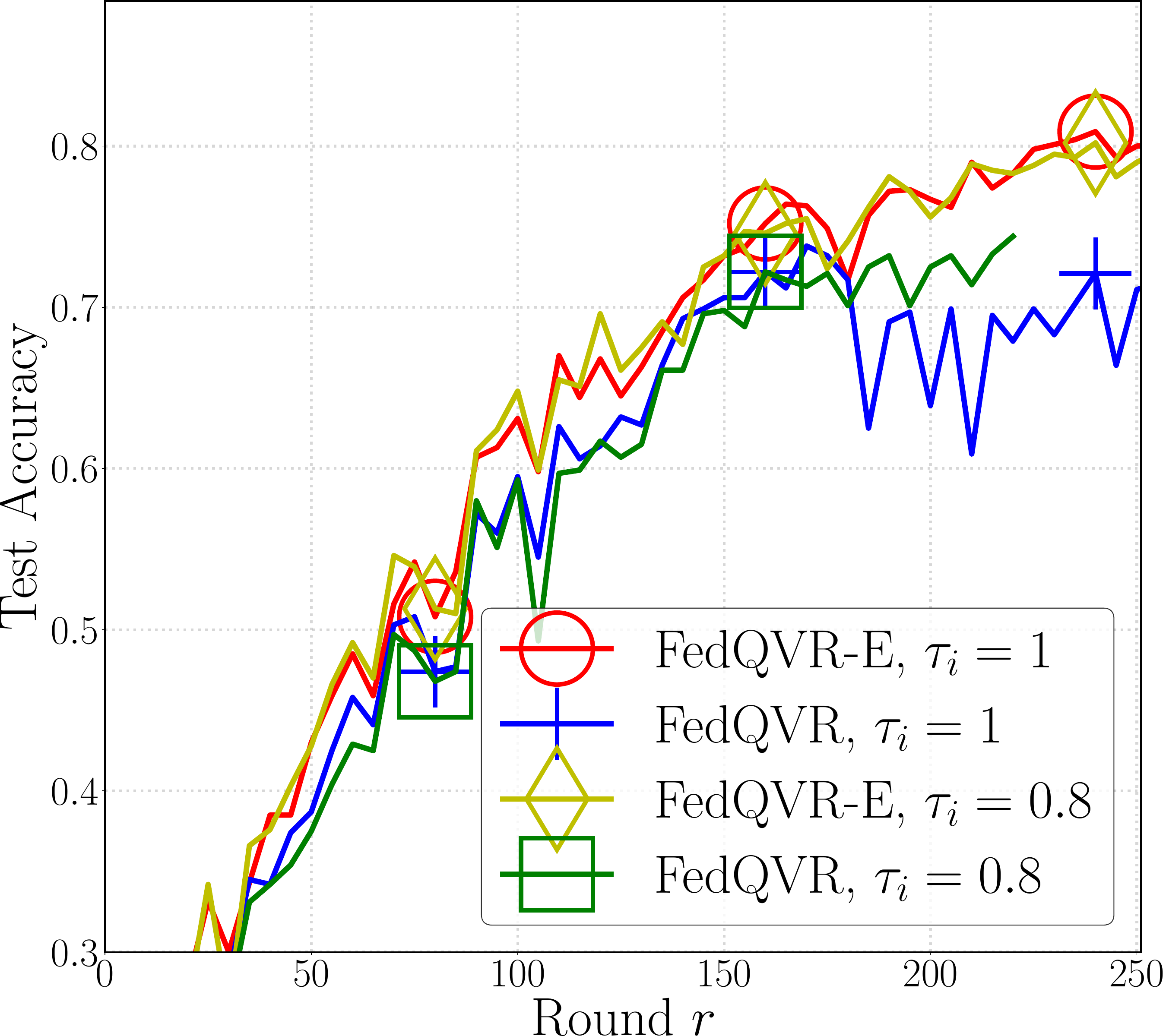} \label{fig: FedQVRE1}} 
        \subfigure[MNIST]{\includegraphics[width=7cm]{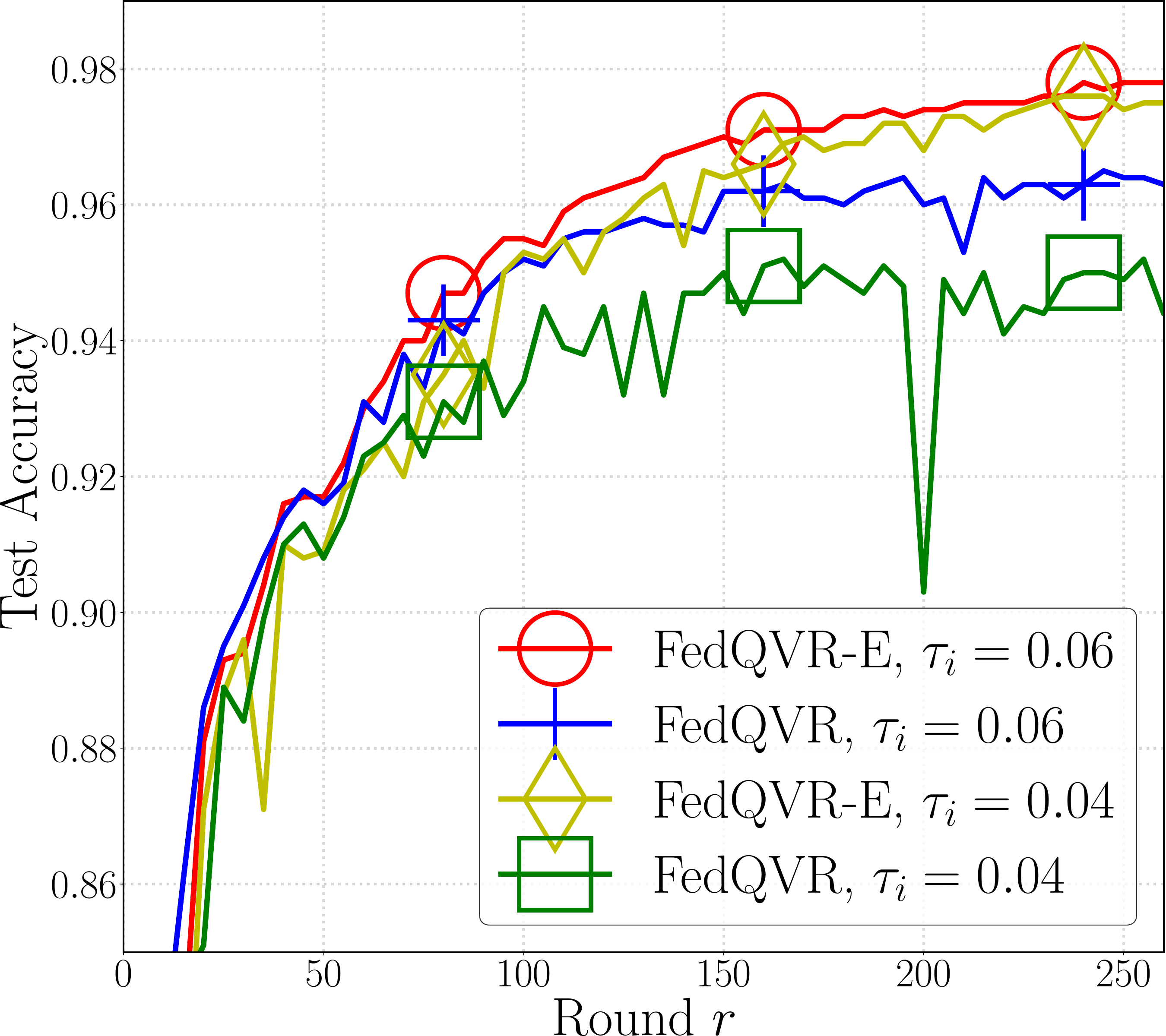} \label{fig: FedQVRE2}} 
	\centering
	\subfigure[CIFAR-10, $\tau_i = 1$]{\includegraphics[width=7cm]{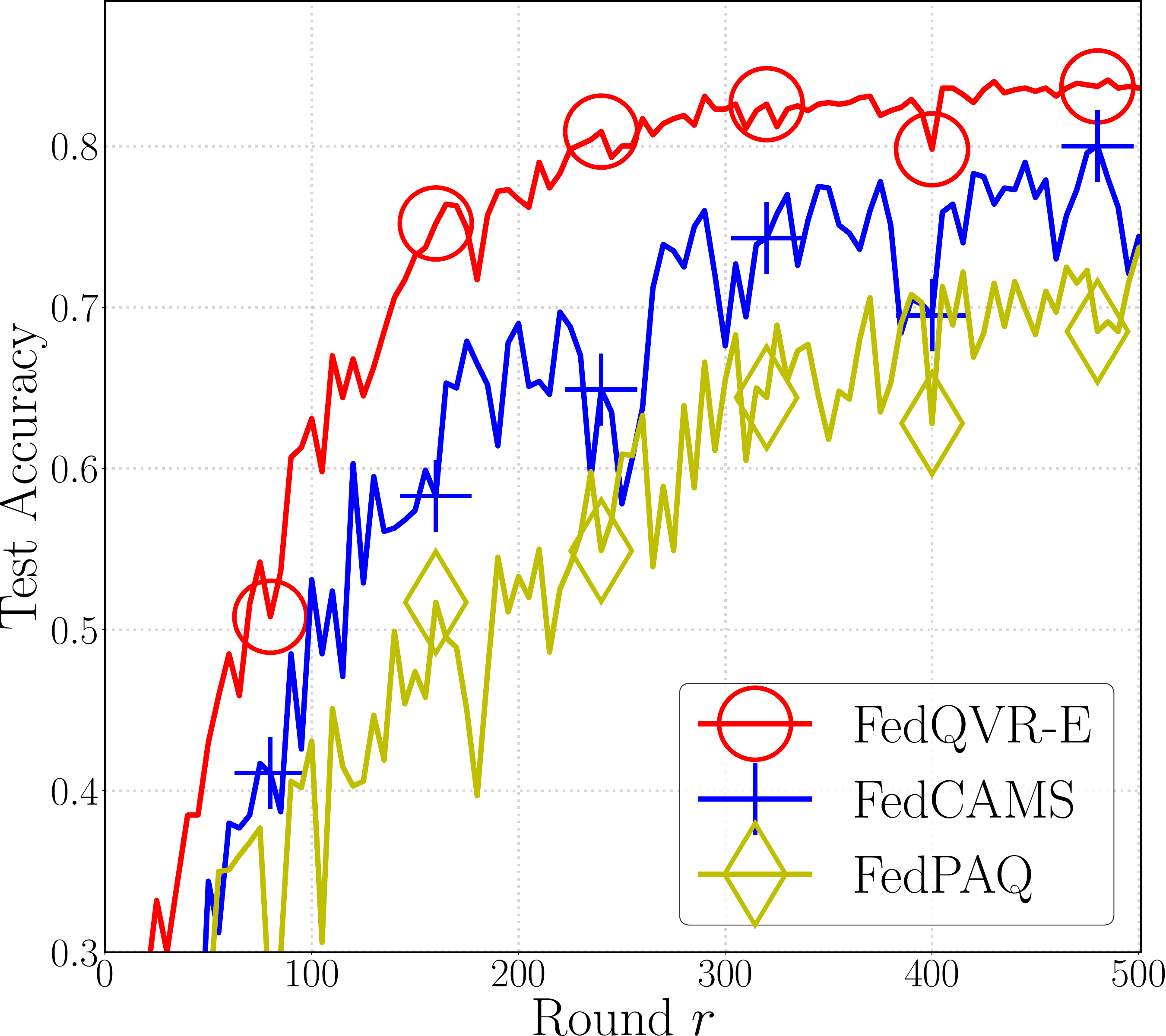} \label{fig: FedQVRE3}} 
        \subfigure[CIFAR-10, $\tau_i = 0.8$]{\includegraphics[width=7cm]{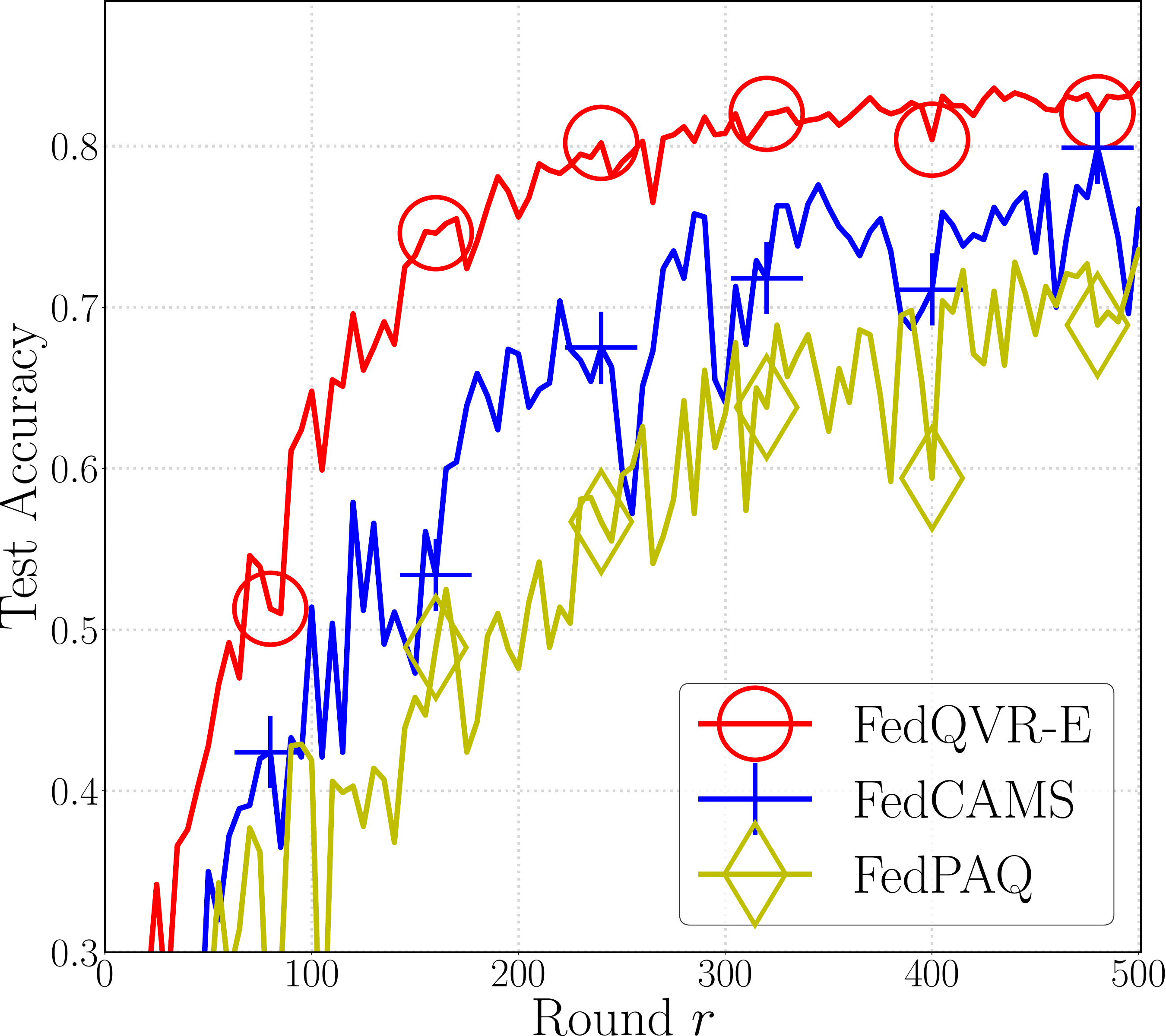} \label{fig: FedQVRE4}} 
	\centering\caption{Convergence performance of the proposed FedQVR-E algorithm and some baseline algorithms under different settings. } 
 \label{fig: performance_FedQVRE}
\end{figure}

\subsection{Performance evaluation of FedQVR-E}

In this subsection, the performance of the proposed FedQVR-E algorithm will be evaluated.
To this end, we assume the channel coefficient is composed by the large-scale path loss and the small-scale Rayleigh fading. In particular, the distance-dependent path loss is modeled by $10^{-3}d_i^{-\alpha}$, where $d_i$ is the Euclidean distance between the transmitter and receiver and $\alpha = 2$ is the path loss exponent; and the variance of the small-scale Rayleigh fading is unity.  The $\alpha$-fairness coefficient in \eqref{eqn: maxmin rra} is set as $0.5$. The system bandwidth is $ W_{total} = 100$ MHz and the noise power density is set to be $\sigma_n^2 = -143$dBm$/$Hz. The transmission power is $P_i = 30$dBm, $\forall i$. 

For FedQVR and all the baseline schemes under test,  the bandwidth is evenly assigned to the active devices, i.e., $W_i = W_{total}/m$, in each communication round. Note that, in the presence of the stringent transmission delay requirement $\tau_i$, the local quantized model may not be successfully transmitted to the edge server in these algorithms due to the deep channel fading (i.e., $|h_i^r|^2$ is too small to make \eqref{eqn: rra2_constraint} satisfied.). In contrast, FedQVR-E adopts the joint bandwidth and
quantization bits allocation scheme by solving problem \eqref{eqn: maxmin rra2}. With optimized $W_i^r$ and $B_i^r, \forall i \in \Ac^r$, the probability of successful transmissions from active edge devices to the edge server can be improved at each round $r$.

Fig. \ref{fig: performance_FedQVRE} depicts the convergence performance of FedQVR-E and the baseline algorithms in the second category with various choices of the tolerable transmission delay  $\tau_i$ for edge device $i$. Note that the baseline algorithms in the first category are not considered because they all have large communication cost per round, and thus always encounter transmission failures under the stringent transmission delay requirements, leading to an awful performance. One can observe from Fig. \ref{fig: performance_FedQVRE} that, the proposed FedQVR-E algorithm indeed enhances the convergence performance of FeQVR in the presence of non-ideal wireless channels, and thereby considerably outperforms FedQVR and other baseline algorithms on all cases under test. 

In particular, Fig. \ref{fig: FedQVRE1} and \ref{fig: FedQVRE2} tell that, on both CIFAR-10 and MNIST datasets, FedQVR, which does not consider non-ideal wireless conditions, would suffer from convergence slowdown and performance deterioration, with a relatively small $\tau_i, \forall i$. This phenomenon becomes more and more apparent as the value of $\tau_i, \forall i$ decreases. However, one can see from Fig. \ref{fig: FedQVRE1} and \ref{fig: FedQVRE2} that FedQVR-E is robust against non-ideal wireless conditions, as its convergence performance has hardly changed as the value of $\tau_i, \forall i$ varies, leading to the efficacy of FedQVR-E over FedQVR. This demonstrates the power of joint bandwidth and quantization bits allocation by solving problem \eqref{eqn: maxmin rra2} in handling non-ideal wireless conditions. Moreover, we present the comparison results between FedQVR-E and some baseline algorithms in Fig. \ref{fig: FedQVRE3} and Fig. \ref{fig: FedQVRE4}. As seen, the proposed FedQVR-E algorithms performs the best and the performance gap is significant. The baseline algorithms FedCAMS and FedPAQ have slower convergence speed and worse application performance. As the transmission delay constraints are more stringent, the influence on their convergence  are more apparent and their convergence curves become heavily fluctuate, while the proposed FedQVR-E algorithm exhibits a consistent performance.

\section{Conclusion}
\label{sec: conclusion}
In this paper, we have investigated communication-efficient FL in wireless edge networks and presented a novel algorithm, FedQVR. In contrast to existing FL algorithms with communication saving strategies, the proposed FedQVR algorithm integrates the strategies of robust inter-device variance reduction, quantized transmission, and HLU to address the challenge of device heterogeneity for algorithm convergence improvement, while reducing the communication cost consumed per round. We have also carried out a comprehensive theoretical analysis for FedQVR. The analysis results have shown that FedQVR owns the capability of inter-device variance reduction in the presence of quantized uplink transmission and HLU, and thereby it is inherently resillient to device heterogeneity. They have also concluded that FedQVR can converge in a sublinear
manner with a comparable rate in the presence of quantized uplink transmission, which enables its significantly improved communication efficiency. 

In addition, we have investigated FL with non-ideal wireless channels and presented an algorithm, FedQVR-E. Inspired by the above analysis results, FedQVR-E enhances FedQVR by performing joint allocation of bandwidth and quantization bits while satisfying the transmission delay constraint at each communication round. Finally, based on the CIFAR-10 and MNIST datasets, the presented experiment results have demonstrated that the proposed algorithms exhibit superior robustness against the device heterogeneity, quantization errors  the number of active devices or transmission delay constraints, as compared
to the existing schemes, thereby enjoying a remarkably improved convergence performance and a better communication efficiency.

As for the future works, it is worthwhile to devise more advanced model quantization or sparsification techniques that can further improve the communication efficiency. While our current work focuses on FL with one edge server, it is equally interesting to consider the case
with numerous edge servers  which can greatly relieve the communication burden for each server, and study the applications of FedQVR and FedQVRE to that FL setting.




 
%

\bibliographystyle{IEEEtran}
\bibliography{reference, refs10, refs20}



\newpage

 




\begin{center}
	\Huge
	\bf Supplementary Materials
	
\end{center}
%
%

\section{Proof of Lemma 1}
\begin{Lemma} \label{lem: local_update}
	For any round $r \geq 0$ and device $i \in \Ac^r$, it holds that $\forall t = 0, \ldots, E_i^r - 1$,
	\begin{align}
		&\thetab_i^{r, t+1}
		=  \frac{1}{1 + \gamma\eta} (\thetab_i^{r, t} - \eta( g_i(\thetab_i^{r,t}) + \cb^r - \cb_i^{r})) + \frac{\gamma\eta}{1 + \gamma\eta} \thetab^r, \label{lem: local_update_xi}\\
		&\E[\cb_i^{r+1}]
		= (1 - a)\E[\cb_i^r] + a \E[G_i^r], \cb^{r+1} = \sum_{i=1}^{N}p_i \cb_i^{r+1}, \label{lem: local_update_dual}
	\end{align}
	where $\bb_i^r \triangleq [b_i^{r, 0}, b_i^{r, 1}, \ldots, b_i^{r, E_i^r-1}]^\top \in \Rbb^{E_i^r}$, $b_i^{r, t} =(\frac{1}{1 + \gamma \eta})^{E_i^r - t}$, $G_i^r = \sum_{t= 0}^{E_i^r - 1}\frac{b_i^{r, t} }{\|\bb_i^r\|_1} g_i(\thetab_i^{r,t})$.
\end{Lemma}
{\bf Proof:} According to the update rule of $\thetab_i$ and $\thetab$, if $i \in \Ac^r$, we have
\begin{align}
	\thetab_i^{r, t+1} 
	=& \frac{1}{1 + \gamma \eta} (\thetab_i^{r,t} - \eta(g_i(\thetab_i^{r,t}) - \cb_i^r)) +  \frac{\gamma \eta}{1 + \gamma \eta}\thetab_0^r \label{lem: local_update_bd1} \\
	= &\frac{1}{1 + \gamma \eta} (\thetab_i^{r,t} - \eta(g_i(\thetab_i^{r,t}) - \cb_i^r))  + \frac{\gamma \eta\big(\thetab^r - \frac{\cb^{r}}{\gamma} \big)}{1 + \gamma \eta}  \label{lem: local_update_bd2}\\
	= & \frac{1}{1 + \gamma \eta} (\thetab_i^{r,t} - \eta(g_i(\thetab_i^{r,t}) + \cb^r - \cb_i^r)) +  \frac{\gamma \eta}{1 + \gamma \eta}\thetab^{r},
\end{align}
where \eqref{lem: local_update_bd2} follows due to the definition of $\thetab_0$. On the other hand, we can also have from \eqref{lem: local_update_bd1} that
\begin{align}
	\thetab_i^{r+1} - \thetab_0^r  
	= & \thetab_i^{r, E_i^r} - \thetab_0^r   \notag \\
	= & \frac{1}{1 + \gamma \eta} (\thetab_i^{r, E_i^r - 1} - \eta(g_i(\thetab_i^{r,E_i^r-1})  - \cb_i^{r})) +  \frac{\gamma \eta}{1 + \gamma \eta} \thetab_0^r - \thetab_{0}^r\notag \\
	=&\frac{1}{1 + \gamma \eta} (\thetab_i^{r, E_i^r - 1} - \thetab_0^r) -  \frac{\eta(g_i(\thetab_i^{r,E_i^r-1})  - \cb_i^{r})}{1 + \gamma \eta} . \label{lem: local_update_bd2_1}
\end{align}
By repeating the above procedure, it holds that
\begin{align}
	\thetab_i^{r+1} - \thetab_0^r 
	= &- \frac{\eta}{1 + \gamma \eta}\sum_{t = 0}^{E_i^r - 1}\big(\frac{1}{1 + \gamma \eta}\big)^{E_i^r - 1 - t}( g_i(\thetab_i^{r,t})   -\cb_i^{r})   \\
	= &- \eta\sum_{t = 0}^{E_i^r - 1}\big(\frac{1}{1 + \gamma \eta}\big)^{E_i^r - t}( g_i(\thetab_i^{r,t})   -\cb_i^{r})   \\
	=& - \eta \wt E_i^r \sum_{t= 0}^{E_i^r - 1}\frac{b_i^{r, t}}{\|\bb_i^r\|_1} (g_i(\thetab_i^{r,t})- \cb_i^r),  \label{lem: local_update_bd3}
\end{align}where
$\bb_i^r \triangleq [b_i^{r, 0}, b_i^{r, 1}, \ldots, b_i^{r, E_i^r-1}]^\top \in \Rbb^{E_i^r}$, $b_i^{r, t} =(\frac{1}{1 + \gamma \eta})^{E_i^r - t}$, $\wt E_i^r = \|\bb_i^r\|_1= \frac{1}{\gamma \eta }(1 - \frac{1}{(1+\gamma\eta)^{E_i^r}})$.
Then, we follow the update rule of $\cb_i$ and obtain
\begin{align}
	\E[\cb_i^{r+1}] =& \E[\cb_i^r] - \frac{a}{\eta \wt E_i^r} \E[\hat \Delta_i^{r+1}] \notag \\
	= & \E[\cb_i^r] + a\E\bigg[\sum_{t= 0}^{E_i^r - 1}\frac{b_i^{r, t} }{\|\bb_i^r\|_1} (g_i(\thetab_i^{r,t})- \cb_i^r)\bigg] \notag \\
	= &(1 - a)\E[\cb_i^r] + a \sum_{t= 0}^{E_i^r - 1}\frac{b_i^{r, t} }{\|\bb_i^r\|_1} \nabla f_i(\thetab_i^{r,t}). \label{lem: local_update_bd4}
\end{align} 
Lastly, as $\cb^0 = \cb_i^0 = \zerob$, then $\cb^0 = \sum_{i = 1}^{N}p_i \cb_i^0$. Suppose $\cb^{r-1} = \sum_{i = 1}^{N}p_i \cb_i^{r-1}, \forall r \geq 1$, then we have
\begin{align}
	\cb^{r} =& \cb^{r-1} - \sum_{i \in \Ac^r} \frac{ap_i}{\eta \wt E_i^r} \Delta_i^{r} 
	= \sum_{i = 1}^{N}p_i \bigg(\cb_i^{r-1} -  \frac{ap_i}{\eta \wt E_i^r} \Delta_i^{r}\bigg) 
	= \sum_{i = 1}^{N}p_i \cb_i^r, \label{lem: local_update_bd5}
\end{align}where \eqref{lem: local_update_bd5} follows by  $\cb_i^{r+1} = \cb_i^{r} -\frac{a}{\eta \wt E_i^r} \Delta_i^{r+1}, \forall r \geq 0$. Then, by mathematical induction, it holds that $\cb^{r} = \sum_{i = 1}^{N}p_i \cb_i^r, \forall r \geq 0$. This completes the proof. \hfill $\blacksquare$

\section{Proof of Theorem 1}
\label{sec: theorem1_proof}

In this section, we present the poof of Theorem 1, which states the convergence of FedQVR.
Before delving into the proof, let us introduce some useful terms for ease of presentation. To deal with the randomness incurred by partial client participation, we define the virtual sequence $ \{(\wt \thetab_i^{r}, \wt \cb_i^{r})\}$ by assuming that all clients are active at round $r$, i.e., $\forall i, 0 \leq t \leq E_i^r - 1$, $\wt \thetab_i^{r, 0} = \thetab_0^{r}, \wt \thetab_{i}^{r+1}=\wt \thetab_i^{r, E_i^r}$,
\begin{subequations}
	\begin{align}
		\wt \thetab_i^{r, t+1} =& \frac{1}{1 + \gamma \eta} (\wt \thetab_i^{r,t} - \eta(g_i(\wt \thetab_i^{r,t}) - \cb_i^r)) +  \frac{\gamma \eta}{1 + \gamma \eta}\thetab_0^r ,\\
		\wt \Delta_i^{r+1} =& \Qc(\wt \thetab_i^{r+1} - \thetab_0^r, B_i^r), \label{eqn: def_virtual_xi}\\
		\wt \cb_{i}^{r+1} =&  \cb_{i}^{r} -  \frac{a}{\eta \wt E_i^r}\wt \Delta_i^{r+1}, \label{eqn: def_virtual_dual}
	\end{align}
\end{subequations}
We also define the following additional terms that will be used in our proof.
\begin{align}
	&\Xi_i^r \triangleq  \E[\| \nabla f_i(\thetab^r) - \cb_i^r\|^2], \\
	&\Psi_i^r \triangleq \E\bigg[\sum_{t= 0}^{E_i^r - 1}\frac{b_i^{r,t}}{\|\bb_i^r\|_1}\| \wt \thetab_i^{r,t} - \thetab^r\|^2\bigg],\label{def: Psi_i}\\
	&{\rm \Bb^r}  \triangleq \sum_{i = 1}^{N}p_i\bigg((1- \gamma \eta \wt E_i^r)\cb_i^r + \gamma \eta \wt E_i^r\sum_{t= 0}^{E_i^r- 1}\frac{b_i^{r, t}}{\|\bb_i^r\|_1} \nabla f_i(\wt \thetab_i^{r,t})\bigg). \label{def: B}
\end{align}

\subsection{Proof Outline}

In this section, we briefly sketch the proof outline of Theorem 1.  It should be noted that the theoretical analysis of the proposed FedQVR algorithm in the typical non-convex setting is not a trivial task at all, instead, it is actually quite complicated and challenging. In order to obtain the convergence property of FedQVR, we follow a similar analysis framework to \cite{SCAFFOLD_2020} and \cite{NESTT_2016}, but provide novel techniques to overcome the challenges brought by the sophisticated algorithm design in FedQVR. Specifically, the analysis relies on the construction of the potential function whose convergence behavior can characterize that of the variables $(\thetab_{i}, \cb_i, \thetab)$ in FedQVR. Then, we can obtain the convergence property of FedQVR by analyzing the potential function as the algorithm proceeds. Details are presented in Sec. \ref{thm: proof}.

\subsection{Key lemmas}

In this section, we will present some key lemmas that will be used in the proof of Theorem 1. The proofs of Lemma \ref{lem: f_descent}, \ref{lem: div_bound}, \ref{lem: opt_bound} and \ref{lem: x0_diff_bound} are respectively presented in the subsequent subsections.

\begin{Lemma} \label{lem: f_descent}
	For any round $r$, it holds that
	\begin{align}
		\E[f(\thetab^{r+1}) - f(\thetab^r)] 
		\leq & \sum_{i=1}^{N}\frac{p_i(1 - \gamma\eta \wt E_i^r)}{2\gamma}\Xi_i^r - \frac{1}{2\gamma} \E[\|\nabla f(\thetab^r)\|^2]- \frac{1}{2\gamma}\E[\|{\rm \Bb^r}\|^2]\notag \\
		&+\sum_{i=1}^{N}\frac{p_i\eta \wt E_i^rL^2}{2}\Psi_i^r + \frac{L}{2} \E[\|\thetab^{r+1} - \thetab^r\|^2] \label{lem: f_descent_bd},
	\end{align}
\end{Lemma}

\begin{Lemma} \label{lem: div_bound}
	For any round $r$ and client $i$, if $\eta \wt E_i^r L < \frac{1}{2}$, it holds that
	\begin{align}
		(1 - 4(\eta \wt E_i^r L)^2)\Psi_i^r 
		\leq & \frac{4}{\gamma^2}\sum_{i=1}^{N}p_i\Xi_i^r
		+ 4(\eta \wt E_i^r)^2\Xi_i^r + \frac{4}{\gamma^2}\E[\|\nabla f(\thetab^r)\|^2]+  \frac{\eta^2\sigma^2\|\bb_i^r\|_2^2}{S}. \label{lem: div_bd}
	\end{align}
\end{Lemma}

\begin{Lemma} \label{lem: opt_bound}
	For any round $r$ and client i, if $0 < a < \min\{\frac{1}{\omega_i^r}, 1\}$, then it holds that
	\begin{align}
		\Xi_i^{r+1} \leq & \frac{2NL^2}{ma(1-a\omega_i^r)} \E[\|\thetab^{r+1} - \thetab^r\|^2]   + \frac{2(N-ma(1-a\omega_i^r))}{2N - ma(1-a\omega_i^r)} \Xi_i^r \notag \\
		&+ \frac{2m a(a\omega_i^r+1)L^2}{2N - ma(1-a\omega_i^r)}\Psi_i^r + \frac{2(1 + \omega_i^r)ma^2\sigma^2\|\bb_i^r\|_2^2}{(2N - ma(1-a\omega_i^r))S\|\bb_i^r\|_1^2}.
	\end{align}
\end{Lemma}

\begin{Lemma} \label{lem: x0_diff_bound}
	For any round $r$, it holds that
	\begin{align}
		\E[\thetab^{r+1} - \thetab^r] =&  -\frac{1}{\gamma} \E[{\rm \Bb^r}],\label{lem: x0_diff1}\\
		\E[\|\thetab^{r+1} - \thetab^r\|^2] 
		\leq &\frac{1}{\gamma^2}\E[\|{\rm \Bb^r}\|^2] +  \frac{\eta^2\sigma^2N}{m S} \sum_{i=1}^{N} (1+\omega_i^r)p_i^2\|\bb_i^r\|_2^2 \notag \\
		& + \frac{2\eta^2N}{m} \sum_{i=1}^{N}(1+\omega_i^r)(p_i\wt E_i^r)^2\Xi_i^r  +  \frac{2\eta^2L^2N}{m} \sum_{i=1}^{N}(1+\omega_i^r)(p_i\wt E_i^r)^2\Psi_i^r.\label{lem: x0_diff2}
	\end{align}
\end{Lemma}

\subsection{Main proof of Theorem 1}
\label{thm: proof}

\noindent {\bf Proof:} We start by combing above lemmas for the construction of the potential function.  In particular, let us respectively scale Lemma \ref{lem: opt_bound} and \eqref{lem: x0_diff2} in Lemma \ref{lem: x0_diff_bound} with positive constants $C_0^r$, $C_1^r = \frac{L}{2} + \frac{2N L^2}{ma(1-a \omegabar^r)} C_0^r$, which gives rise to
\begin{align}
	&C_0^r\Xi_i^{r+1} 	\leq  \frac{2NL^2 C_0^r}{ma(1-a\omegabar^r)} \E[\|\thetab^{r+1} - \thetab^r\|^2]  + \frac{2(N-ma(1-a\omega_i^r))C_0^r}{2N - ma(1-a\omega_i^r)} \Xi_i^r \notag \\
	&~~~~~~~~~~~~~+ \frac{2ma(a\omega_i^r+1) L^2C_0^r}{2N - ma(1-a\omega_i^r)}\Psi_i^r + \frac{2(1+\omega_i^r)ma^2\sigma^2\|\bb_i^r\|_2^2 C_0^r}{(2N - ma(1-a\omega_i^r))S\|\bb_i^r\|_1^2}, \label{thm:  bd1}\\
	&C_1^r \E[\|\thetab^{r+1} - \thetab^r\|^2]
	\leq  \frac{C_1^r}{\gamma^2}\E[\|{\rm \Bb^r}\|^2] +  \frac{\eta^2\sigma^2NC_1^r}{m S}\sum_{i = 1}^{N}(1+\omega_i^r)p_i^2\|\bb_i^r\|_2^2\notag \\
	&~~~~~~~~~~~~~~~~~~~~~~~~~~~~+ \sum_{i=1}^{N} \frac{2(1 + \omega_i^r)N(p_i\eta \Ebar^r L)^2C_1^r}{m}\Psi_i^r  +  \sum_{i=1}^{N}\frac{2(1 + \omega_i^r)N(p_i\eta \Ebar^r)^2C_1^r}{m}\Xi_i^r.\label{thm:  bd2}
\end{align} where $\Ebar^r \triangleq \max\limits_{i} \wt E_i^r$, $\ul E^r \triangleq \max\limits_{i} \wt E_i^r$, and $\omegabar^r = \max\limits_{i} \omega_i^r$. Then, taking average of \eqref{thm:  bd1} with respect to all clients, and summing it with \eqref{thm:  bd2} and \eqref{lem: f_descent_bd} in Lemma \ref{lem: f_descent} yields
\begin{align}
	&\E[f(\thetab^{r+1}) - f(\thetab^r)]  + C_0^r \sum_{i=1}^{N}p_i\Xi_i^{r+1} - C_0^r\sum_{i=1}^{N}p_i\Xi_i^{r}\notag \\
	\leq & \sum_{i=1}^{N}p_i\bigg(\frac{1 - \gamma\eta \ul E^r}{2\gamma} + \frac{2Np_i(1+\omega_i^r)(\eta \Ebar^r)^2C_1^r}{m} + \frac{2(N-ma(1-a\omega_i^r))C_0^r}{2N - ma(1-a\omega_i^r)}\bigg) \Xi_i^r \notag \\
	&- \frac{1}{2\gamma} \E[\|\nabla f(\thetab^r)\|^2]- \frac{\gamma - 2C_1^r}{2\gamma^2}\E[\|{\rm \Bb^r}\|^2]\notag \\
	&+\sum_{i=1}^{N}p_i\bigg(\frac{\eta \Ebar^rL^2}{2} +\frac{2Np_i(1+\omega_i^r)(\eta \Ebar^rL)^2C_1^r}{m} +  \frac{2m a(a\omega_i^r+1) L^2C_0^r}{2N - ma(1-a\omega_i^r)}\bigg)\Psi_i^r \notag \\
	& + \sum_{i=1}^{N} \frac{2p_i(1+\omega_i^r)ma^2\sigma^2 \|\bb_i^r\|_2^2 C_0^r}{(2N - ma(1-a\omega_i^r))S\|\bb_i^r\|_1^2}  + \frac{\eta^2\sigma^2NC_1^r}{m S}\sum_{i = 1}^{N} (1+\omega_i^r)p_i^2\|\bb_i^r\|_2^2.\label{thm:  bd3}
\end{align}
Next, we scale \eqref{lem: div_bd} in Lemma \ref{lem: div_bound} with $C_2^r > 0$, taking average of it with respect to all clients, and add it to \eqref{thm:  bd3}, which yields
\begin{align}
	&\E[f(\thetab^{r+1}) - f(\thetab^r)]  +  C_0^r\sum_{i=1}^{N}p_i\Xi_i^{r+1} -  C_0^r\sum_{i=1}^{N}p_i\Xi_i^{r}\notag \\
	\leq & -\sum_{i=1}^{N}D_0^{r} p_i\Xi_i^r - \frac{\gamma - 8C_2^r}{2\gamma^2} \E[\|\nabla f(\thetab^r)\|^2]- \frac{\gamma - 2C_1^r}{2\gamma^2}\E[\|{\rm \Bb^r}\|^2]-\sum_{i=1}^{N}D_1^{r}p_i\Psi_i^r   \notag \\
	&+ \sum_{i=1}^{N} \frac{2p_i(1+\omega_i^r)ma^2\sigma^2 \|\bb_i^r\|_2^2 C_0^r}{(2N - ma(1-a\omega_i^r))S\|\bb_i\|_1^2}  + \frac{\eta^2\sigma^2NC_1^r}{mS}\sum_{i = 1}^{N} (1+\omega_i^r)p_i^2\|\bb_i^r\|_2^2 + \frac{\eta^2\sigma^2C_2^r}{S}\sum_{i=1}^{N}p_i\|\bb_i^r\|_2^2, \label{thm:  bd4}
\end{align}where $\ol p = \max_{i} p_i$ and
\begin{align}
	D_0^{r} \triangleq &\frac{ma(1-a\omegabar^r)C_0^r}{2N - ma(1-a\omegabar^r)} - \frac{1}{2\gamma} - \frac{2N \ol p (1+ \omegabar^r)(\eta \Ebar^r)^2C_1^r}{m} - \frac{4C_2^r}{\gamma^2} - 4(\eta\Ebar^r)^2 C_2^r, \label{thm:  def_D0}\\
	D_1^{r} \triangleq & (1 - 4(\eta \Ebar^r L)^2) C_2^r -\frac{\eta \Ebar^rL^2}{2} -  \frac{2m a(a\omegabar^r+1) L^2C_0^r}{2N - ma(1-a\omegabar^r)}-\frac{2N \ol p (1+\omegabar^r)(\eta \Ebar^rL)^2C_1^r}{m}. \label{thm:  def_D1}
\end{align}

Let us define $P^r \triangleq \E[f(\thetab^r)] + C_0^r\sum_{i=1}^{N}p_i\Xi_i^r$, and we have from \eqref{thm:  bd4} that
\begin{align}
	&P^{r+1} - P^{r}\notag \\
	\leq & -\sum_{i=1}^{N}D_0^{r}p_i \Xi_i^r - \frac{\gamma - 8C_2^r}{2\gamma^2} \E[\|\nabla f(\thetab^r)\|^2]- \frac{\gamma - 2C_1^r}{2\gamma^2}\E[\|{\rm \Bb^r}\|^2]-\sum_{i=1}^{N}D_1^{r}p_i\Psi_i^r   \notag \\
	& + \sum_{i=1}^{N} \frac{2p_i(1+\omega_i^r)ma^2\sigma^2 \|\bb_i^r\|_2^2 C_0^r}{(2N - ma(1-a\omega_i^r))S\|\bb_i\|_1^2}  + \frac{\eta^2\sigma^2NC_1^r}{mS}\sum_{i = 1}^{N} (1+\omega_i^r)p_i^2\|\bb_i^r\|_2^2 + \frac{\eta^2\sigma^2C_2^r}{S}\sum_{i=1}^{N}p_i\|\bb_i^r\|_2^2.  \label{thm:  bd5}
\end{align}
It is natural to consider $P^r$ as the potential function because the inequality \eqref{thm: bd5} just delineates its progress made in each round. By letting $D_0^{r} = D_1^{r} = 0$, we have
\begin{align}
	&\frac{ma(1-a\omegabar^r)C_0^r}{2N - ma(1-a\omegabar^r)} - \frac{1}{2\gamma} - \frac{2N\ol p (1+\omegabar^r)(\eta \Ebar^r)^2C_1^r}{m} - \frac{4C_2^r}{\gamma^2} - 4(\eta\Ebar^r)^2 C_2^r = 0,\label{thm:  D0=0}\\
	&(1 - 4(\eta \Ebar^r L)^2) C_2^r -\frac{\eta \Ebar^rL^2}{2} -  \frac{2m a(a\omegabar^r+1) L^2C_0^r}{2N - ma(1-a\omegabar^r) }-\frac{2N \ol p (1+\omegabar^r)(\eta \Ebar^rL)^2C_1^r}{m} = 0. \label{thm: D1=0}
\end{align}
Using $D_0^{r} L^2 - D_1^{r} = 0$, we have
\begin{align}
	C_2^r = \frac{\gamma^2}{\gamma^2 + 4L^2}\bigg(\frac{ma(a\omegabar^r+3) L^2 C_0^r}{2N - ma(1-a\omegabar^r)} - \frac{(1 - \gamma \eta \Ebar^r)L^2}{2\gamma} \bigg). \label{thm: D1D2_bd1}
\end{align}
Substituting \eqref{thm: D1D2_bd1} into  \eqref{thm:  def_D0} yields
\begin{align}
	&\frac{m a(1-a\omegabar^r) C_0^r}{2N - ma(1-a\omegabar^r)} - \frac{1}{2\gamma} - \frac{2N\ol p (1+\omegabar^r)(\eta \Ebar^r)^2}{m}\bigg(\frac{L}{2} + \frac{2NL^2C_0^r}{ma(1-a\omegabar^r)}\bigg) \notag \\
	= & \frac{4(1 + (\gamma \eta \Ebar^r)^2)L^2}{\gamma^2 + 4L^2}\bigg(\frac{ma(a\omegabar^r+3) C_0^r}{2N - ma(1-a\omegabar^r)} -\frac{1 - \gamma \eta \Ebar^r}{2\gamma}\bigg),
\end{align}which implies that
\begin{align}
	C_0^r =& \frac{ma(1-a\omegabar^r)(2N-ma(1-a\omegabar^r))z_1^{r}}{2\gamma z_2^{r}},\label{thm: def_C0} \\
	C_2^r = & \frac{\gamma L^2 \big(m^2a^2(1-a\omegabar^r)(a\omegabar^r+3)z_1^{r} - (1 - \gamma \eta \Ebar^r) z_2^{r}\big)}{2(\gamma^2 + 4L^2) z_2^{r}}, \label{thm: def_C2}
\end{align} where
\begin{align}
	z_1^{r} =& (\gamma^2 + 4L^2)\big(m
	+2\gamma N \ol p (1+\omegabar^r)(\eta \Ebar^r)^2L \big) - 4m(1 + (\gamma \eta \Ebar^r)^2)(1 -  \gamma \eta \Ebar^r) L^2, \\
	z_2^{r} = & (\gamma^2 + 4L^2) \big(m^3a^2(1-a\omegabar^r)^2 - 4(1+\omegabar^r)N^2 \ol p (\eta \Ebar^r L)^2 (2N - ma(1-a\omegabar^r))\big) \notag \\
	&-4m^3a^2(a\omegabar^r+3)(1-a\omegabar^r)(1 + (\gamma \eta \Ebar^r)^2) L^2.
\end{align}
In order to get the desired result, we need the following Lemma.
\begin{Lemma} \label{lem: utility}
	If the parameters $\eta$, $\gamma$ and $a$ satisfy, $\forall r, 0 < a < \min\{\frac{1}{\omega_i^r}, 1\}$,
	\begin{align}
		&\eta \leq \min\bigg\{\frac{1}{2\gamma\Ebar^r\sqrt{N\ol p(1+\omegabar^r)}}, \frac{ma(1-a\omegabar^r)\sqrt{m}}{3\sqrt{(1+\omegabar^r)N(2N - ma(1-a\omegabar^r))}}\bigg\}, \\
		& \gamma \geq \max\bigg\{8L, L\sqrt{\frac{30(a\omegabar^r+3)}{(1-a\omegabar^r)} - 4}, \frac{2L\sqrt{N(2N-ma(1-a\omegabar^r))}}{ma(1-a\omegabar^r)} \bigg\},
	\end{align}where $\Ebar^r = \max\limits_{r} \wt E_{i}^r$, then the following results hold.
	\begin{align}
		&z_1^r > \gamma^2m, z_1^r \leq \frac{17m}{16}(\gamma^2 + 4L^2),  z_2^r > \frac{1}{3}m^3a^2(1-a\omegabar^r)^2(\gamma^2 + 4L^2), \\
		& m^2a^2(1-a\omegabar^r)(a\omegabar^r+3)z_1^r > (1 - \gamma\eta\Ebar^r)z_2^r, \gamma \geq 16C_2^r, \gamma \geq 2C_1^r, 
	\end{align}where $z_1^r, z_2^r, C_1^r, C_2^r$ are defined above.
\end{Lemma}

By applying Lemma \ref{lem: utility}, we can have from \eqref{thm:  bd5} that
\begin{align}
	&\frac{1}{4\gamma}	\E[\|\nabla f(\thetab^r)\|^2]  \notag \\
	\leq & P^r - P^{r+1} + \sum_{i=1}^{N} \frac{2p_i(1+\omega_i^r)ma^2\sigma^2 \|\bb_i^r\|_2^2 C_0^r}{(2N - ma(1-a\omega_i^r))S\|\bb_i\|_1^2}  + \frac{\eta^2\sigma^2NC_1^r}{mS}\sum_{i = 1}^{N} (1+\omega_i^r)p_i^2\|\bb_i^r\|_2^2 \notag \\
	&+ \frac{\eta^2\sigma^2C_2^r}{S}\sum_{i=1}^{N}p_i\|\bb_i^r\|_2^2.  \label{thm:  bd6}
\end{align}
Summing \eqref{thm:  bd6} up from $r = 0$ to $R - 1$, and then dividing it by $R$ yields
\begin{align}
	&\frac{1}{R}\sum_{r= 0}^{R-1}	\E[\|\nabla f(\thetab^r)\|^2]  \notag \\
	\leq & \frac{4\gamma(P^0 - P^{R})}{R} + \frac{8\gamma ma^2\sigma^2}{SR} \sum_{r = 0}^{R-1}\sum_{i=1}^{N} \frac{p_i(1+\omega_i^r)\|\bb_i^r\|_2^2C_0^r}{(2N - ma(1-a\omega_i^r))\|\bb_i^r\|_1^2} \notag \\
	&+ \frac{4\gamma \eta^2\sigma^2N}{m SR}\sum_{r = 0}^{R-1}\sum_{i = 1}^{N} (1+\omega_i^r)p_i^2\|\bb_i^r\|_2^2 C_1^r+ \frac{4\gamma \eta^2\sigma^2}{SR}\sum_{r = 0}^{R-1}\sum_{i=1}^{N}p_i\|\bb_i^r\|_2^2C_2^r \\
	\leq & \frac{4\gamma(P^0 - \ul f)}{R} + \frac{8\gamma ma^2\sigma^2}{SR} \sum_{r = 0}^{R-1}\sum_{i=1}^{N} \frac{p_i(1+\omega_i^r)\|\bb_i^r\|_2^2C_0^r}{(2N - ma(1-a\omega_i^r))\|\bb_i^r\|_1^2} \notag \\
	&+ \frac{4\gamma \eta^2\sigma^2N}{m SR}\sum_{r = 0}^{R-1}\sum_{i = 1}^{N} (1+\omega_i^r)p_i^2\|\bb_i^r\|_2^2 C_1^r+ \frac{4\gamma \eta^2\sigma^2}{SR}\sum_{r = 0}^{R-1}\sum_{i=1}^{N}p_i\|\bb_i^r\|_2^2C_2^r.  \label{thm:  bd6}
\end{align}
Then, we need to analyze the terms $C_0^r, C_1^r, C_2^r$ in the RHS of \eqref{thm: bd6}. First, using Lemma \ref{lem: utility}, we get
\begin{align}
	C_0^r =& \frac{ma(1-a\omegabar^r)(2N-ma(1-a\omegabar^r))z_1^{r}}{2\gamma z_2^{r}} \notag \\
	\leq & \frac{3m^2a(1-a\omegabar^r)(2N-ma(1-a\omegabar^r))}{2\gamma m^3a^2(1-a\omegabar^r)^2(\gamma^2 + 4L^2)} \frac{17}{16} (\gamma^2 + 4L^2) \label{thm: bd7} \\
	\leq & \frac{51(2N-ma(1-a\omegabar^r))}{32\gamma ma(1-a\omegabar^r)}.  \label{thm: bd8}
\end{align}
Based on \eqref{thm: bd8}, we can also bound $C_1^r$ by
\begin{align}
	C_1^r =& \frac{L}{2} + \frac{2N L^2}{ma(1-a\omegabar^r)} C_0^r 
	\leq  \frac{L}{2} + \frac{51N(2N-ma(1-a\omegabar^r))L^2}{16\gamma m^2a^2(1-a\omegabar^r)^2}. \label{thm: bd9}
\end{align}
For the term $C_2^r$, we get
\begin{align}
	C_2^r = & \frac{\gamma L^2 \big(m^2a^2(1-a\omegabar^r)(a\omegabar^r+3)z_1^{r} - (1 - \gamma \eta \Qbar^r) z_2^{r}\big)}{2(\gamma^2 + 4L^2) z_2^{r}} \notag \\
	\leq  & \frac{\gamma L^2 \big(m^2a^2(1-a\omegabar^r)(a\omegabar^r+3)z_1^{r} - \frac{1}{2} z_2^{r}\big)}{2(\gamma^2 + 4L^2) z_2^{r}} \\
	\leq &\frac{\gamma L^2 \big(\frac{17}{16}m^3a^2(1-a\omegabar^r)(a\omegabar^r+3)(\gamma^2 + 4L^2) - \frac{1}{2} z_2^{r}\big)}{2(\gamma^2 + 4L^2) z_2^{r}} \\
	\leq & \frac{3\gamma L^2  \big(\frac{17}{16}(a\omegabar^r+3) - \frac{1}{6}(1-a\omegabar^r)\big)}{2(1-a\omegabar^r)} \\
	=  &\frac{\gamma L^2(59a\omegabar^r+145)}{32(1-a\omegabar^r)}. \label{thm: bd10}
\end{align}
Substituting \eqref{thm: bd8},\eqref{thm: bd9} and \eqref{thm: bd10} into the RHS of \eqref{thm: bd6} yields
\begin{align}
	&\frac{1}{R}\sum_{r = 0}^{R -1} \E[\|\nabla f(\thetab^r)\|^2] \notag \\
	\leq& \frac{4\gamma(P^0 - \ul f)}{R} + \frac{51  a\sigma^2 }{4SR }\sum_{r = 0}^{R - 1}\sum_{i=1}^{N} \frac{p_i(1+ \omega_i^r)\|\bb_i^r\|_2^2}{(1-a\omegabar^r) \|\bb_i^r\|_1^2} + \frac{\gamma^2 L^2 \eta^2\sigma^2}{8SR}\sum_{r = 0}^{R - 1}\sum_{i=1}^{N}\frac{p_i(59\omegabar^r + 145)\|\bb_i^r\|_2^2}{1-a\omegabar^r}\notag \\
	& +\sum_{r = 0}^{R - 1}\sum_{i = 1}^{N} \frac{4\gamma (1+ \omega_i^r)\eta^2\sigma^2N p_i\|\bb_i^r\|_2^2}{m SR} \bigg(\frac{L}{2} + \frac{51N(2N-ma(1-a\omegabar^r))L^2}{16\gamma m^2a^2(1-a\omegabar^r)^2}\bigg)  \notag \\
	= & \frac{4\gamma(P^0 - \ul f)}{R} + \frac{51  a\sigma^2 }{4SR }\sum_{r = 0}^{R - 1}\sum_{i=1}^{N} \frac{p_i(1+ \omega_i^r)\|\bb_i^r\|_2^2}{(1-a\omegabar^r) \|\bb_i^r\|_1^2} + \frac{\gamma^2 L^2 \eta^2\sigma^2}{8SR}\sum_{r = 0}^{R - 1}\sum_{i=1}^{N}\frac{p_i(59\omegabar^r + 145)\|\bb_i^r\|_2^2}{1-a\omegabar^r} \notag \\
	&+ \frac{2\gamma \eta^2\sigma^2 L}{mSR}\sum_{r = 0}^{R}\sum_{i=0}^{N} p_i(1+ \omega_i^r) \|\bb_i^r\|_2^2+ \frac{51N\eta^2\sigma^2L^2}{4m^3a^2SR}\sum_{r = 0}^{R}\sum_{i=0}^{N} \frac{p_i(1+ \omega_i^r)(2N - ma(1-a\omegabar^r))\|\bb_i^r\|_2^2 }{(1-a\omegabar^r)^2}  \notag \\
	\leq & \frac{4\gamma(P^0 - \ul f)}{R} + \frac{51  a\sigma^2 }{4SR }\sum_{r = 0}^{R - 1}\sum_{i=1}^{N} \frac{p_i(1+ \omega_i^r)\|\bb_i^r\|_2^2}{(1-a\omegabar^r) \|\bb_i^r\|_1^2} + \frac{\gamma^2 L^2 \eta^2\sigma^2}{8SR}\sum_{r = 0}^{R - 1}\sum_{i=1}^{N}\frac{p_i(59\omegabar^r + 145)\|\bb_i^r\|_2^2}{1-a\omegabar^r} \notag \\
	&+ \frac{2\gamma \eta^2\sigma^2 L}{mSR}\sum_{r = 0}^{R}\sum_{i=0}^{N} p_i(1+ \omega_i^r) \|\bb_i^r\|_2^2+ \frac{51N^2\eta^2\sigma^2L^2}{2m^3a^2SR}\sum_{r = 0}^{R}\sum_{i=0}^{N} \frac{p_i(1+ \omega_i^r)\|\bb_i^r\|_2^2 }{(1-a\omegabar^r)^2} 
\end{align}
Thus, we complete the proof. $\hfill\blacksquare$

\section{Proof of Corollary 2}
\label{appendix: proof_corly2}

\begin{Corollary} \label{corly: com_complexity}
	Suppose $\eta = \Oc(\frac{\sqrt{K_1}}{\sqrt{K_2}})$ and $S = \Oc(\frac{K_1}{\epsilon})$, where $K_1$ and $K_2$ are respectively defined in \eqref{corly: def_K1} and \eqref{corly: def_K2}, the communication complexity  of FedQVR to reach $\epsilon$-accuracy, i.e., $\|\nabla f(\thetab^r)\|^2 \leq \epsilon  $ for some $r \in (1, R)$, is of the order $\Oc(\frac{1}{\epsilon})$, and accordingly, the communication cost (the total number of bits uploaded to the server\footnote{Only the uplink communication cost is considered since it is the primary
		bottleneck when the number of edge devices is large.}) is of the order $\Oc(\frac{m(d(B+1) + \mu)}{\epsilon})$.
\end{Corollary}

\noindent {\bf Proof:} To achieve $\epsilon$-accuracy, we need $\frac{1}{R}\sum_{r= 0}^{R-1}	\E[\|\nabla f(\thetab^r)\|^2]  = \Oc(\epsilon)$, which holds if the following conditions are valid.
\begin{align}
	\frac{4\gamma(P^0 - \ul f)}{R} = \frac{\epsilon}{2}, \frac{\sigma^2 K_1}{S} = \frac{\epsilon}{4},  \frac{\eta^2 \sigma^2 K_2}{S} = \frac{\epsilon}{4},  
\end{align}where 
\begin{align} 
	K_1 =& \frac{51  a }{4R }\sum_{r = 0}^{R - 1}\sum_{i=1}^{N} \frac{p_i(1 + \omega_i^r)\|\bb_i^r\|_2^2}{(1-a\omegabar^r) \|\bb_i^r\|_1^2}, \label{corly: def_K1}\\
	K_2 =&  \frac{2\gamma  L}{mR}\sum_{r = 0}^{R}\sum_{i=0}^{N} p_i(1+ \omega_i^r) \|\bb_i^r\|_2^2 + \frac{51N^2L^2}{2m^3a^2R}\sum_{r = 0}^{R}\sum_{i=0}^{N} \frac{p_i(1+ \omega_i^r)\|\bb_i^r\|_2^2 }{(1-a\omegabar^r)^2} \notag \\
	& + \frac{\gamma^2 L^2 }{8R}\sum_{r = 0}^{R - 1}\sum_{i=1}^{N}\frac{p_i(59\omegabar^r + 145)\|\bb_i^r\|_2^2}{1-a\omegabar^r}.\label{corly: def_K2}
\end{align}
Then, we have
\begin{align}
	R = \frac{4\gamma(P^0 - \ul f)}{8\epsilon}, S = \frac{4K_1\sigma^2}{\epsilon}, \eta = \sqrt{\frac{K_1}{K_2}}.
\end{align}
As the number of bits uploaded to the server for each communication round is $m(d(B+1) +\mu)$, the order of the uplink communication cost is $\Oc(\frac{m(d(B+1) + \mu)}{\epsilon})$. This completes the proof. $\hfill\blacksquare$

\section{Proof of Lemma \ref{lem: utility}}

First of all, we start to prove the results on $z_1^r$ and $z_2^r$. Since $0 < \gamma \eta \Ebar^r \leq \frac{1}{2}$, we have
\begin{align}
	z_1^r >& (\gamma^2 + 4L^2)m - 4mL^2(1-(\gamma\eta\Ebar^r)^2) 
	> \gamma^2m.
\end{align}
Using $\gamma\eta \Ebar^r \leq \frac{1}{2\sqrt{N \ol p (1+\omegabar^r)}}$, we get
\begin{align}
	z_1^r \leq& (\gamma^2 + 4L^2)\big(m + \frac{2N \ol p (1+\omegabar^r)(\gamma \eta \Ebar^r)^2L}{\gamma}\big)- 2mL^2 \notag \\
	\leq & (\gamma^2 + 4L^2)\big(m + \frac{L}{2\gamma}\big)- 2mL^2 
	\leq  \big(\frac{17\gamma^2}{16} + \frac{9L^2}{4}\big)m 
	\leq  \frac{17m}{16}(\gamma^2 + 4L^2).
\end{align}
For $z_2^r$, we obtain
\begin{align}
	z_2^r > &\frac{1}{2}(\gamma^2 + 4L^2)m^3a^2(1-a\omegabar^r)^2 - 4m^3a^2(a\omegabar^r+3)(1-a\omegabar^r)(1+ (\gamma\eta\Ebar^r)^2)L^2 \notag \\
	\geq  & \frac{1}{2}m^3a^2(1-a\omegabar^r) ((\gamma^2 + 4L^2) (1-a\omegabar^r) - 10(a\omegabar^r+3)L^2) \notag \\
	\geq  & \frac{1}{3}m^3a^2(1-a\omegabar^r)^2(\gamma^2 + 4L^2), \label{lem: util_bd1}
\end{align}where \eqref{lem: util_bd1} follows because $\gamma \geq L\sqrt{\frac{30(a\omegabar^r + 3)}{(1-a\omegabar^r)} - 4}$. Similarly, using above results, we have
\begin{align}
	& m^2a^2(1-a\omegabar^r)(a\omegabar^r+3)z_1^r - (1 - \gamma\eta\Ebar^r)z_2^r\notag \\
	> & \gamma^2m^3a^2(1-a\omegabar^r)(a\omegabar^r+3)- z_2^r \notag \\
	> & 3\gamma^2m^3a^2(1-a\omegabar^r)^2 - (\gamma^2 + 4L^2)m^3a^2(1-a\omegabar^r)^2 \\
	= & (2\gamma^2 - 4L^2)m^3a^2(1-a\omegabar^r)^2 
	\geq 0. 
\end{align}
Then, we proceed to prove the results on $C_1^r$ and $C_2^r$. In particular, we have
\begin{align}
	&\gamma - 16C_2^r \notag \\
	= & \frac{\gamma}{(\gamma^2 + 4L^2)z_2^r}\big[(\gamma^2 + 4L^2 + 8L^2(1-\gamma\eta \Qbar^r))z_2^r - 8L^2m^2a^2(1-a\omegabar^r)(a\omegabar^r +3)z_1^r\big] \notag \\
	\geq & \frac{\gamma }{(\gamma^2 + 4L^2) z_2^r}\big[(\gamma^2 + 8L^2)z_2^r - 8L^2m^2a^2(1-a\omegabar^r)(a\omegabar^r+3)z_1^r\big] \label{lem: util_bd2}\\
	\geq & \frac{\gamma }{(\gamma^2 + 4L^2) z_2^r} \big[(\gamma^2 + 8L^2)z_2^r -\frac{17}{2}(\gamma^2 + 4L^2)L^2m^3a^2(1-a\omegabar^r)(a\omegabar^r+3)\big] \label{lem: util_bd3}\\
	\geq & \frac{\gamma(\frac{3\gamma^2}{2} + 46L^2)L^2m^3a^2(1-a\omegabar^r)(a\omegabar^r+3)}{(\gamma^2 + 4L^2) z_2^r} > 0, 
\end{align}where \eqref{lem: util_bd2} holds by $\gamma \eta \Ebar^r \leq \frac{1}{2}$; \eqref{lem: util_bd3} follows because $z_2^r \geq   10m^3a^2(1-a\omegabar^r)(a\omegabar^r+3)L^2$ and $z_1^r \leq \frac{17m}{16}(\gamma^2 + 4L^2)$. The result $\gamma \geq 2C_1^r$ is also true because
\begin{align}
	&\gamma - 2C_1^r \notag \\
	= & \gamma - L - \frac{N(2N-ma(1-a\omegabar^r))L^2 z_1^r}{\gamma z_2^r} \notag \\
	= & \frac{\gamma(\gamma - L)z_2^r - N(2N-ma(1-a\omegabar^r))L^2z_1^r}{\gamma z_2^r} 
	\notag \\
	\geq & \frac{1}{3\gamma z_2^r}\bigg[\gamma(\gamma - L)m^3a^2(1-a\omegabar^r)^2(\gamma^2 + 4L^2) - \frac{51mN}{16}(2N-ma(1-a\omegabar^r))L^2(\gamma^2 + 4L^2)\bigg] \label{lem: util_bd4} \\
	= & \frac{m(\gamma^2 + 4L^2)}{3\gamma z_2^r}\bigg[\gamma(\gamma - L)m^2a^2(1-a\omegabar^r)^2 - \frac{51}{16}N(2N-ma(1-a\omegabar^r))L^2\bigg] \notag \\
	\geq & \frac{m(\gamma^2 + 4L^2)}{3\gamma z_2^r}\bigg[\frac{7}{8}\gamma^2m^2a^2(1-a\omegabar^r)^2 - \frac{51}{16}N(2N-ma(1-a\omegabar^r))L^2\bigg] \label{lem: util_bd4}\\
	\geq & 0, \label{lem: util_bd4}
\end{align}where  \eqref{lem: util_bd4} holds because $z_2^r > \frac{1}{3}m^3a^2(1-a\omegabar^r)^2(\gamma^2 + 4L^2)$ and $z_1^r \leq \frac{17m}{16}(\gamma^2 + 4L^2)$; \eqref{lem: util_bd4} follows because $\gamma \geq \frac{2L\sqrt{N(2N-ma(1-a\omegabar^r))}}{ma(1-a\omegabar^r)}$. 
Thus, we complete the proof. $\hfill\blacksquare$

\section{Proof of Lemma \ref{lem: f_descent}}
By the definition of $f(\cdot)$, we have
\begin{align}
	\E[f(\thetab^{r+1}) - f(\thetab^r)] 
	\leq & \E[\langle\nabla f(\thetab^r),\thetab^{r+1} -\thetab^r\rangle]+\frac{L}{2} \E[\|\thetab^{r+1} - \thetab^r\|^2], \label{lem: one_round_obj_bd2}
\end{align}
where \eqref{lem: one_round_obj_bd2} follows by Assumption 3. Then, we proceed to bound the terms in the right hand side (RHS) of \eqref{lem: one_round_obj_bd2} with the help of Lemma \ref{lem: x0_diff_bound}.

By applying Lemma \ref{lem: x0_diff_bound}, we get
\begin{align}
	&\E[\langle\nabla f(\thetab^r),\thetab^{r+1} -\thetab^r\rangle]	\notag \\
	=& - \frac{1}{\gamma}\E[\langle\nabla f(\thetab^r),{\rm \Bb^r}\rangle] \notag \\
	= &\frac{1}{2\gamma}\E[\|\nabla f(\thetab^r)-{\rm \Bb^r}\|^2] - \frac{1}{2\gamma} \E[\|\nabla f(\thetab^r)\|^2] - \frac{1}{2\gamma}\E[\|{\rm \Bb^r}\|^2] \label{lem: one_round_obj_bd3}\\
	= &\frac{1}{2\gamma}\E\bigg[\bigg\|\sum_{i=1}^{N}p_i\bigg((1-\gamma\eta \wt E_i^r)(\nabla f_i(\thetab^r)-\cb_i^r)+ \gamma\eta \wt E_i^r\sum_{t= 0}^{E_i^r - 1}\frac{b_i^{r, t}}{\|\bb_i^r\|_1} (\nabla f_i(\thetab^r)-\nabla f_i(\wt \thetab_i^{r,t}))\bigg)\bigg\|^2\bigg] \notag \\
	&- \frac{1}{2\gamma} \E[\|\nabla f(\thetab^r)\|^2]  - \frac{1}{2\gamma}\E[\|{\rm \Bb^r}\|^2]  \notag \\
	\leq & \sum_{i=1}^{N}\frac{p_i(1 - \gamma\eta \wt E_i^r)}{2\gamma N}\Xi_i^r - \frac{1}{2\gamma} \E[\|\nabla f(\thetab^r)\|^2]- \frac{1}{2\gamma}\E[\|{\rm\Bb^r}\|^2]\notag \\
	&+\sum_{i=1}^{N}\frac{p_i(\gamma\eta \wt E_i^r)}{2\gamma }\sum_{t= 0}^{E_i^r - 1}\frac{b_i^{r, t}}{\|\bb_i^r\|_1}\E[\|\nabla f_i(\thetab^r)-\nabla f_i(\wt \thetab_i^{r,t})\|^2], \label{lem: one_round_obj_bd4} \\
	\leq & \sum_{i=1}^{N}\frac{p_i(1 - \gamma\eta \wt E_i^r)}{2\gamma}\Xi_i^r - \frac{1}{2\gamma} \E[\|\nabla f(\thetab^r)\|^2]- \frac{1}{2\gamma}\E[\|{\rm \Bb^r}\|^2]+\sum_{i=1}^{N}p_i\frac{\eta \wt E_i^rL^2}{2}\Psi_i^r, \label{lem: one_round_obj_bd5}		
\end{align}
where \eqref{lem: one_round_obj_bd3} follows because $\langle \vb_1, \vb_2 \rangle = \frac{1}{2}\|\vb_1\|^2 + \frac{1}{2}\|\vb_2\|^2 - \frac{1}{2}\|\vb_1 - \vb_2\|^2, \forall \vb_1, \vb_2 \in \Rbb^n$; \eqref{lem: one_round_obj_bd4} holds by the convexity of $\|\cdot\|^2$; \eqref{lem: one_round_obj_bd5} holds by Assumption 3.
Substituting the results of \eqref{lem: one_round_obj_bd5} and \eqref{lem: x0_diff2} into \eqref{lem: one_round_obj_bd2} yields
\begin{align}
	\E[f(\thetab^{r+1}) - f(\thetab^r)] 
	\leq & \sum_{i=1}^{N}\frac{p_i(1 - \gamma\eta \wt E_i^r)}{2\gamma}\Xi_i^r - \frac{1}{2\gamma} \E[\|\nabla f(\thetab^r)\|^2]- \frac{1}{2\gamma}\E[\|{\rm \Bb^r}\|^2]\notag \\
	&+\sum_{i=1}^{N}\frac{p_i\eta \wt E_i^rL^2}{2}\Psi_i^r + \frac{L}{2} \E[\|\thetab^{r+1} - \thetab^r\|^2],
\end{align}
This completes the proof. \hfill $\blacksquare$

\section{Proof of Lemma \ref{lem: div_bound}}
According to the update of $\thetab_{i}^{r, t}$, we have
\begin{align}
	&\E[\|\thetab^r-\wt \thetab_i^{r,t}\|^2] \notag \\
	= &\E[\|\thetab^r- \thetab_0^r + \thetab_{0}^r - \wt \thetab_i^{r,t}\|^2] \\
	= & \E\bigg[\bigg\|\frac{\cb^r}{\gamma}+ \thetab_{0}^r - \wt \thetab_i^{r,t}\bigg\|^2\bigg] \\
	= &  \frac{1}{\gamma^2} \E\bigg[\bigg\|\cb^r + \gamma\eta \sum_{k = 0}^{t - 1}b_i^{r, E_i^r + k - t}(g_i(\wt \thetab_i^{r, k}) - \cb_i^r)\bigg\|^2\bigg] \\
	= & \frac{1}{\gamma^2} \E\bigg[\bigg\|\cb^r  - \nabla f(\thetab^r)+ \nabla f(\thetab^r) \notag \\
	&~~~~~~~~+ \gamma\eta \sum_{k = 0}^{t - 1}b_i^{r, E_i^r + k - t}(g_i(\wt \thetab_i^{r, k}) - \nabla f_i(\wt \thetab_{i}^{r,k}) + \nabla f_i(\wt \thetab_{i}^{r,k}) -\nabla f_i(\thetab^r) + \nabla f_i(\thetab^r) - \cb_i^r)\bigg\|^2\bigg] \\
	\leq &  \frac{4}{\gamma^2} \E[\|\cb^r  - \nabla f(\thetab^r)\|^2] + \frac{4}{\gamma^2}\E[\|\nabla f(\thetab^r)\|^2] + \frac{1}{\gamma^2} \E\bigg[\bigg\|\gamma\eta \sum_{k = 0}^{t - 1}b_i^{r, E_i^r + k - t}(g_i(\wt \thetab_i^{r, k}) - \nabla f_i(\wt \thetab_{i}^{r,k}))\bigg\|^2\bigg] \notag \\
	& + \frac{4}{\gamma^2} \E\bigg[\bigg\|\gamma\eta \sum_{k = 0}^{t - 1}b_i^{r, E_i^r + k - t}( \nabla f_i(\wt \thetab_{i}^{r,k})-\nabla f_i(\thetab^r))\bigg\|^2\bigg]  + \frac{4}{\gamma^2} \E\bigg[\bigg\|\gamma\eta \sum_{k = 0}^{t - 1}b_i^{r, E_i^r + k - t}(\nabla f_i(\thetab^r) - \cb_i^r)\bigg\|^2\bigg]\\
	\leq &  \frac{4}{\gamma^2} \E[\|\cb^r  - \nabla f(\thetab^r)\|^2] + \frac{4}{\gamma^2}\E[\|\nabla f(\thetab^r)\|^2] + \eta^2\sum_{k = 0}^{t - 1}(b_i^{r, E_i^r + k - t})^2 \E[\| g_i(\wt \thetab_i^{r, k}) - \nabla f_i(\wt \thetab_{i}^{r,k})\|^2] \notag \\
	& + 4\eta^2\bigg(\sum_{k = 0}^{t - 1}b_i^{r, E_i^r + k - t}\bigg)\sum_{k = 0}^{t - 1}b_i^{r, E_i^r + k - t}\E[\| \nabla f_i(\wt \thetab_{i}^{r,k})-\nabla f_i(\thetab^r)\|^2]  \notag \\
	& + 4\eta^2\bigg(\sum_{k = 0}^{t - 1}b_i^{r, E_i^r + k - t}\bigg)\sum_{k = 0}^{t - 1}b_i^{r, E_i^r + k - t} \E[\|\nabla f_i(\thetab^r) - \cb_i^r\|^2]\\
	\leq &  \frac{4}{\gamma^2}\sum_{i=1}^{N}p_i\Xi_i^r + \frac{4}{\gamma^2}\E[\|\nabla f(\thetab^r)\|^2]  + 4\eta^2L^2\bigg(\sum_{k = 0}^{t - 1}b_i^{r, E_i^r + k - t}\bigg)\sum_{k = 0}^{t - 1}b_i^{r, E_i^r + k - t}\E[\| \wt \thetab_{i}^{r,k}-\thetab^r\|^2] \notag \\
	&+ 4\eta^2\bigg(\sum_{k = 0}^{t - 1}b_i^{r, E_i^r + k - t}\bigg)^2 \E[\|\nabla f_i(\thetab^r) - \cb_i^r\|^2]+ \frac{\eta^2\sigma^2}{S}\sum_{k = 0}^{t - 1}(b_i^{r, E_i^r + k - t})^2. \label{lem： div_bd1}
\end{align}
Taking weighted average over the two sides of \eqref{lem： div_bd1} with respect to $t$ yields
\begin{small}
	\begin{align}
		\Psi_i^{r} 
		\leq &\frac{4}{\gamma^2}\sum_{i=1}^{N}p_i\Xi_i^r + \frac{4}{\gamma^2}\E[\|\nabla f(\thetab^r)\|^2]+  4\eta^2L^2\sum_{t=0}^{E_i^r - 1} \frac{b_i^{r, t}}{\|\bb_i^r\|_1}\bigg(\sum_{k = 0}^{t - 1}b_i^{r, E_i^r + k - t}\bigg)\sum_{k = 0}^{E_i^r - 1}b_i^{r, k}\E[\| \wt \thetab_{i}^{r,k}-\thetab^r\|^2] \notag \\
		& +  4\eta^2\sum_{t=0}^{E_i^r - 1} \frac{b_i^{r, t}}{\|\bb_i^r\|_1}\bigg(\sum_{k = 0}^{t - 1}b_i^{r, E_i^r + k - t}\bigg)^2 \E[\|\nabla f_i(\thetab^r) - \cb_i^r\|^2]  + \frac{\eta^2\sigma^2}{S}\sum_{t = 0}^{E_i^r - 1}\frac{b_i^{r,t}}{\|\bb_i^r\|_1}\sum_{k = 0}^{t - 1}(b_i^{r, E_i^r + k - t})^2.
	\end{align}
\end{small}
Furthermore, note that
\begin{align}
	\sum_{t= 0}^{E_i^r - 1} \frac{b_i^{r,t}}{\|\bb_i^r\|_1}\sum_{k = 0}^{t - 1}b_i^{r, E_i^r + k - t} 
	\leq & \sum_{t= 0}^{E_i^r - 1} \frac{b_i^{r,t}}{\|\bb_i^r\|_1}\sum_{k = 0}^{E_i^r - 2}b_i^{r, k +1}
	= \|\bb_i^r\|_1 - b_i^{r, 0} \leq  \wt E_i^r, \\
	\sum_{t= 0}^{E_i^r - 1} \frac{b_i^{r, t}}{\|\bb_i^r\|_1}\sum_{k = 0}^{t - 1}(b_i^{r, E_i^r + k - t})^2 
	\leq& \sum_{t= 0}^{E_i^r - 1} \frac{b_i^{r, t}}{\|\bb_i^r\|_1}\sum_{k = 0}^{E_i^r - 2}(b_i^{r, k+1})^2 
	=  \|\bb_i^r\|_2^2 - (b_i^{r, 0})^2\leq  \|\bb_i^r\|_2^2, \\
	\sum_{t=0}^{E_i^r - 1} \frac{b_i^{r, t}}{\|\bb_i^r\|_1}\bigg(\sum_{k = 0}^{t - 1}b^{r, E_i^r + k - t}\bigg)^2 
	\leq &\sum_{t=0}^{E_i^r - 1} \frac{b_i^{r, t}}{\|\bb_i^r\|_1}\bigg(\sum_{k = 0}^{E_i^r - 2}b_i^{r, k + 1}\bigg)^2 
	=  (\|\bb_i^r\|_1 - b_i^{r, 0})^2 \leq  (\wt E_i^r)^2.
\end{align}
Therefore, we have
\begin{align}
	\Psi_i^r
	\leq & \frac{4}{\gamma^2}\sum_{i=1}^{N}p_i\Xi_i^r
	+ 4(\eta \wt E_i^r L)^2\Psi_i^r  + 4(\eta \wt E_i^r)^2\Xi_i^r + \frac{4}{\gamma^2}\E[\|\nabla f(\thetab^r)\|^2]+  \frac{\eta^2\sigma^2\|\bb_i^r\|_2^2}{S} .
\end{align}
Rearranging the two sides of the above inequality yields
\begin{align}
	(1 - 4(\eta \wt E_i^r L)^2)\Psi_i^r 
	\leq & \frac{4}{\gamma^2}\sum_{i=1}^{N}p_i\Xi_i^r
	+ 4(\eta \wt E_i^r)^2\Xi_i^r + \frac{4}{\gamma^2}\E[\|\nabla f(\thetab^r)\|^2]+  \frac{\eta^2\sigma^2\|\bb_i^r\|_2^2}{S}.
\end{align}
This completes the proof.\hfill $\blacksquare$

\section{Proof of Lemma \ref{lem: opt_bound}}
First, we have
\begin{align}
	\Xi_i^{r+1} 
	= &	\E[\|\nabla f_i(\thetab^{r+1}) - \nabla f_i(\thetab^r) + \nabla f_i(\thetab^r) - \cb_i^{r+1}\|^2] \\
	\leq & (1 + 1/\epsilon_i^r) \E[\|\nabla f_i(\thetab^{r+1}) - \nabla f_i(\thetab^r) \|^2] + (1 + \epsilon_i^r) \E[\|\nabla f_i(\thetab^r) - \cb_i^{r+1}\|^2] \label{lem: opt_bd0}\\
	\leq & (1 + 1/\epsilon_i^r)L^2 \E[\|\thetab^{r+1} - \thetab^r\|^2] + (1 + \epsilon_i^r) \frac{m}{N} \E[\|\nabla f_i(\thetab^r) - \wt \cb_i^{r+1}\|^2]   + (1 + \epsilon_i^r) (1 - \frac{m}{N}) \Xi_i^r, \label{lem: opt_bd1}
\end{align}where $\epsilon_i^r > 0$ is a constant, and \eqref{lem: opt_bd0} follows by the fact that $(z_1 + z_2)^2 \leq (1 + \frac{1}{c}) z_1^2 + (1 + c)z_2^2, \forall c > 0$.  
Then, by using \eqref{eqn: def_virtual_dual}, we can bound $\E[\| \nabla f_i(\thetab^r) - \wt \cb_i^{r+1}\|^2] $ by
\begin{small}
	\begin{align}
		&\E[\| \nabla f_i(\thetab^r) - \wt \cb_i^{r+1}\|^2] \notag \\
		= &\E\bigg[\bigg\| \nabla f_i(\thetab^r) - \cb_i^{r} + \frac{a}{\eta \wt E_i^r}\wt \Delta_i^{r+1}\bigg\|^2\bigg] \notag \\
		=  &\E\bigg[\bigg\| \nabla f_i(\thetab^r) - \cb_i^{r} - \frac{a}{\eta \wt E_i^r}(\thetab_0^r - \wt \thetab_i^{r+1}) + \frac{a}{\eta \wt E_i^r}(\thetab_0^r - \wt \thetab_i^{r+1} +\wt \Delta_i^{r+1})\bigg\|^2\bigg] \notag \\
		=&\E\bigg[\bigg\| \nabla f_i(\thetab^r) -(1 - a)\cb_i^r - a\sum_{t= 0}^{E_i^r - 1}\frac{b_i^{r,t}}{\|\bb_i^r\|_1} g_i(\wt \thetab_i^{r,t}) \bigg\|^2\bigg] + \frac{a^2}{(\eta \wt E_i^r)^2}\E[\|\wt \Delta_i^{r+1} - (\wt \thetab_i^{r+1} - \thetab_0^r)\|^2] \label{lem: opt_bd2_0}\\
		\leq & \E\bigg[\bigg\| \nabla f_i(\thetab^r) -(1 - a)\cb_i^r - a\sum_{t= 0}^{E_i^r - 1}\frac{b_i^{r,t}}{\|\bb_i^r\|_1} g_i(\wt \thetab_i^{r,t})\bigg\|^2\bigg]  +\frac{a^2 \omega_i^r }{(\eta\wt E_i^r)^2}\E[\|\wt \thetab_i^{r+1} - \thetab_0^r\|^2]\label{lem: opt_bd2_1}\\
		= & \E\bigg[\bigg\| \nabla f_i(\thetab^r) -(1 - a)\cb_i^r - a\sum_{t= 0}^{E_i^r - 1}\frac{b_i^{r,t}}{\|\bb_i^r\|_1} \nabla f_i(\wt \thetab_i^{r,t})\bigg\|^2\bigg]  + \E\bigg[\bigg\| a\sum_{t= 0}^{E_i^r - 1}\frac{b_i^{r,t}}{\|\bb_i^r\|_1} (g_i(\wt \thetab_{i}^{r, t}) -\nabla f_i(\wt \thetab_i^{r,t}))\bigg\|^2\bigg] \notag \\
		&+ \frac{a^2 \omega_i^r}{(\eta\wt E_i^r)^2}\E[\|\wt \thetab_i^{r+1} - \thetab_0^r\|^2] \label{lem: opt_bd2} \\
		= & \E\bigg[\bigg\| \nabla f_i(\thetab^r) -(1 - a)\cb_i^r - a\sum_{t= 0}^{E_i^r - 1}\frac{b_i^{r,t}}{\|\bb_i^r\|_1} \nabla f_i(\wt \thetab_i^{r,t})\bigg\|^2\bigg]  + a^2\sum_{t= 0}^{E_i^r - 1}\frac{(b_i^{r,t})^2}{\|\bb_i^r\|_1^2}\E[\| g_i(\wt \thetab_{i}^{r, t}) -\nabla f_i(\wt \thetab_i^{r,t})\|^2] \notag \\
		&+ \frac{a^2 \omega_i^r }{(\eta\wt E_i^r)^2}\E[\|\wt \thetab_i^{r+1} - \thetab_0^r\|^2] \label{lem: opt_bd3}\\
		\leq & \E\bigg[\bigg\| \nabla f_i(\thetab^r) -(1 - a)\cb_i^r - a\sum_{t= 0}^{E_i^r - 1}\frac{b_i^{r,t}}{\|\bb_i^r\|_1} \nabla f_i(\wt \thetab_i^{r,t})\bigg\|^2\bigg]  + \frac{(1+\omega_i^r)a^2\sigma^2 \|\bb_i^r\|_2^2}{S\|\bb_i^r\|_1^2} +   a^2\omega_i^r (L^2 \Psi_i^r +  \Xi_i^r)\label{lem: opt_bd4}\ \\ 
		\leq & (1- a)\Xi_i^r + a\sum_{t= 0}^{E_i^r - 1}\frac{b_i^{r,t}}{\|\bb_i^r\|_1}\E[\|\nabla f_i(\thetab^r)  - \nabla f_i(\wt \thetab_i^{r,t})\|^2] 
		+ \frac{(1+\omega_i^r)a^2\sigma^2 \|\bb_i^r\|_2^2}{S\|\bb_i^r\|_1^2}+   a^2\omega_i^r (L^2 \Psi_i^r +  \Xi_i^r) \label{lem: opt_bd5}\\
		\leq &(1- a + a^2\omega_i^r)\Xi_i^r + a(a\omega_i^r +1)L^2\Psi_i^r +\frac{(1+\omega_i^r)a^2\sigma^2\|\bb_i^r\|_2^2}{S\|\bb_i^r\|_1^2}, \label{lem: opt_bd6}
	\end{align}
\end{small}where \eqref{lem: opt_bd2_0} holds by Assumption 3; \eqref{lem: opt_bd2_1} and \eqref{lem: opt_bd2} follow by \cite[Lemma 4]{SCAFFOLD_2020}; \eqref{lem: opt_bd3} follows by the independence of data sampling used to obtain local SGs;  \eqref{lem: opt_bd5} follows because of the convexity of $\|\cdot\|^2$; \eqref{lem: opt_bd6} holds due to Assumption 1; \eqref{lem: opt_bd4} follows because of Assumption 2 and 
\begin{align}
	&\E[\|\wt \thetab_i^{r+1} - \thetab_0^r\|^2] \notag \\
	= & (\eta \wt E_i^r)^2\E\bigg[\bigg\|\sum_{t= 0}^{E_i^r - 1}\frac{b_i^{r,t}}{\|\bb_i^r\|_1}(g_i(\wt\thetab_i^{r,t}) - \cb_i^r)\bigg\|^2\bigg] \\
	=&  (\eta \wt E_i^r)^2\E\bigg[\bigg\|\sum_{t= 0}^{E_i^r - 1}\frac{b_i^{r, t}}{\|\bb_i^r\|_1} (g_i(\wt \thetab_i^{r,t}) - \nabla f_i(\wt \thetab_i^{r,t}) +\nabla f_i(\wt \thetab_i^{r,t}) - \nabla f_i(\thetab^{r} )+\nabla f_i(\thetab^{r} ) - \cb_i^r)\bigg\|^2\bigg] \\
	\leq & (\eta \wt E_i^r)^2 \E\bigg[\bigg\|\sum_{t= 0}^{E_i^r - 1}\frac{b_i^{r, t}}{\|\bb_i^r\|_1} (g_i(\wt \thetab_i^{r,t}) - \nabla f_i(\wt \thetab_i^{r,t}))\bigg\|\bigg] + (\eta \wt E_i^r)^2\E\bigg[\bigg\|\sum_{t= 0}^{E_i^r - 1}\frac{b_i^{r, t}}{\|\bb_i^r\|_1} (\nabla f_i(\wt \thetab_i^{r,t}) - \nabla f_i(\thetab^r))\bigg\|^2\bigg] \notag \\
	& + (\eta \wt E_i^r)^2\E[\| \nabla f_i(\thetab^{r}) - \cb_i^r\|^2]\\
	\leq &  \frac{(\eta \wt E_i^r)^2\sigma^2 \|\bb_i^r\|_2^2}{S\|\bb_i^r\|_1^2} + (\eta \wt E_i^r)^2L^2 \Psi_i^r + (\eta \wt E_i^r)^2 \Xi_i^r.
\end{align}
Then, substituting \eqref{lem: opt_bd3} into \eqref{lem: opt_bd1} yields
\begin{align}
	\Xi_i^{r+1} 
	\leq & (1 + 1/\epsilon_i^r)L^2 \E[\|\thetab^{r+1} - \thetab^r\|^2]+ (1 + \epsilon_i^r) (1 - \frac{m}{N})\Xi_i^r \notag \\
	&+ (1 + \epsilon_i^r) \frac{m}{N} \bigg((1- a + a^2\omega_i^r)\Xi_i^r + a(a\omega_i^r+1) L^2\Psi_i^r\bigg)  + (1 + \epsilon_i^r) \frac{m}{N}\frac{(1+\omega_i^r) a^2\sigma^2\|\bb_i^r\|_2^2}{S\|\bb_i^r\|_1^2} \label{lem: opt_bd7}\\
	=& (1 + 1/\epsilon_i^r)L^2 \E[\|\thetab^{r+1} - \thetab^r\|^2] + (1 + \epsilon_i^r)\frac{(1+\omega_i^r) m a^2 \sigma^2\|\bb_i^r\|_2^2}{N S\|\bb_i^r\|_1^2}\notag \\
	& + (1 + \epsilon_i^r) \bigg(1 - \frac{m a(1-a\omega_i^r)}{N}\bigg) \Xi_i^r + (1 + \epsilon_i^r) \frac{m a(a\omega_i^r +1) L^2}{N} \Psi_i^r.\label{lem: opt_bd8}
\end{align}
Let us pick $\epsilon_i^r = \frac{ma(1-a\omega_i^r)}{2N - ma(1-a\omega_i^r)}$, then \eqref{lem: opt_bd8} can be simplified to	
\begin{align}
	\Xi_i^{r+1} 
	\leq & \bigg(1 + \frac{2N - ma(1-a\omega_i^r)}{ma(1-a\omega_i^r)}\bigg)L^2 \E[\|\thetab^{r+1} - \thetab^r\|^2]    \notag \\
	& + \bigg(1 +  \frac{ma(1-a\omega_i^r)}{2N - ma(1-a\omega_i^r)}\bigg) \bigg(1 - \frac{ma(1-a\omega_i^r)}{N}\bigg) \Xi_i^r \notag \\
	&+ \frac{2(1+\omega_i^r)ma^2\sigma^2\|\bb_i^r\|_2^2}{(2N - ma(1-a\omega_i^r))S\|\bb_i^r\|_1^2}+ \frac{2 m a(a\omega_i^r+1)L^2}{2N - ma(1-a\omega_i^r)}\Psi_i^r\notag \\
	= & \frac{2NL^2}{ma(1-a\omega_i^r)} \E[\|\thetab^{r+1} - \thetab^r\|^2]  + \frac{2(1 + \omega_i^r)ma^2\sigma^2\|\bb_i^r\|_2^2}{(2N - ma(1-a\omega_i^r))S\|\bb_i^r\|_1^2} \notag \\
	& + \frac{2(N-ma(1-a\omega_i^r))}{2N - ma(1-a\omega_i^r)} \Xi_i^r + \frac{2m a(a\omega_i^r+1)L^2}{2N - ma(1-a\omega_i^r)}\Psi_i^r.
\end{align}
This completes the proof.
\hfill $\blacksquare$

	

\section{Proof of Lemma \ref{lem: x0_diff_bound}}
By the definition of $\thetab^{r+1}$, we have
\begin{align}
	\E[\thetab^{r+1} - \thetab^r] 
	= & \E[\thetab^{r+1} - \thetab_0^r + \thetab_0^r - \thetab^r] 
	=  \E\bigg[\frac{N}{m} \sum\limits_{i \in  \Ac^r} p_i\Delta_i^{r+1} - \frac{1}{\gamma} \cb^{r}\bigg] \\
	= & - \frac{1}{\gamma} \E\bigg[\cb^{r} + \frac{\gamma N}{m} \sum\limits_{i \in  \Ac^r} (\thetab_0^r-\thetab_i^{r+1})\bigg] 
	=  - \frac{1}{\gamma} \E\bigg[ \cb^r +\gamma\sum\limits_{i =1}^{N} p_i (\thetab_0^r - \wt \thetab_i^{r+1}) \bigg], \label{lem: x0_diff_bd1}
\end{align}where \eqref{lem: x0_diff_bd1} holds by Assumption 3, and the fact that $\Prob(i \in \Ac^r) = \frac{m}{N}$.
Note that
\begin{align}
	\wt \thetab_i^{r+1} - \thetab_0^r = - \eta \wt E_i^r \sum_{t= 0}^{E_i^r - 1}\frac{b_i^{r, t}}{\|\bb_i^r\|_1} (g_i(\wt \thetab_i^{r,t})- \cb_i^r).\label{lem: x0_diff_bd3}
\end{align}
Thus, we have from \eqref{lem: x0_diff_bd1} that
\begin{align}
	\E[\thetab^{r+1} - \thetab^r] 
	=& - \frac{1}{\gamma} \E\bigg[ \cb^r +\frac{\gamma}{N}\sum\limits_{i =1}^N  \eta \wt E_i^r \sum_{t= 0}^{E_i^r - 1}\frac{b_i^{r+1}}{\|\bb_i\|_1} (g_i(\wt \xb_i^{r,t})- \cb_i^r) \bigg]\\
	= & - \frac{1}{\gamma} \E\bigg[\frac{1}{N}\sum_{i = 1}^{N}\bigg(\cb_i^r + \gamma \eta \wt E_i^r \sum_{t= 0}^{E_i^r- 1}\frac{b_i^{r, t}}{\|\bb_i^r\|_1} (g_i(\wt \thetab_i^{r,t})- \cb_i^r)\bigg)\bigg]\notag \\
	= & - \frac{1}{\gamma} \E\bigg[\frac{1}{N}\sum_{i = 1}^{N}\bigg((1- \gamma \eta \wt E_i^r)\cb_i^r + \gamma \eta \wt E_i^r\sum_{t= 0}^{E_i^r- 1}\frac{b_i^{r, t}}{\|\bb_i^r\|_1} \nabla f_i(\wt \thetab_i^{r,t})\bigg)\bigg] \\
	= &-\frac{1}{\gamma} \E[{\rm \Bb^r}].
\end{align}

On the other hand, 	
by applying \cite[Lemma 4]{SCAFFOLD_2020}, we have
\begin{align}
	&\E[\|\thetab^{r+1} - \thetab^r\|^2] \notag \\
	=&\frac{1}{\gamma^2}\E\bigg[\bigg\|\cb^r -\frac{\gamma}{m}\sum\limits_{i \in \Ac^r}\Delta_i^{r+1}\bigg\|^2\bigg]	\notag \\
	=&\frac{1}{\gamma^2}\E\bigg[\bigg\|\cb^r + \frac{\gamma N}{m}\sum\limits_{i \in \Ac^r}p_i(\thetab_0^r - \thetab_i^{r+1}) + \frac{\gamma N}{m}\sum\limits_{i \in \Ac^r}p_i(\thetab_i^{r+1} - \thetab_0^r - \Delta_i^{r+1})\bigg\|^2\bigg]	\notag \\
	=&\frac{1}{\gamma^2}\E\bigg[\bigg\|\cb^r + \frac{\gamma N}{m}\sum\limits_{i \in \Ac^r}p_i(\thetab_0^r - \thetab_i^{r+1})\bigg\|^2\bigg] + \frac{1}{\gamma^2}\E\bigg[\bigg\|\frac{\gamma N}{m}\sum\limits_{i \in \Ac^r}p_i(\thetab_i^{r+1} - \thetab_0^r - \Delta_i^{r+1})\bigg\|^2\bigg] \label{lem: x0_diff_bd4_0}\\
	= & \frac{1}{\gamma^2}\E[\|{\rm \Bb^r}\|^2] + \E\bigg[\Var_{r}\bigg[\frac{N}{m}\sum_{i \in \Ac^r}p_i(\thetab_0^r - \wt \thetab_{i}^{r+1})\bigg]\bigg]  + \frac{N^2}{m^2}\E\bigg[\bigg\|\sum\limits_{i \in \Ac^r}p_i(\thetab_i^{r+1} - \thetab_0^r - \Delta_i^{r+1})\bigg\|^2\bigg] \label{lem: x0_diff_bd4} \\
	= & \frac{1}{\gamma^2}\E[\|{\rm \Bb^r}\|^2] + \E\bigg[\E_{r}	\bigg[\bigg\|\frac{N}{m}\sum_{i \in \Ac^r}p_i(\thetab_{0}^r - \wt \thetab_{i}^{r+1})\bigg\|^2\bigg]  - \bigg\|\E_{r}\bigg[\frac{N}{m}\sum_{i \in \Ac^r}p_i(\thetab_{0}^r - \wt \thetab_{i}^{r+1})\bigg]\bigg\|^2\bigg] \notag \\
	&+ \frac{N^2}{m^2}\E\bigg[\bigg\|\sum\limits_{i \in \Ac^r}p_i(\thetab_i^{r+1} - \thetab_0^r - \Delta_i^{r+1})\bigg\|^2\bigg] \label{lem: x0_diff_bd5} \\
	= &  \frac{1}{\gamma^2}\E[\|{\rm \Bb^r}\|^2] + \E\bigg[\bigg\|\frac{N}{m}\sum_{i \in \Ac^r}p_i(\thetab_{0}^r - \wt \thetab_{i}^{r+1})\bigg\|^2\bigg] - \E \bigg[\bigg\|\eta\sum_{i=1}^{N} p_i\wt E_i^r\sum_{t= 0}^{E_i^r - 1}\frac{b_i^{r, t}}{\|\bb_i^r\|_1} (\nabla f_i(\wt \thetab_i^{r,t})- \cb_i^r)\bigg\|^2\bigg] \notag \\
	&+ \frac{N^2}{m^2}\E\bigg[\bigg\|\sum\limits_{i \in \Ac^r}p_i(\thetab_i^{r+1} - \thetab_0^r - \Delta_i^{r+1})\bigg\|^2\bigg], \label{lem: x0_diff_bd6}
\end{align}where $\Var_{r}$ and $\E_{r}$ denotes the variance and expectation conditional on the history up to round $r$; \eqref{lem: x0_diff_bd4_0} holds due to the independence between client sampling and random quantization and Assumption 3; \eqref{lem: x0_diff_bd6} follows because of \eqref{lem: x0_diff_bd3} and Assumption 2. 

To bound the second term in the RHS of \eqref{lem: x0_diff_bd6}, we have
\begin{align}
	\E\bigg[\bigg\|\frac{N}{m}\sum_{i \in \Ac^r}p_i(\thetab_{0}^r - \wt \thetab_{i}^{r+1})\bigg\|^2\bigg] 
	= &	\frac{N^2}{m^2} \E\bigg[\bigg\|\sum_{i=1}^{N}\oneb_{i}^{\Ac^r}p_i\eta \wt E_i^r\sum_{t= 0}^{E_i^r - 1}\frac{b_i^{r, t}}{\|\bb_i^r\|_1} (g_i(\wt \thetab_i^{r,t})- \cb_i^r) \bigg\|^2\bigg] \notag \\
	= &	\frac{\eta^2N^2}{m^2} \E\bigg[\bigg\|\sum_{i=1}^{N}p_i\oneb_{i}^{\Ac^r}\wt E_i^r\sum_{t= 0}^{E_i^r - 1}\frac{b_i^{r, t}}{\|\bb_i^r\|_1} (g_i(\wt \thetab_i^{r,t})- \cb_i^r)\bigg)\bigg\|^2\bigg] \notag \\
	= &\frac{\eta^2N^2}{m^2} \underbrace{ \E\bigg[\bigg\|\sum_{i=1}^{N}p_i\oneb_{i}^{\Ac^r}\wt E_i^r\sum_{t= 0}^{E_i^r - 1}\frac{b_i^{r, t}}{\|\bb_i^r\|_1} (g_i(\wt \thetab_i^{r,t})- \nabla f_i(\wt \thetab_i^{r,t})\bigg\|^2\bigg] }_{\rm (a)}\notag \\
	& + \frac{\eta^2N^2}{m^2} \underbrace{ \E\bigg[\bigg\|\sum_{i=1}^{N}p_i\oneb_{i}^{\Ac^r}\wt E_i^r\sum_{t= 0}^{E_i^r - 1}\frac{b_i^{r, t}}{\|\bb_i^r\|_1} (\nabla f_i(\wt \thetab_i^{r,t}) - \cb_{i}^r)\bigg\|^2\bigg]}_{\rm (b)}, \label{lem: x0_diff_bd7}
\end{align}where $\oneb_{i}^{\Ac^r}$ is the indicator function which equals to one if the event $i \in \Ac^r$ is true and zero otherwise;\eqref{lem: x0_diff_bd7} follows by Assumption 2. Then we proceed to bound the two terms $\rm (a)$, $\rm (b)$ and $\rm (c)$. In particular, we can bound $\rm (a)$ by
\begin{small}
	\begin{align}
		{\rm (a)}
		= &\E \bigg[\sum_{i=1}^{N}\bigg\|p_i\oneb_{i}^{\Ac^r}\wt E_i^r\sum_{t= 0}^{E_i^r - 1}\frac{b_i^{r, t}}{\|\bb_i^r\|_1} (g_i(\wt \thetab_i^{r,t})- \nabla f_i(\wt \thetab_i^{r,t}))\bigg\|^2  \notag \\
		&~~+\sum_{i \ne j } \bigg\langle\oneb_{i}^{\Ac^r}p_i\wt E_i^r\sum_{t= 0}^{E_i^r - 1}\frac{b_i^{r, t}}{\|\bb_i^r\|_1} (g_i(\wt \thetab_i^{r,t})- \nabla f_i(\wt \thetab_i^{r,t}), \oneb_{j}^{\Ac^r}p_i\wt E_i^r \sum_{t= 0}^{E_j^r - 1}\frac{b_j^{r, t}}{\|\bb_j^r\|_1} (g_j(\wt \thetab_j^{r,t})- \nabla f_j(\wt \thetab_j^{r,t})\bigg \rangle \bigg]  \\	
		= & \E \bigg[\sum_{i=1}^{N}\oneb_{i}^{\Ac^r}\bigg\|p_i\wt E_i^r\sum_{t= 0}^{E_i^r - 1}\frac{b_i^{r, t}}{\|\bb_i^r\|_1} (g_i(\wt \thetab_i^{r,t})- \nabla f_i(\wt \thetab_i^{r,t}))\bigg\|^2\bigg] \\
		= & \frac{m}{N} \sum_{i=1}^{N} (p_i\wt E_i^r)^2\sum_{t= 0}^{E_i^r - 1}\frac{(b_i^{r, t})^2}{\|\bb_i^r\|_1^2} \E[\|g_i(\wt \thetab_i^{r,t})- \nabla f_i(\wt \thetab_i^{r,t})\|^2]\\
		\leq & \frac{m\sigma^2}{N S} \sum_{i=1}^{N} (p_i\wt E_i^r)^2\sum_{t= 0}^{E_i^r - 1}\frac{(b_i^{r, t})^2}{\|\bb_i^r\|_1^2} \\
		= & \frac{m\sigma^2}{N S} \sum_{i=1}^{N} p_i^2\|\bb_i^r\|_2^2, \label{lem: x0_diff_bd8}
	\end{align}
\end{small}where \eqref{lem: x0_diff_bd8} follows because $\wt E_i^r = \|\bb_i^r\|_1$.
We can also bound $\rm (b)$ by
\begin{small}
	\begin{align}
		{\rm (b)} 
		= &\E \bigg[\sum_{i=1}^{N}\bigg\|p_i\oneb_{i}^{\Ac^r}\wt E_i^r\sum_{t= 0}^{E_i^r - 1}\frac{b_i^{r, t}}{\|\bb_i^r\|_1} (\nabla f_i(\wt \thetab_i^{r,t}) - \cb_i^r)\bigg\|^2  \notag \\
		&~~+\sum_{i \ne j } \bigg\langle p_i\oneb_{i}^{\Ac^r}\wt E_i^r\sum_{t= 0}^{E_i^r - 1}\frac{b_i^{r, t}}{\|\bb_i^r\|_1} ( \nabla f_i(\wt \thetab_i^{r,t}) - \cb_i^r, p_j\oneb_{j}^{\Ac^r}\wt E_i^r\sum_{t= 0}^{E_j^r - 1}\frac{b_j^{r, t}}{\|\bb_j^r\|_1} ( \nabla f_j(\wt \thetab_j^{r,t} - \cb_j^r)\bigg \rangle \bigg] \notag \\	
		= & \frac{m}{N} \sum_{i=1}^{N}(p_i\wt E_i^r)^2 \E \bigg[\bigg\|\sum_{t= 0}^{E_i^r - 1}\frac{b_i^{r, t}}{\|\bb_i^r\|_1} (\nabla f_i(\wt \thetab_i^{r,t}) - \cb_i^r)\bigg\|^2  \notag \\
		&~~+\sum_{i \ne j } \Pr(i, j \in \Ac^r)\bigg\langle p_i\wt E_i^r\sum_{t= 0}^{E_i^r - 1}\frac{b_i^{r, t}}{\|\bb_i^r\|_1} ( \nabla f_i(\wt \thetab_i^{r,t}) - \cb_i^r), p_j\wt E_j^r\sum_{t= 0}^{E_j^r - 1}\frac{b_j^{r, t}}{\|\bb_j^r\|_1} ( \nabla f_j(\wt \thetab_j^{r,t} - \cb_j^r)\bigg \rangle \bigg] \notag \\	
		= & \frac{m}{N} \sum_{i=1}^{N} (p_i\wt E_i^r)^2\E \bigg[\bigg\|\sum_{t= 0}^{E_i^r - 1}\frac{b_i^{r, t}}{\|\bb_i^r\|_1} (\nabla f_i(\wt \thetab_i^{r,t}) - \cb_i^r)\bigg\|^2  \notag \\
		&~~+\frac{m(m-1)}{N(N-1)}\sum_{i \ne j } \bigg\langle p_i\wt E_i^r\sum_{t= 0}^{E_i^r - 1}\frac{b_i^{r, t}}{\|\bb_i^r\|_1} ( \nabla f_i(\wt \thetab_i^{r,t}) - \cb_i^r), p_j\wt E_j^r\sum_{t= 0}^{E_j^r - 1}\frac{b_j^{r, t}}{\|\bb_j^r\|_1}  \nabla f_j(\wt \thetab_j^{r,t} - \cb_j^r)\bigg \rangle \bigg]\label{lem: x0_diff_bd9} \\
		= & \frac{m}{N} \sum_{i=1}^{N} (p_i\wt E_i^r)^2\E \bigg[\bigg\|\sum_{t= 0}^{E_i^r - 1}\frac{b_i^{r, t}}{\|\bb_i^r\|_1} (\nabla f_i(\wt \thetab_i^{r,t}) - \cb_i^r)\bigg\|^2  \notag \\
		&~~+\frac{m(m-1)}{N(N-1)}\sum_{i \ne j } \bigg(\frac{1}{2}\bigg\|p_i\wt E_i^r\sum_{t= 0}^{E_i^r - 1}\frac{b_i^{r, t}}{\|\bb_i^r\|_1} ( \nabla f_i(\wt \thetab_i^{r,t}) - \cb_i^r)\bigg\|^2 +\frac{1}{2}\bigg\|p_j\wt E_j^r\sum_{t= 0}^{E_j^r - 1}\frac{b_j^{r, t}}{\|\bb_j^r\|_1} ( \nabla f_j(\wt \thetab_j^{r,t}) - \cb_j^r)\bigg\|^2\notag \\
		&~~~~~~~~-\frac{1}{2}\bigg\|p_i\wt E_i^r\sum_{t= 0}^{E_i^r - 1}\frac{b_i^{r, t}}{\|\bb_i^r\|_1} ( \nabla f_i(\wt \thetab_i^{r,t}) - \cb_i^r) - p_j\wt E_j^r\sum_{t= 0}^{E_j^r - 1}\frac{b_j^{r, t}}{\|\bb_j^r\|_1} ( \nabla f_j(\wt \thetab_j^{r,t}) - \cb_j^r)\bigg\|^2\bigg) \bigg] \label{lem: x0_diff_bd10}\\
		= & \frac{m^2}{N} \sum_{i=1}^{N} (p_i\wt E_i^r)^2\E \bigg[\bigg\|\sum_{t= 0}^{E_i^r - 1}\frac{b_i^{r, t}}{\|\bb_i^r\|_1} (\nabla f_i(\wt \thetab_i^{r,t}) - \cb_i^r)\bigg\|^2 \notag \\
		& - \frac{m(m-1)}{2N(N-1)} \sum_{i \ne j}\bigg\|p_i\wt E_i^r\sum_{t= 0}^{E_i^r - 1}\frac{b_i^{r, t}}{\|\bb_i^r\|_1} ( \nabla f_i(\wt \thetab_i^{r,t}) - \cb_i^r) -p_j\wt E_j^r \sum_{t= 0}^{E_j^r - 1}\frac{b_j^{r, t}}{\|\bb_j^r\|_1} ( \nabla f_j(\wt \thetab_j^{r,t}) - \cb_j^r)\bigg\|^2 \bigg] \label{lem: x0_diff_bd11}\\
		= & \frac{m(N-m)}{N(N-1)} \sum_{i=1}^{N} (p_i\wt E_i^r)^2\E \bigg[\bigg\|\sum_{t= 0}^{E_i^r - 1}\frac{b_i^{r, t}}{\|\bb_i^r\|_1} (\nabla f_i(\wt \thetab_i^{r,t}) - \cb_i^r)\bigg\|^2\bigg] \notag \\
		&+ \frac{m(m-1)}{N(N-1)} \E\bigg[\bigg\|\sum_{i=1}^{N}p_i\wt E_i^r\sum_{t= 0}^{E_i^r - 1}\frac{b_i^{r, t}}{\|\bb_i^r\|_1} (\nabla f_i(\wt \thetab_i^{r,t}) - \cb_i^r)\bigg\|^2\bigg], \label{lem: x0_diff_bd12}
	\end{align}
\end{small}where \eqref{lem: x0_diff_bd9} holds because $\Pr(i, j \in \Ac^r) = \frac{m(m-1)}{N(N-1)}$; \eqref{lem: x0_diff_bd10} holds because $\langle \ab, \bb \rangle  = \frac{1}{2}\|\ab\|^2 + \frac{1}{2}\|\bb\|^2 - \frac{1}{2} \|\ab - \bb\|^2, \forall \ab, \bb \in \Rbb^n$; \eqref{lem: x0_diff_bd11}  holds because $\sum_{i \ne j}^{N} (\|\ab_i\|^2 + \|\ab_j\|^2) = 2 (N-1)\sum_{i = 1}^{N} \|\ab_i\|^2$; \eqref{lem: x0_diff_bd12} holds because $\frac{1}{2} \sum_{i \ne j}^{N} \|\ab_i - \ab_j\|^2 = N\sum_{i = 1}^{N}\|\ab_i\|^2 - \|\sum_{i=1}^{N}\ab_i\|^2$. Note that 
\begin{align}
	&\E \bigg[\bigg\|\sum_{t= 0}^{E_i^r - 1}\frac{b_i^{r, t}}{\|\bb_i^r\|_1} (\nabla f_i(\wt \thetab_i^{r,t}) - \cb_i^r)\bigg\|^2\bigg] \notag \\
	= & \E \bigg[\bigg\|\sum_{t= 0}^{E_i^r - 1}\frac{b_i^{r, t}}{\|\bb_i^r\|_1} (\nabla f_i(\wt \thetab_i^{r,t}) - \nabla f_i(\thetab^{r}) + \nabla f_i(\thetab^{r}) - \cb_i^r)\bigg\|^2\bigg] \notag \\
	\leq & 2 \E \bigg[\bigg\|\sum_{t= 0}^{E_i^r - 1}\frac{b_i^{r, t}}{\|\bb_i^r\|_1} (\nabla f_i(\wt \thetab_i^{r,t}) - \nabla f_i(\thetab^{r}))\bigg\|^2\bigg]  + 2\E [\|\nabla f_i(\thetab^{r}) - \cb_i^r\|^2] \label{lem: x0_diff_bd13}\\
	\leq & 2 L^2 \sum_{t= 0}^{E_i^r - 1}\frac{b_i^{r, t}}{\|\bb_i^r\|_1} \E[\| \wt \thetab_i^{r,t} - \thetab^{r}\|^2 + 2 \Xi_i^r \label{lem: x0_diff_bd14}\\
	= &2 L^2 \Psi_i^r + 2 \Xi_i^r, \label{lem: x0_diff_bd15}
\end{align}where \eqref{lem: x0_diff_bd13} holds due the Jensen's Inequality; \eqref{lem: x0_diff_bd14} holds because of the convexity of $\|\cdot\|^2$ and Assumption 2. Substituting \eqref{lem: x0_diff_bd15} into \eqref{lem: x0_diff_bd12} yields
\begin{align}
	{\rm (b)} 
	\leq&  \frac{m(m-1)}{N(N-1)} \E\bigg[\bigg\|\sum_{i=1}^{N}p_i\wt E_i^r\sum_{t= 0}^{E_i^r - 1}\frac{b_i^{r, t}}{\|\bb_i^r\|_1} (\nabla f_i(\wt \thetab_i^{r,t}) - \cb_i^r)\bigg\|^2\bigg] \notag \\
	&+\frac{2 m(N-m)L^2}{N(N-1)} \sum_{i=1}^{N}(p_i\wt E_i^r)^2\Psi_i^r + \frac{2 m(N-m)}{N(N-1)} \sum_{i=1}^{N} (p_i\wt E_i^r)^2 \Xi_i^r.  \label{lem: x0_diff_bd16}
\end{align} Using the bounds of the terms $\rm (a)$, $\rm (b)$ and $\rm (c)$, we have from \eqref{lem: x0_diff_bd7} that
\begin{align}
	&\E\bigg[\bigg\|\frac{N}{m}\sum_{i \in \Ac^r}p_i(\thetab_{0}^r - \wt \thetab_{i}^{r+1})\bigg\|^2\bigg] \notag \\
	\leq & \frac{\eta^2\sigma^2N}{m S} \sum_{i=1}^{N} p_i^2\|\bb_i^r\|_2^2 + \frac{2N (N-m)\eta^2}{m(N-1)} \sum_{i=1}^{N}(p_i\wt E_i^r)^2\Xi_i^r+ \frac{2N(N-m)L^2\eta^2}{m(N-1)} \sum_{i=1}^{N}(p_i\wt E_i^r)^2\Psi_i^r\notag \\
	&+ \frac{(m-1)N\eta^2}{m(N-1)} \E\bigg[\bigg\|\sum_{i=1}^{N}p_i\wt E_i^r\sum_{t= 0}^{E_i^r - 1}\frac{b_i^{r, t}}{\|\bb_i^r\|_1} (\nabla f_i(\wt \thetab_i^{r,t}) - \cb_i^r)\bigg\|^2\bigg]. \label{lem: x0_diff_bd17}
\end{align}

To bound the third term in the RHS of \eqref{lem: x0_diff_bd6}, we have
\begin{align}
	&\frac{N^2}{m^2}\E\bigg[\bigg\|\sum\limits_{i \in \Ac^r}p_i(\thetab_i^{r+1} - \thetab_0^r - \Delta_i^{r+1})\bigg\|^2\bigg] \notag \\
	= &\frac{N^2}{m^2}\E\bigg[\bigg\|\sum\limits_{i =1}^{N}\oneb_{i}^{\Ac^r}p_i(\wt\thetab_i^{r+1} - \thetab_0^r - \wt \Delta_i^{r+1})\bigg\|^2\bigg] \\
	= & \frac{N^2}{m^2}\E\bigg[\sum\limits_{i =1}^{N}\oneb_{i}^{\Ac^r} p_i^2\|\wt\thetab_i^{r+1} - \thetab_0^r - \wt \Delta_i^{r+1}\|^2\bigg]\label{lem: x0_diff_bd18}\\
	= & \frac{N}{m} \sum_{i=1}^{N}\omega_i^rp_i^2\E[\|\wt\thetab_i^{r+1}-\thetab_0^r\|^2] \label{lem: x0_diff_bd18_1}\\
	= & \frac{ \eta^2N}{m} \sum_{i=1}^{N}\omega_i^r(p_i\wt E_i^r)^2\E\bigg[\bigg\|\sum_{t= 0}^{E_i^r - 1}\frac{b_i^{r, t}}{\|\bb_i^r\|_1} (g_i(\wt \thetab_i^{r,t}) - \cb_i^r)\bigg\|^2\bigg] \label{lem: x0_diff_bd18_2}\\
	= & \frac{\eta^2N}{m} \sum_{i=1}^{N}\omega_i^r(p_i\wt E_i^r)^2\E\bigg[\bigg\|\sum_{t= 0}^{E_i^r - 1}\frac{b_i^{r, t}}{\|\bb_i^r\|_1} (g_i(\wt \thetab_i^{r,t}) - \nabla f_i(\wt \thetab_i^{r,t}) +\nabla f_i(\wt \thetab_i^{r,t}) - \nabla f_i(\thetab^{r} )+\nabla f_i(\wt \thetab^{r} ) - \cb_i^r)\bigg\|^2\bigg] \notag \\
	\leq & \frac{\eta^2N}{m} \sum_{i=1}^{N}\omega_i^r(p_i\wt E_i^r)^2\E\bigg[\bigg\|\sum_{t= 0}^{E_i^r - 1}\frac{b_i^{r, t}}{\|\bb_i^r\|_1} (g_i(\wt \thetab_i^{r,t}) - \nabla f_i(\wt \thetab_i^{r,t}))\bigg\|\bigg] \notag \\
	& + \frac{2 \eta^2}{mN} \sum_{i=1}^{N}\omega_i^r(p_i\wt E_i^r)^2\E\bigg[\bigg\|\sum_{t= 0}^{E_i^r - 1}\frac{b_i^{r, t}}{\|\bb_i^r\|_1} (\nabla f_i(\wt \thetab_i^{r,t}) - \nabla f_i(\thetab^r))\bigg\|^2\bigg] \notag \\
	& + \frac{2\eta^2}{mN} \sum_{i=1}^{N}\omega_i^r(p_i\wt E_i^r)^2\E[\| \nabla f_i(\thetab^{r}) - \cb_i^r\|^2]\label{lem: x0_diff_bd18_3}\\
	\leq & \frac{\eta^2\sigma^2N}{mS}\sum_{i=1}^{N}\omega_i^rp_i^2\|\bb_i^r\|_2^2 + \frac{2N\eta^2L^2}{m}\sum_{i=1}^{N}\omega_i^r(p_i\wt E_i^r)^2 \Psi_i^r + \frac{2N\eta^2}{m}\sum_{i=1}^{N}\omega_i^r(p_i\wt E_i^r)^2 \Xi_i^r, \label{lem: x0_diff_bd19}
\end{align}where \eqref{lem: x0_diff_bd18} holds by Assumption 3; \eqref{lem: x0_diff_bd18_1} holds because $\Pr(i \in \Ac^r) = \frac{m}{N}$; \eqref{lem: x0_diff_bd18_2} and \eqref{lem: x0_diff_bd19} follow thanks to Assumption 2.
Lastly, we substitute \eqref{lem: x0_diff_bd17} and \eqref{lem: x0_diff_bd19} into \eqref{lem: x0_diff_bd6}, which yields
\begin{align}
	&\E[\|\thetab^{r+1} - \thetab^r\|^2] \notag \\
	\leq & \frac{1}{\gamma^2}\E[\|{\rm \Bb^r}\|^2] +  \frac{\eta^2\sigma^2N}{m S} \sum_{i=1}^{N}(1 + \omega_i^r)p_i^2 \|\bb_i^r\|_2^2 + \frac{(m-N)\eta^2}{mN(N-1)} \E\bigg[\bigg\|\sum_{i=1}^{N}p_i\wt E_i^r\sum_{t= 0}^{E_i^r - 1}\frac{b_i^{r, t}}{\|\bb_i^r\|_1} (\nabla f_i(\wt \xb_i^{r,t}) - \cb_i^r)\bigg\|^2\bigg] \notag \\
	& + \eta^2\sum_{i=1}^{N}\bigg(\frac{2(1+\omega_i^r)N}{m} - \frac{2(m-1)N}{m(N-1)}\bigg) (p_i\wt E_i^r)^2\Xi_i^r  +  \eta^2L^2\sum_{i=1}^{N}\bigg(\frac{2(1+\omega_i^r)N}{m} - \frac{2(m-1)N}{m(N-1)}\bigg) (p_i\wt E_i^r)^2\Psi_i^r \notag \\
	\leq & \frac{1}{\gamma^2}\E[\|{\rm \Bb^r}\|^2] +  \frac{\eta^2\sigma^2N}{m S} \sum_{i=1}^{N} (1+\omega_i^r)p_i^2\|\bb_i^r\|_2^2  + \frac{2\eta^2N}{m} \sum_{i=1}^{N}(1+\omega_i^r)(p_i\wt E_i^r)^2\Xi_i^r \notag \\
	& +  \frac{2\eta^2L^2N}{m} \sum_{i=1}^{N}(1+\omega_i^r)(p_i\wt E_i^r)^2\Psi_i^r.
\end{align}
This completes the proof.
\hfill $\blacksquare$

\vfill

\end{document}